\pdfoutput=1

\documentclass[camera,letterpaper,nomarginnotes,nonarrowgutter]{jpaper}
    
\PassOptionsToPackage{hyphens}{url}
\usepackage[bookmarks=true,breaklinks=true,colorlinks,linkcolor=black,citecolor=black,urlcolor=blue]{hyperref}

\usepackage{marginnote}

\usepackage[normalem]{ulem}
\usepackage{listings}
\usepackage{subcaption}
\usepackage{tikz}
\usepackage{flushend}
\usepackage[noadjust]{cite}
\usepackage{xargs} 
\usepackage{mathtools}
\usepackage{enumitem}
\usepackage{booktabs}
\usepackage{multirow}
\usepackage[many]{tcolorbox}
\usepackage{longfbox}
\usepackage{dblfloatfix} 
\usepackage[ruled,vlined,linesnumbered]{algorithm2e}

\usepackage{amsmath,amssymb}
\usepackage{algorithmic}
\usepackage{graphicx}
\usepackage{textcomp}
\usepackage[table]{xcolor}
\usepackage{pifont}
\usepackage[acronym,nonumberlist,nowarn]{glossaries}
\glsdisablehyper
\newacronym{ADI}{ADI}{Alternating Direction Implicit}
\newacronym{AGU}{AGU}{Address Generation Unit}
\newacronym[longplural={Access History Tables}]{AHT}{AHT}{Access History Table}
\newacronym{AHT-R}{AHT-R}{Access History Table - Read}
\newacronym{AHT-W}{AHT-W}{Access History Table - Write-Back}
\newacronym{AIP}{AIP}{Access Interval Predictor}
\newacronym{AL}{AL}{Added Latency for column accesses}
\newacronym{ASIC}{ASIC}{Application Specific Integrated Circuit}
\newacronym{AVX}{AVX}{Advanced Vector Extensions}
\newacronym[longplural={Basic Block Vectors}]{BBV}{BBV}{Basic Block Vector}
\newacronym{BHT}{BHT}{Branch History Table}
\newacronym{MABM}{MABM}{mat-aware bit mapping}

\newacronym{HBM}{HBM}{Hybrid Bandwidth Memory}
\newacronym{DDR4}{DDR4}{Double-Data Rate 4}
\newacronym{MIMD}{MIMD}{multiple-instruction multiple-data}
\newacronym{SIMT}{SIMT}{Single Instruction Multiple Threads}
\newacronym{SLP}{SALP}{subarray-level parallelism}
\newacronym{BLP}{BLP}{bank-level parallelism}

\newacronym{NN}{NN}{neural network}
\newacronym{BTB}{BTB}{Branch Target Buffer}
\newacronym{CAS}{CAS}{Column Address Strobe}
\newacronym{AMAT}{AMAT}{Average Memory Access Time}
\newacronym{CCD}{CCD}{Column to Column Delay}
\newacronym{AI}{AI}{Arithmetic Intensity}
\newacronym[longplural={Columnar Database Systems}]{CDBS}{CDBS}{Columnar Database System}
\newacronym[longplural={Chip Multiprocessors}]{CMP}{CMP}{Chip Multiprocessor}
\newacronym{CMT}{CMT}{Chip Multithreading}
\newacronym[longplural={Coarse-Grain Reconfigurable Arrays}]{CGRA}{CGRA}{Coarse-Grain Reconfigurable Array}
\newacronym{C-RAM}{C-RAM}{Computational-RAM}
\newacronym{CWD}{CWD}{Column Write Delay}
\newacronym{DBI}{DBI}{Dynamic Binary Instrumentation}
\newacronym[longplural={Database Management Systems}]{DBMS}{DBMS}{Database Management System}
\newacronym{DDR}{DDR}{double data rate}
\newacronym{VLIW}{VLIW}{very long instruction word}

\newacronym{DEWP}{DEWP}{Dead Line and Early Write-Back Predictor}
\newacronym{DRAM}{DRAM}{Dynamic Random Access Memory}
\newacronym{DSBP}{DSBP}{Dead Sub-Block Predictor}
\newacronym{DSM}{DSM}{Decomposed Storage Manage}
\newacronym{FAW}{FAW}{Four row Activation Window}
\newacronym{FP}{FP}{Floating-Point}
\newacronym{EDP}{EDP}{Energy-Delay Product}
\newacronym{FCFS}{FCFS}{First-Come First-Serve}
\newacronym{FIFO}{FIFO}{Fist In, First Out}
\newacronym{FSM}{FSM}{finite state machine}
\newacronym{FPGA}{FPGA}{Field-Programmable Gate Array}
\newacronym[longplural={Functional Units}]{FU}{FU}{Functional Unit}
\newacronym{GAg}{GAg}{Global Adaptive branch prediction using one Global PHT}
\newacronym{GAs}{GAs}{Global Adaptive branch prediction using per-Set PHT}
\newacronym{GCC}{GCC}{GNU Compiler Collection}
\newacronym{GEMS}{GEMS}{General Execution-driven Multiprocessor Simulator}
\newacronym[longplural={General Purpose Processors}]{GPP}{GPP}{General Purpose Processor}
\newacronym{HMC}{HMC}{Hybrid Memory Cube}
\newacronym{HIVE}{HIVE}{HMC Instruction Vector Extensions}
\newacronym{IATAC}{IATAC}{Inter-Access Time per Access Count}
\newacronym{ILP}{ILP}{Instruction Level Parallelism}
\newacronym{IPC}{IPC}{Instructions per Cycle}
\newacronym{ISA}{ISA}{Instruction Set Architecture}
\newacronym{IRAM}{IRAM}{Intelligent RAM}
\newacronym{KIPS}{KIPS}{Kilo Instructions per Second}
\newacronym{LDS}{LDS}{Linked Data Structure}
\newacronym{LFSR}{LFSR}{Linear Feedback Shift Register}
\newacronym{LFMR}{LFMR}{Last-to-First Miss-Ratio}
\newacronym{MLP}{MLP}{Memory-Level Parallelism}

\newacronym{LLC}{LLC}{last-level cache}
\newacronym{LRU}{LRU}{Least Recently Used}
\newacronym{LSU}{LSU}{Load Store Unit}
\newacronym{LTP}{LTP}{Last-Touch Predictor}
\newacronym{LvP}{LvP}{Live-time Predictor}
\newacronym{LWP}{LWP}{Last Write Predictor}
\newacronym{MARSS}{MARSS}{Micro Architectural and System Simulator}
\newacronym{McPAT}{McPAT}{Multi-core Power, Area, and Timing}
\newacronym{MIPS}{MIPS}{Microprocessor without Interlocked Pipeline Stages}
\newacronym{MOB}{MOB}{Memory Order Buffer}
\newacronym{MMU}{MMU}{memory management unit}
\newacronym{PuD}{PUD}{processing-using-DRAM}
\newacronym{MPKI}{MPKI}{Misses per Kilo-Instruction}
\newacronym{MSHR}{MSHR}{Miss-Status Handling Registers}
\newacronym{NAS}{NAS}{Numerical Aerodynamic Simulation}
\newacronym{NDP}{NDP}{Near-Data Processing}
\newacronym{NMP}{NMP}{Near Memory Processor}
\newacronym{NoC}{NoC}{Network-on-Chip}
\newacronym{NPB}{NPB}{NAS Parallel Benchmark}
\newacronym{NUCA}{NUCA}{Non-Uniform Cache Architecture}
\newacronym{NUMA}{NUMA}{Non-Uniform Memory Access}
\newacronym{OoO}{OoO}{Out-of-Order}
\newacronym{OpenMP}{OpenMP}{Open Multi-Processing}
\newacronym{OS}{OS}{operating system}
\newacronym{PAg}{PAg}{Per-address Adaptive branch prediction using one Global PHT}
\newacronym{PAs}{PAs}{Per-address Adaptive branch prediction using per-Set PHT}
\newacronym{PC}{PC}{Program Counter}
\newacronym[longplural={Pointer-Chasing Engines}]{PCE}{PCE}{Pointer-Chasing Engine}
\newacronym{PCM}{PCM}{Performance Counter Monitor}
\newacronym{PIM}{PIM}{processing-in-memory}
\newacronym[longplural={Pattern History Tables}]{PHT}{PHT}{Pattern History Table}
\newacronym{RAPL}{RAPL}{Running Average Power Limit}
\newacronym{RAS}{RAS}{Row Address Strobe}
\newacronym{RAT}{RAT}{Registers Alias Table}
\newacronym{RC}{RC}{Row Cycle}
\newacronym{RCD}{RCD}{RAS to CAS Delay}
\newacronym[longplural={Row-based Database Systems}]{RDBS}{RDBS}{Row-based Database System}
\newacronym{ROB}{ROB}{Reorder Buffer}
\newacronym{RRD}{RRD}{Row to Row activation Delay}
\newacronym{DNN}{DNN}{Deep Neural Network}
\newacronym{ANN}{ANN}{Artificial Neural Network}
\newacronym{PnM}{PNM}{processing-near-memory}
\newacronym{PuM}{PUM}{processing-using-memory}
\newacronym{FFD}{FFD}{first-fit-decreasing}
\newacronym{HFF}{HFF}{helper flip-flop}
\newacronym{OBPS}{OBPS}{one-bit per-subarray}
\newacronym{ABOS}{ABOS}{all-bits in one-subarray}
\newacronym{ABPS}{APBS}{all-bits per-subarray}
\newacronym{GEMV}{GEMV}{general matrix-vector product}

\newacronym{RBR}{RBR}{redundant binary representation}
\newacronym{RP}{RP}{Row Precharge}
\newacronym{RTL}{RTL}{Register Transfer Level}
\newacronym{RTP}{RTP}{Read To Precharge}
\newacronym{RVU}{RVU}{Reconfigurable Vector Unit}
\newacronym{SDP}{SDP}{Skewed Dead-Block Predictor}
\newacronym{SESC}{SESC}{Superescalar Simulator}
\newacronym{SSE}{SSE}{Streaming SIMD Extensions}
\newacronym{SFP}{SFP}{Spatial Footprint Predictor}
\newacronym{SiNUCA}{SiNUCA}{Simulator of Non-Uniform Cache Architectures}
\newacronym{SMT}{SMT}{Simultaneous Multi-Threading}
\newacronym{SIMD}{SIMD}{single-instruction multiple-data}
\newacronym{LUT}{LUT}{lookup table}

\newacronym{SPEC}{SPEC}{Standard Performance Evaluation Corporation}
\newacronym{SoC}{SoC}{System-on-Chip}
\newacronym{SPP}{SPP}{Spatial Pattern Predictor}
\newacronym{SRAM}{SRAM}{Static Random Access Memory}
\newacronym{SSOR}{SSOR}{Symmetric Successive Over-Relaxation}
\newacronym{SSV}{SSV}{Search Set Vector}
\newacronym{TLB}{TLB}{Translation Look-aside Buffer}
\newacronym{TLP}{TLP}{Thread Level Parallelism}
\newacronym[longplural={Through-Silicon Vias}]{TSV}{TSV}{Through-Silicon Via}
\newacronym{VWQ}{VWQ}{Virtual Write Queue}
\newacronym{WTR}{WTR}{Write Recovery time}
\newacronym{WR}{WR}{Write To Read delay time}
\newacronym{TRA}{TRA}{triple row activation}

\usepackage{xspace}
\usepackage[binary-units=true]{siunitx}
\usepackage{todonotes}
\usepackage{soul}
\usepackage{makecell}
\usepackage{enumitem}
\usepackage{setspace}
\usepackage[compact]{titlesec}
\usepackage[export]{adjustbox}
\usepackage{balance}
\usepackage[us,12hr]{datetime}
\usepackage[en-GB, useregional=numeric]{datetime2}
\usepackage{fancyhdr}

\hypersetup{
  colorlinks = true,
  urlcolor   = blue,
  linkcolor  = blue,
  citecolor  = black
}

\usepackage{cleveref}

\crefformat{section}{\S#2#1#3}
\crefformat{subsection}{\S#2#1#3}
\crefformat{subsubsection}{\S#2#1#3}
\definecolor{commandbg}{gray}{0.96}
\definecolor{commandrule}{gray}{0.55}

\hypersetup{
  colorlinks=true,
  urlcolor=blue,
  linkcolor=black
}

\lstdefinestyle{shell}{
  basicstyle=\ttfamily\footnotesize,
  breaklines=true,
  columns=fullflexible,
  backgroundcolor=\color{commandbg},
  frame=l,
  rulecolor=\color{commandrule},
  framerule=1.2pt,
  framesep=5pt,
  xleftmargin=7pt,
  framexleftmargin=3pt,
  xrightmargin=3pt,
  aboveskip=4pt,
  belowskip=4pt,
  showstringspaces=false
}
\newcommand{\artifact}{\texttt{artifact\_ae}}

\newcommand{\versionnum}[0]{2.0}

\definecolor{MidnightBlue}{rgb}{0.1, 0.1, 0.44}
\definecolor{dollarbill}{rgb}{0.52, 0.73, 0.4}

\newif\ifcameraready
\camerareadytrue

\ifcameraready
    \newcommand{\gfcr}[1]{\textcolor{black}{#1}}
    \newcommand{\gfcri}[1]{\textcolor{black}{#1}} 
     
    \newcommand{\gfcriii}[1]{\textcolor{black}{#1}}

    \newcommand{\omcri}[1]{\textcolor{black}{#1}}
    \newcommand{\omcrii}[1]{\textcolor{black}{#1}}

\else 
    \newcommand{\gfcr}[1]{\textcolor{black}{#1}} 
    \newcommand{\gfcri}[1]{\textcolor{black}{#1}} 
     
    \newcommand{\gfcriii}[1]{\textcolor{black}{#1}}

    \newcommand{\omcri}[1]{\textcolor{BlueViolet}{#1}}
    \newcommand{\omcrii}[1]{\textcolor{red}{#1}}

    \let\marginpar\marginnote
\fi

\newcommand{\li}{(\textit{i})}
\newcommand{\lii}{(\textit{ii})}
\newcommand{\liii}{(\textit{iii})}
\newcommand{\liv}{(\textit{iv})}
\newcommand{\lv}{(\textit{v})}

\definecolor{blush}{rgb}{0.87, 0.36, 0.51}

\newcommand{\prop}{\emph{PipeDRAM}\xspace}

\newcommand\bbop{\emph{bbop}\xspace}

\newcommand\aaps{\texttt{AAP}s/\texttt{AP}s\xspace}

\newcommand\uprog{\textmu{}Program\xspace}
\newcommand\uprogs{\textmu{}Programs\xspace}

\newcommand{\paratitle}[1]{\vspace{4pt}\noindent\textbf{#1.}}

\definecolor{airforceblue}{rgb}{0.36, 0.54, 0.66}
\definecolor{dodgerblue}{rgb}{0.12, 0.56, 1.0}
\definecolor{brandeisblue}{rgb}{0.0, 0.44, 1.0}
\definecolor{brickred}{rgb}{0.8, 0.25, 0.33}
\definecolor{eggplant}{rgb}{0.38, 0.25, 0.32}
\definecolor{byzantium}{rgb}{0.44, 0.16, 0.39}
\definecolor{ddgreen}{rgb}{0.00, 0.50, 0.00}

\definecolor{mygreen}{rgb}{0,0.6,0}
\definecolor{mygray}{rgb}{0.5,0.5,0.5}
\definecolor{mymauve}{rgb}{0.58,0,0.82}

\definecolor{bluehl}{rgb}{0.8,0.874,1}
\definecolor{pinkhl}{rgb}{0.992156863,0.847058824,1}
\definecolor{macaroniandcheese}{rgb}{1.0, 0.74, 0.53}
\definecolor{mossgreen}{rgb}{0.68, 0.87, 0.68}
\definecolor{greenhl}{rgb}{0.835,0.996,0.939}
\definecolor{yellowhl}{rgb}{0.996,0.957,0.8}
\definecolor{palecerulean}{rgb}{0.61, 0.77, 0.89}
\definecolor{gray(x11gray)}{rgb}{0.75, 0.75, 0.75}
\definecolor{amethyst}{rgb}{0.6, 0.4, 0.8}
\definecolor{ao}{rgb}{0.0, 0.5, 0.0}
\definecolor{burntorange}{rgb}{0.8, 0.33, 0.0}

\definecolor{cadmiumorange}{rgb}{0.93, 0.53, 0.18}

\definecolor{frenchlilac}{rgb}{0.53, 0.38, 0.56}
\definecolor{heliotrope}{rgb}{0.87, 0.45, 1.0}
\definecolor{peridot}{rgb}{0.9, 0.89, 0.0}
\definecolor{saffron}{rgb}{0.96, 0.77, 0.19}
\definecolor{tuscanred}{rgb}{0.51, 0.21, 0.21}
\definecolor{uscgold}{rgb}{1.0, 0.8, 0.0}
\definecolor{tangerineyellow}{rgb}{1.0, 0.8, 0.0}
\definecolor{rufous}{rgb}{0.66, 0.11, 0.03}
\definecolor{safetyorange}{rgb}{1.0, 0.4, 0.0}
\newcommand{\tempcommand}[1]{\renewcommand{\arraystretch}{#1}}

\newcommand\ignore[1]{ }
\newcommand{\revdel}[1]{}
\newcommand{\sgdel}[1]{}

 \newcommand{\cmark}{\ding{51}}

\newif\ifmicrosubmission
\microsubmissiontrue
\ifmicrosubmission 
    \newcommand{\gfmicro}[1]{\textcolor{black}{#1}}

    \newcommand{\agymicrocomment}[1]{}
\else
    \newcommand{\gfmicro}[1]{\textcolor{blue}{#1}}

    \newcommand{\agymicrocomment}[1]{\textcolor{red}{\textbf{!!!~Giray:} #1}}
    
\fi

\newif\ifmicrorevision
\microrevisionfalse
\ifmicrorevision 
    \newcommand{\revA}[1]{\textcolor{red}{#1}}
    \newcommand{\revB}[1]{\textcolor{heliotrope}{#1}}
    \newcommand{\revC}[1]{\textcolor{safetyorange}{#1}}
    \newcommand{\revD}[1]{\textcolor{mygreen}{#1}}
    \newcommand{\revE}[1]{\textcolor{rufous}{#1}}
    \newcommand{\revCommon}[1]{\textcolor{blue}{#1}}

    \newcommandx{\changeCM}[2][1=]{\todo[linecolor=blue,backgroundcolor=blue!25,bordercolor=blue,#1,size=\scriptsize]{\revCommon{\textbf{#2}}}}
    
    \newcommandx{\changeA}[2][1=]{\todo[linecolor=red,backgroundcolor=red!25,bordercolor=red,#1,size=\scriptsize]{\revA{\textbf{#2}}}}
    
    \newcommandx{\changeB}[2][1=]{\todo[linecolor=heliotrope,backgroundcolor=heliotrope!25,bordercolor=heliotrope,#1,size=\scriptsize]{\revB{\textbf{#2}}}}
    
    \newcommandx{\changeC}[2][1=]{\todo[linecolor=safetyorange,backgroundcolor=safetyorange!25,bordercolor=safetyorange,#1,size=\scriptsize]{\revC{\textbf{#2}}}}
    
    \newcommandx{\changeD}[2][1=]{\todo[linecolor=mygreen,backgroundcolor=mygreen!25,bordercolor=mygreen,#1,size=\scriptsize]{\revD{\textbf{#2}}}}
    
    \newcommandx{\changeE}[2][1=]{\todo[linecolor=rufous,backgroundcolor=rufous!25,bordercolor=rufous,#1,size=\scriptsize]{\revE{\textbf{#2}}}}

    \newcommand{\revdelmrev}[1]{}
\else
    \newcommand{\revA}[1]{\textcolor{black}{#1}}
    \newcommand{\revB}[1]{\textcolor{black}{#1}}
    \newcommand{\revC}[1]{\textcolor{black}{#1}}
    \newcommand{\revD}[1]{\textcolor{black}{#1}}
    \newcommand{\revE}[1]{\textcolor{black}{#1}}
    \newcommand{\revCommon}[1]{\textcolor{black}{#1}}

    \newcommandx{\changeCM}[2][1=]{\todo[disable,#1]{#2}}
    \newcommandx{\changeA}[2][1=]{\todo[disable,#1]{#2}}
    \newcommandx{\changeB}[2][1=]{\todo[disable,#1]{#2}}
    \newcommandx{\changeC}[2][1=]{\todo[disable,#1]{#2}}
    \newcommandx{\changeD}[2][1=]{\todo[disable,#1]{#2}}
    \newcommandx{\changeE}[2][1=]{\todo[disable,#1]{#2}}

    \newcommand{\revdelmrev}[1]{#1}

\fi

\newif\ifcut
\cutfalse

\ifcut
   \newcommand{\gfcut}[1]{} 
\else
    \newcommand{\gfcut}[1]{#1}
\fi

\newif\ifiscasubmission
\iscasubmissiontrue

\ifiscasubmission
    \newcommand{\gfisca}[1]{#1}
    \newcommand{\gfbisca}[1]{}
    \newcommand{\sg}[1]{#1}
    \newcommand{\sgi}[1]{#1}
    
\else
    \newcommand{\gfbisca}[1]{\textcolor{blue}{\textit{GF: #1}}}
    \newcommand{\gfisca}[1]{\textcolor{blue}{#1}}

    \newcommand{\sg}[1]{\textcolor{red}{#1}}
    \newcommand{\sgi}[1]{\textcolor{brickred}{#1}}

\fi

\newif\ifsubmission
\submissiontrue

\ifsubmission
    
    \newcommand{\juan}[1]{#1}
    \newcommand{\gf}[1]{#1}

    \newcommand{\jgl}[1]{}

    \newcommand{\gfb}[1]{}
    \newcommand{\mayank}[1]{}
    
    \newcommand{\agy}[1]{#1}
    \newcommand{\agycomment}[1]{}
\else
    \newcommand{\jgl}[1]{\textcolor{brickred}{\textit{JGL: #1}}}
    \newcommand{\gfb}[1]{\textcolor{blue}{\textit{GF: #1}}}
    \newcommand{\juan}[1]{\textcolor{brickred}{#1}}
    \newcommand{\mayank}[1]{\textcolor{green}{\textit{Mayank: #1}}}
    
    \newcommand{\gf}[1]{\textcolor{blue}{#1}}

    \newcommand{\agy}[1]{\textcolor{orange}{#1}}
    \newcommand{\agycomment}[1]{\agy{\textbf{[@gy:} #1\textbf{]}}}

\fi

\newif\ifmicroshort
\microshortfalse
\newcommand{\revdelm}[1]{\ifmicroshort #1\fi}

\newcommand{\circlediv}[1]{\tikz[baseline=(char.base)]{\node[shape=circle,draw,inner sep=0pt,fill=black, text=white] (char) {#1};}}

\DontPrintSemicolon
\SetKwInOut{KwIn}{Input}
\SetKwInOut{KwOut}{Output}

\newsavebox{\circledbox}
\newcommand{\circledsize}{1.0em}  
\newcommand{\circledH}{0.64}       
\newcommand{\circledW}{0.80}       
\newcommand{\circledsqueeze}{0.92} 
\newcommand{\circledfont}{\normalfont} 

\NewDocumentCommand{\circled}{O{black} O{white} m}{%
  \begingroup
  \sbox{\circledbox}{\color{#2}\circledfont #3}%
  \edef\circledD{\the\dimexpr\circledsize\relax}%
  \edef\circledsh{\fpeval{\circledH*(\circledD)/(\the\ht\circledbox)}}%
  \edef\circledsw{\fpeval{\circledW*(\circledD)/(\the\wd\circledbox)}}%
  \edef\circledsy{\fpeval{min(\circledsh, \circledsw/\circledsqueeze)}}%
  \edef\circledsx{\fpeval{min(\circledsy, \circledsw)}}%
  \tikz[baseline=(char.base)]{%
    \node[shape=circle, draw=#1, fill=#1,
          line width=0.3pt, inner sep=0pt,
          minimum size=\circledD] (char)
      {\makebox[0pt]{\scalebox{\circledsx}[\circledsy]{\usebox{\circledbox}}}};}%
  \endgroup}

\newcommand\pimdef{\cite{ghose.ibmjrd19, mutlu2020modern,deoliveira2021IEEE,pim-book,mutlu2019processing,mutlu2019enabling,mutlu2015research,mutlu2013memory,loh2013processing,Near-Data,stone1970logic,Miss_Mem_Wall_1996,Kautz1969,mutlu2025memory,mutlu2024memory}\xspace}

\newcommand\pnm{\cite{farmahini2015nda,babarinsa2015jafar,devaux2019true,ghiasi2022genstore,gomez2021benchmarkingcut,gomezluna2021benchmarking,gomez2022benchmarking,syncron,singh2020nero,skhynixpim,ke2021near,giannoula2022sparsep,shin2018mcdram,cho2020mcdram,denzler2021casper,asghari2016chameleon,IRAM_Micro_1997,C_RAM_1999,CASES_MVX,Xi_2015,sun2021abc,matam2019graphssd,gokhale1995processing,hall1999mapping,MEMSYS_MVX,lockerman2020livia,ahn2015scalable,nai2017graphpim,boroumand2018google,lazypim,top-pim,gao2016hrl,kim2018grim,drumond2017mondrian,RVU,NIM,PEI,gao2017tetris,Kim2016,gu2016leveraging,boroumand2019conda,hsieh2016transparent,cali2020genasm,NDC_ISPASS_2014,pattnaik2016scheduling,akin2015data,hsieh2016accelerating,lee2015bssync,boroumand2021mitigating,boroumand2021google,boroumand2022polynesia,boroumand2021polynesia,amiraliphd,besta2021sisa,fernandez2020natsa,singh2019napel,kwon202125,lee2021hardware,niu2022184qps,Sparse_MM_LiM,azarkhish2016logic,azarkhish2018neurostream,guo20143d,de2018design,akin2014hamlet,huang2020heterogeneous,dai2018graphh,liu2018processing,tsai:micro:2018:ams,gu2020ipim,DRAMA_CAL_2014,Asghari-Moghaddam_2016,huang2019active,kersey2017lightweight,li2019pims,kim2018grim,boroumand2017lazypim,zhuo2019graphq,zhang2018graphp,lim2017triple,smc_sim,HIVE,jang2019charon,IBM_ActiveCube,hadidi2017cairo,santos2018processing,yang2026dcc,zhao2026cosm,barkhordar2025alpha,giannoula2024pygim,he2025papi,gu2025pim,rhyner2024pim,10.1093/bioinformatics/btae631,Chi2016,gao2015practical,hashemi2016accelerating,hashemi2016continuous,hassan2015near,liu2017concurrent,herruzo2021enabling,asgarifafnir,upmem2018,Near-Data,jacob2016compiling,lloyd2015memory,nair2015evolution,lenjani2020fulcrum,loh2013processing,acm,hadidi2017demystifying,gu2020dlux,asgari2020mahasim,baskaran2020decentralized,ahmed2019compiler,picorel2017near,min2019neuralhmc,zhou2022flexidram,he2020newton,pugsley2014comparing,devic2022pim,subramaniyan2017parallel,IRAM_WML_1997,NMP_2005,ke2019recnmp,kim2017heterogeneous,zhao2024pim,gu.isca16,kim.sc17,morad.taco15,gomez2022machine,fernandez2022exploiting,oliveira2022heterogeneous,lloyd2018dse,gokhale2015rearr,rodrigues2016scattergather,lloyd2017keyvalue,landgraf2021combining,kim2021aquabolt,lee2022improving,siddique2024architectural,jaiyeoba2023acts,lenjani2022pulley,lenjani2021supporting,zhou2021ultra,sadredini2021sunder,mosanu2022pimulator,singh2021fpga,singh2021accelerating,dai2022dimmining,gomez2023evaluating,gupta2023evaluating,oliveira2023dappa,park2024attacc,seo2024ianus,li2024pim,lopes2024pim,leepresto2024,baekpsyncpim2024,wangndsearch2024,liuisca2024,yueisca2024,tianndpbridge2024,ghiasimegis2024,li2024stream,schwedock2024leviathan,lee2024pim,huo2024pifs,ham2024low,mahapatra2024storage,heo2024neupims}\xspace}

\newcommand\pum{\cite{Chi2016,Shafiee2016,seshadri2017ambit,seshadri2019dram,li2017drisa,seshadri2013rowclone,seshadri2016processing,deng2018dracc,xin2020elp2im,song2018graphr,song2017pipelayer,gao2019computedram,eckert2018neural,aga2017compute,dualitycache,besta2021sisa,seshadri2016buddy,seshadri.bookchapter17,seshadri2018rowclone,seshadri2015fast,li2016pinatubo,ferreira2021pluto,ferreira2022pluto,imani2019floatpim,he2020sparse,flashcosmos,truong2022adapting,truong2021racer,olgun2021quactrng,kim2019d,kim2018dram,bostanci2022dr,olgun2022pidram,ali2019memory,angizi2019graphide,li2018scope,subramaniyan2017parallel,zha2020hyper,fujiki2018memory,orosa2021codic,sharad2013ultra,rezaei2020nom,chang2016low,chang2017understandingphd,hajinazarsimdram,gao2021parabit,lee20223d,si2019dual,simon2020blade,nag2019gencache,wang2019bit,wang2023infinity,deng2019lacc,peng2023chopper,mimdramextended,missingnot,yuksel2024simultaneous,yavits2021giraf,seshadri.thesis16,choi2020flash,han2019novel,merrikh2017high,wang2018three,lue2019optimal,kim2021behemoth,wang2022memcore,han2021flash,kang2021s,lee2020neuromorphic,al2020towards,kang.icassp14,kim2021colonnade,jiang2020c3sram,jeloka201628,kang2015energy,imani2020dual,sutradhar2021look,sutradhar2020ppim,shahroodi2023swordfish,yavits2023drama,jahshan2024majork,khalifa2023clapim,garzon2022aida,hanhan2022edam,morad2016resistive,wu2022dramcam_generalpurpose,sadredini2020flexamata,sadredini2019eap,angstadt2018aspen,wang2016sequential,wang2015association,angizi2018pima,angizi2018cmp,angizi2019dna,levy.microelec14,kvatinsky.tcasii14,kvatinsky.iccd11,kvatinsky.tvlsi14,gaillardon2016plim,bhattacharjee2017revamp,hamdioui2015memristor,xie2015fast,hamdioui2017myth,yu2018memristive,xi2020memory,zheng2016tcam,ma20232,slesazeck20192tnc,wang20211t2c, tokuda2026pudghost, olgun2025dram,oliveira2025proteus,soysal2025mars,mutlu2024memory,yuksel2025pudhammer,tokuda2026clutch}\xspace}

\newcommand\drampum{\cite{ali2019memory,angizi2019graphide,besta2021sisa,bostanci2022dr,deng2018dracc,deng2019lacc,ferreira2021pluto,ferreira2022pluto,gao2019computedram,hajinazarsimdram,jahshan2024majork,li2017drisa,li2018scope,mimdramextended,missingnot,mutlu2024memory,olgun2021quactrng,olgun2022pidram,olgun2025dram,oliveira2025proteus,orosa2021codic,peng2023chopper,seshadri.bookchapter17,seshadri.thesis16,seshadri2013rowclone,seshadri2015fast,seshadri2016buddy,seshadri2016processing,seshadri2017ambit,seshadri2018rowclone,seshadri2019dram,soysal2025mars,tokuda2026pudghost,
wu2022dramcam_generalpurpose,xin2020elp2im,yavits2023drama,yuksel2024simultaneous,yuksel2025pudhammer,tokuda2026clutch}\xspace}

\newcommandx{\unsure}[2][1=]{\todo[linecolor=red,backgroundcolor=red!25,bordercolor=red,#1, size=\tiny]{#2}}
\newcommandx{\change}[2][1=]{\todo[linecolor=blue,backgroundcolor=blue!25,bordercolor=blue,#1,size=\tiny]{\textbf{#2}}}
\newcommandx{\feedback}[2][1=]{\todo[linecolor=yellow,backgroundcolor=yellow!25,bordercolor=yellow,#1]{#2}}
\newcommandx{\improvement}[2][1=]{\todo[linecolor=Plum,backgroundcolor=Plum!25,bordercolor=Plum,#1]{#2}}
\newcommandx{\thiswillnotshow}[2][1=]{\todo[disable,#1]{#2}}
\newcommandx{\completedRevision}[2][1=]{\todo[disable,backgroundcolor=red,#1]{#2}}
\newcommandx{\dataSource}[2][1=]{\todo[disable,backgroundcolor=red,#1]{#2}}
\newcommandx{\info}[2][1=]{\todo[linecolor=dollarbill,backgroundcolor=dollarbill!25,bordercolor=dollarbill,#1, size=\tiny]{#2}}

\newcommand{\boxbegin} {
	\begin{tcolorbox}[enhanced, frame hidden, colback=gray!50, breakable]
}

\newcommand{\boxend} {
	\end{tcolorbox}
}

\ifcameraready
    \fancypagestyle{cameraready}{%
      \fancyhf{}%
      \fancyfoot[C]{\thepage}%
    }
\else
    \fancypagestyle{firstpage}{%
        \fancyhead{}%
        \fancyhead[C]{\textcolor{red}{CONFIDENTIAL DRAFT -- DO NOT DISTRIBUTE -- TO APPEAR IN MICRO'26} \\
                      \textcolor{MidnightBlue}{\emph{Version \versionnum~---~\today, \ampmtime}}}%
    }
\fi

\makeatletter
\def\bstctlcite{\@ifnextchar[{\@bstctlcite}{\@bstctlcite[@auxout]}}
\def\@bstctlcite[#1]#2{\@bsphack
  \@for\@citeb:=#2\do{%
    \edef\@citeb{\expandafter\@firstofone\@citeb}%
    \if@filesw\immediate\write\csname #1\endcsname{\string\citation{\@citeb}}\fi}%
  \@esphack}
\makeatother 

\begin{document}
\bstctlcite{IEEEexample:BSTcontrol}

\title{\scalebox{0.945}{\prop: A Data-Transposition-Free Processing-Using-DRAM}\\\scalebox{0.945}{Architecture with Hardware/Software Pipelining}}

\author{
\scalebox{0.9}{Geraldo F. Oliveira$^\dagger$}~\quad
\scalebox{0.9}{Ataberk Olgun$^\ddagger$$^\star$}~\quad
\scalebox{0.9}{Ismail Emir Yüksel$^\ddagger$$^\star$}~\quad
\scalebox{0.9}{F. Nisa Bostancı$^\ddagger$$^\star$}\\
\scalebox{0.9}{Pedro H. E. Becker$^\dagger$}~\quad
\scalebox{0.9}{Mohammad Sadrosadati$^\ddagger$$^\star$}~\quad
\scalebox{0.9}{Saugata Ghose$^\star$}~\quad
\scalebox{0.9}{Juan Gómez-Luna$^\star$}~\quad 
\scalebox{0.9}{Onur Mutlu$^\ddagger$$^\star$}\vspace{10pt}
\\
\scalebox{0.9}{$^\dagger$~\emph{Huawei Research, Zürich}} \qquad 
\scalebox{0.9}{$^\ddagger$~\emph{ETH Zürich}} \qquad 
\scalebox{0.9}{$^\star$~\emph{SAFARI Research Group}} 
\vspace{5pt}
}

\maketitle

\ifcameraready
  \thispagestyle{cameraready}
\else
  \thispagestyle{firstpage}
\fi

\begin{abstract}



\revdelmrev{\Gls{PuD} architectures exploit the analog operational properties of DRAM to perform bulk bitwise \gf{Boolean} and arithmetic operations inside memory arrays by organizing data in a vertical layout, where operand bits are stacked along DRAM columns. 
However, modern computing systems natively employ a horizontal data layout that preserves the cache line abstraction\agy{, leverages spatial locality in row buffers,} and enables high memory throughput. This fundamental mismatch forces existing \gls{PuD} architectures to \agy{frequently} perform 
data layout transformations between horizontal and vertical formats, incurring significant performance, energy, and system integration overheads. Our \emph{goal}
is to eliminate data transposition overheads in \gls{PuD} systems at low cost.
%
To this end,} \omcri{we} propose PipeDRAM, a \gls{PuD} architecture that eliminates the need for runtime data layout transformation, enabling 
\agy{\gls{PuD} operations} directly over horizontally laid-out data.
\prop's key ideas are to 
\li~deterministically reorganize bits \omcri{inside} each memory request to enable a \gls{PuD}-friendly data placement within a DRAM array in a horizontal data layout, and 
\lii~employ a pipeline-based execution model that overlaps bit-dependent and bit-independent in-DRAM operations to exploit bit-level parallelism across the memory array. 
\revdelm{Based on these key ideas, PipeDRAM implements three main components.
First, \prop's \emph{mat-aware bit mapping (MABM) mechanism}, placed in the memory controller, performs deterministic bit-level data reorganization while preserving the conventional cache line abstraction and avoiding additional data movement. 
Second, PipeDRAM deploys a software-assisted pipeline execution model that statically schedules bit-serial/bit-parallel \gls{PuD} primitives across the memory array to sustain high throughput with simple control logic. 
Third, PipeDRAM's \emph{hierarchical in-memory data copy scheme}, within a DRAM bank, separates latency-critical carry propagation from non-critical intermediate data transfers during \gls{PuD} execution, ensuring that communication does \emph{not} dominate execution time.}
We compare PipeDRAM to different \revdelm{processor-centric and memory-centric }computing platforms.
PipeDRAM provides 
\li~11.8$\times$, 11.8$\times$, and 80.4$\times$ \omcri{higher} performance and
\lii~25.4$\times$, 3.0$\times$, and 38.0$\times$ lower energy consumption than three state-of-the-art \gls{PuD} systems\revdelm{ (i.e., SIMDRAM, MIMDRAM, and \emph{Proteus}, respectively), on average across twelve real-world applications}.
PipeDRAM incurs low area cost on top of a DRAM chip (1.86\%) and CPU die (0.05\%).
\omcrii{To enable further research on \gls{PuD} systems, we open-source PipeDRAM at \url{https://github.com/CMU-SAFARI/PipeDRAM}.}

\end{abstract}

\glsresetall

\glsresetall

\section{Introduction}
\label{sec:introduction}

\Gls{PIM}~\omcri{\pimdef} is a paradigm that alleviates the ever-growing cost of moving data\revdel{ back-and-forth} \juan{between} computing (e.g., CPU, GPU) and memory (e.g., DRAM) elements. 
In \gls{PIM} architectures, computation is done by adding logic \emph{near} memory arrays, i.e., \gls{PnM}~\omcri{\pnm}, or by \emph{using} the analog properties of the memory arrays, i.e., \gls{PuM}~\omcri{\pum}). 
Many prior works~\omcri{\drampum} demonstrate the 
\juan{feasibility of} \gls{PuD}, which uses DRAM cells to implement 
\li~simple \emph{\gls{PuD} primitives}, such as in-DRAM row copy~\omcri{\cite{seshadri2013rowclone,seshadri2018rowclone,gao2019computedram,olgun2022pidram,yuksel2024simultaneous}} and Boolean/majority~\omcri{\cite{li2017drisa,seshadri2017ambit,xin2020elp2im,seshadri2015fast,seshadri2019dram,li2016pinatubo, missingnot}} logic, and 
\lii~complex \emph{\gls{PuD} operations}, such as arithmetic~\omcri{\cite{hajinazarsimdram,deng2018dracc,angizi2019graphide,li2018scope,peng2023chopper,mimdramextended, oliveira2025proteus}} and \gls{LUT}-based~\omcri{\cite{ferreira2021pluto, ferreira2022pluto,tokuda2026clutch,deng2019lacc,sutradhar2020ppim,zhou2022red,zhou2022lt}} operations, implemented by composing \gls{PuD} primitives.
\gls{PuD} architectures commonly employ a bit-serial \gf{\gls{SIMD}} execution model to implement arithmetic and logic operations inside DRAM arrays. 
To enable efficient carry propagation and bitwise computation, such architectures organize data in a vertical layout: \gf{for an input array with $M$ data elements, each of which with $N$ bits, this layout spans $M$ columns and $N$ rows, where} each DRAM row corresponds to a single bit-position of all $M$ data elements.
Multiple inputs arrays are placed in different DRAM rows and aligned within the same DRAM column. 

\revdelmrev{\gf{The vertical data layout mitigates two key issues that arise when performing arithmetic \gls{PuD} operations under the conventional horizontal data organization that modern computing systems natively employ to enable high memory throughput.
First, conventional DRAM interleaving schemes distribute the bits of an $N$-bit data word evenly across the DRAM hierarchy~\cite{zhang2000permutation,10.1145/115953.115961,Hsu1993PerformanceOC,11443109,plin2026knockknock} (i.e., rank $\rightarrow$ chip $\rightarrow$ bank $\rightarrow$ subarray $\rightarrow$ mat), scattering the operand bits that must interact during in-DRAM execution. 
Since \gls{PuD} computation is driven by the sense amplifiers of a DRAM mat~\omcri{\cite{seshadri2017ambit,seshadri2013rowclone,seshadri2018rowclone, gao2019computedram, gao2022fracdram, missingnot, yuksel2024simultaneous,seshadri2015fast,mutlu2024memory}} (the basic building block of \gls{PuD} computation), this distribution creates a \emph{data locality} problem, where the bits of a given \gls{PuD} operand are spread across the DRAM hierarchy rather than being centralized within a single DRAM mat. 
The vertical layout resolves this by confining all $N$ bits of each data word to a single DRAM column,
thereby localizing the full operand within one mat.
Second, \gls{PuD} computation is inherently \emph{row-wise}, where sense amplifiers operate across all columns of an activated DRAM row, and \emph{no} native interconnect exists to propagate data \emph{column-wise} across rows~\cite{hajinazarsimdram}. 
As a result, \gls{PuD} operations that require intermediate values to propagate across bit-positions (e.g., carry propagation in a bit-serial adder) \emph{cannot} be performed under a horizontal data layout without costly hardware modifications to the DRAM array~\cite{hajinazarsimdram, mimdramextended}, where the $N$ bits of an operand reside in $N$ different columns. 
The vertical layout resolves this by enabling carry propagation via in-DRAM row copy operations~\cite{seshadri2013rowclone}: $C_{in}^{i+1} \leftarrow C_{out}^{i}$ simply translates to copying row $i$ to row $i+1$.}}

\gf{However, the vertical data layout is \emph{fundamentally} incompatible with the horizontal data organization that modern memory systems rely on\revdelm{for high throughput and cache line abstraction}, introducing three major limitations. 
First, \gls{PuD} architectures incur \emph{costly runtime data layout conversion overheads}~\omcri{\cite{hajinazarsimdram}}.\revdelm{Before (after) \gls{PuD} execution, data must be transposed from horizontal to vertical (vertical to horizontal) layout, distributing (collecting) $M$ 1-bit elements across $M$ DRAM columns, $N$ times. 
These data layout transformations add significant performance and energy overheads, especially for large data volumes.
We quantify the impact of this data layout transformation by evaluating SIMDRAM-based \gls{PuD} execution across twelve real-world applications.}
Our evaluation results  (\cref{sec:methodology}) show that, on average, data layout conversion \emph{reduces} performance and energy efficiency by 3.9$\times$ and 3.7$\times$, respectively, relative to an idealized \gls{PuD} system with \emph{no} transposition requirements.
Second, data layout transformation \emph{amplifies data movement and interferes with the cache hierarchy}. 
For example, prior works~\cite{hajinazarsimdram, mimdramextended, oliveira2025proteus} store vertically laid-out data as object slices spanning multiple cache lines, such that accessing or evicting a single data element triggers \omcri{writeback} of an entire slice. 
This amplifies cache line traffic, increases writeback pressure, and entangles \gls{PuD} execution with \omcri{CPU} cache hierarchy behavior.
Third, vertical-layout \gls{PuD} architectures introduce a \emph{dual data-layout view of memory}, where an application needs to distinguish between horizontally and vertically organized objects and coordinate data layout transformations\revdelm{, complicating programming models, requiring specialized allocation routines, and tightly coupling \gls{PuD} execution to explicit data-format management}.}

\gf{Our \omcri{\textbf{goal}} in this work is to eliminate the need for runtime data layout transformation while preserving \omcri{\gls{PuD} execution} compatibility with \omcri{conventional} horizontally-organized memory systems. 
To this end, we propose \prop, a \gls{PuD} architecture that enables bit-serial \gls{PuD} execution
directly over horizontally laid-out data\revdelm{, allowing efficient in-DRAM computation while removing data transposition costs}. The \emph{key ideas} \omcri{of} \prop are twofold.}
\gf{First, \prop leverages a \emph{\gls{MABM} mechanism} that \emph{deterministically} reorganizes bits within each cache line to enable a horizontal, yet \gls{PuD}-friendly, data layout within DRAM without incurring data transposition overheads.
Placed at the memory controller, \gls{MABM} applies a deterministic bit-level permutation to each cache line on writes (and its inverse on reads), such that 
\li~each data word is confined within a \emph{single} DRAM chip, and 
\lii~its $N$ bits are \emph{horizontally} distributed  \emph{evenly} across the mats of that chip, with each mat holding one bit-position within a single DRAM row. 
This directly addresses the \emph{data locality} problem \omcri{by keeping the operand data inside a single subarray (multiple mats)}. 
\revdelm{Rather than being scattered across the DRAM hierarchy, the $N$ bits of a data element are localized within a \emph{single} DRAM chip, where the $i$-th bit of $M$ data elements are placed in a DRAM row within a \emph{single} DRAM mat of that DRAM chip.   
Data elements within a cache line remain interleaved across DRAM chips and columns as in a conventional system, preserving memory parallelism.} 
\gls{MABM} introduces \emph{no} memory throughput loss or data format conversion overhead.\revdelm{, since
\li~the amount of data transferred to/from DRAM is exactly one cache line,
\lii~bit reorganization is confined to a single cache line at a time, and
\liii~it is entirely transparent to the DRAM subsystem, incurring \emph{no} additional latency on the DRAM side}}
\gf{Second, \prop combines \emph{fine-grained DRAM row access}~\omcri{\cite{cooper2010fine,udipi2010rethinking,zhang2014half,ha2016improving,lee2017partial,olgun2022sectored,o2021energy,oconnor2017fine, olgun2024sectored}} with a \emph{hardware/software pipeline} to exploit bit-level parallelism~\cite{oliveira2025proteus} in \gls{PuD} execution. 
Since \gls{MABM} assigns each bit-position to a dedicated DRAM mat, carry values must propagate mat-to-mat rather than row-to-row. 
\prop enables this via the inter-mat interconnect of MIMDRAM~\cite{mimdramextended}, a fine-grained \gls{PuD} architecture that allows individual DRAM mats to be activated independently and provides low-cost data movement across mats.}

\gf{The ability to individually activate mats and move data between them opens a new opportunity: \emph{to turn each DRAM mat into a pipeline stage}, where each stage processes a distinct bit-position of the same \gls{PuD} operation. 
Such pipelining is possible because the \gls{PuD} primitives of a \gls{PuD} operation naturally partition into \emph{bit-dependent} primitives (those on the carry-dependency chain) and \emph{bit-independent} primitives (those that can execute before or after dependency resolution). 
Hence, while bit-dependent \gls{PuD} primitives advance in dependency order with carry values flowing mat-to-mat, bit-independent \gls{PuD} primitives from successive bit-positions can overlap across stages, keeping multiple mats busy simultaneously. 
To orchestrate this pipeline without costly runtime hardware, \prop statically assigns \gls{PuD} primitives to pipeline
stages using a classical modulo scheduling algorithm~\cite{ramakrishna1994ims, ramakrishna1981some, ramakrishna1992code}, which computes a fixed, repeating schedule offline\revdelm{ that sustains high throughput with simple control logic}.
To keep the pipeline busy, \prop introduces a \emph{hierarchical in-DRAM data copy scheme} that prevents latency-critical carry propagation from stalling pipeline stages. 
\revdelm{The key challenge is that the inter-mat interconnect, while low-cost, is very narrow, requiring many sequential data transfers to propagate a carry value across mats, each moving only a few bits at a time.
If bit-independent \gls{PuD} primitives are ready to execute, these sequential data transfers directly stall the pipeline. 
To avoid this, \prop leverages the wider inter-subarray interconnect of~\cite{chang2016low}, which can transfer an entire
DRAM row at once, to \emph{quickly} move the carry value out of the \emph{computing subarray} and into a neighboring \emph{copying subarray}. The inter-mat propagation to the destination mat then happens asynchronously within the copying subarray, while the pipeline in the computing subarray is immediately free to continue executing bit-independent \gls{PuD} primitives. Once the carry reaches its destination, it is transferred back to the computing subarray, and the dependent pipeline stage can proceed. 
This hierarchical organization removes blocking data movement from the critical path of \gls{PuD} execution while incurring low hardware overhead.}} 
\gf{Beyond performance, \prop eliminates the dual data-layout view of memory entirely by operating natively over horizontally laid-out data. 
Since data does \emph{not} need to be reorganized into a vertical format, applications require \emph{no} explicit data-format management, and \emph{no} coordination of layout transformations, addressing the the dual data-layout view of memory limitation we described.}

We compare the benefits of \prop to different state-of-the-art processor-centric (CPU, GPU) and memory-centric architectures (SIMDRAM~\cite{hajinazarsimdram}, MIMDRAM~\cite{mimdramextended}, \emph{Proteus}~\cite{oliveira2025proteus}). 
\revdelm{We comprehensively evaluate \prop performance, energy, and energy efficiency (performance per Watt) for twelve real-world applications.} Using a single DRAM rank, \prop provides              
\li~11.8$\times$, 11.8$\times$, and 80.4$\times$ \omcri{higher} performance\omcri{,}  
\lii~25.4$\times$, 3.0$\times$, and 38.0$\times$ lower energy consumption than SIMDRAM, MIMDRAM, and \emph{Proteus}; and 
\liii~356$\times$ and 11.7$\times$ \omcri{higher} energy efficiency \omcri{than a} CPU and GPU, respectively, on average across twelve real-world applications. \prop incurs low area cost on top of a DRAM chip (1.86\%) and CPU die (0.05\%). 


\noindent \gfcr{We make the following key contributions:}
\begin{itemize}
[noitemsep,topsep=0pt,parsep=0pt,partopsep=0pt,labelindent=0pt,itemindent=0pt,leftmargin=*]
\item \gfcr{We propose \prop, a \omcri{new} \gls{PuD} architecture that enables in-DRAM computation directly over horizontally laid-out data, eliminating runtime data layout transformation and removing the need for \omcri{two} physical data layouts. To our knowledge, \prop is the first architecture to support pipeline bit-serial \gls{PuD} execution without requiring vertical data organization.}

\item \gfcr{\prop deploys a \omcri{new} pipeline-based execution model, which decomposes \gls{PuD} operations into bit-dependent and bit-independent primitives and exploits fine-grained DRAM to overlap their execution. This execution model leverages the distributed placement of operand bits across \omcri{multiple} DRAM mats to enable steady-state pipelined computation over horizontally laid out data.}

\gfcut{\item \gfcr{\prop introduces two novel components. First, \prop's \emph{\gls{MABM} mechanism} deterministically reorganizes bits within each memory request to enable \gls{PuD}-friendly \omcri{data} placement. Second, \prop's \emph{hierarchical in-DRAM data copy scheme} decouples carry-dependent communication from computation, removing \omcri{data} communication from the critical path of \gls{PuD} execution. }}

\item \gfcr{We extensively evaluate \prop \agy{for} twelve real-world applications\agy{, showing} that \prop 
outperforms state-of-the-art \gls{PuD} \omcri{frameworks} and processor-centric \omcri{designs} while incurring low area \agy{cost} to the system.}

\item \gfcr{We open-source \prop at \url{https://github.com/CMU-SAFARI/PipeDRAM}.}
\end{itemize}

\section{Background}
\label{sec:background}


\gfcut{\gfcr{We first briefly explain the architecture of a typical DRAM chip. Next, we describe prior DRAM enhancements and \gls{PuD} works that \prop builds on.}}

\subsection{DRAM Organization \& Operation} 
\label{sec:background:organization}

\begin{figure}[!ht]
    \centering
    \includegraphics[width=\linewidth]{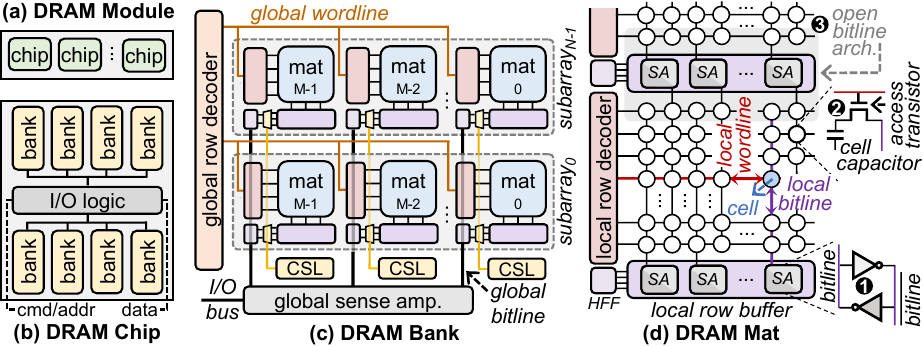}
    \caption{DRAM organization. Adapted from~\cite{ferreira2022pluto,mimdramextended}.}
    \label{fig_subarray_dram}
\end{figure}

\paratitle{\gfisca{DRAM Organization}} 
\sgi{Fig.~\ref{fig_subarray_dram} shows the hierarchy of a DRAM system.}
\sgdel{A DRAM system comprises a hierarchy of components, as Fig.~\ref{fig_subarray_dram} illustrates.}
A \emph{DRAM module} (Fig.~\ref{fig_subarray_dram}a) has several (e.g., 8--16) DRAM chips. 
A \emph{DRAM rank} is a group of DRAM chips that operate in lockstep (not shown).
A \emph{DRAM chip} (Fig.~\ref{fig_subarray_dram}b) has multiple banks (e.g., 8--16). 
A \emph{DRAM bank} (Fig.~\ref{fig_subarray_dram}c) has 
\li~multiple (e.g., 64--128) 2D arrays of DRAM cells known as \emph{DRAM subarrays}~\omcri{\cite{kim2012case, Tiered-Latency_LEE, yauglikcci2022hira}};
\lii~a \emph{global row decoder} and a \emph{global address latch} that select a row of cells in a subarray through \emph{global wordlines};
\liii~\emph{column select logic} (CSL) that selects portions (e.g., 64-bit) of the row; and
\liv~a \sgdel{\emph{global sense amplifier} \sg{(i.e., a \emph{global row buffer}, GRB)}}%
\emph{global sense amplifier} that transfers the selected fraction of the data from the row through \emph{global bitlines}.
\sgi{Each subarray contains multiples (e.g., 16--32) \emph{DRAM mats} (Fig.~\ref{fig_subarray_dram}d), each of which with
\li~multiple rows (e.g., 512--1024) and columns (e.g., 2--8~kB~\cite{kim2018solar, lee2017design, kim2002adaptive}\gfisca{)} of DRAM cells,
\lii~a \emph{local row decoder} that activates a \emph{local wordline}, 
\liii~a \emph{local row buffer} containing a row of \emph{sense amplifiers} (SAs; \omcri{\circlediv{1}} in Fig.~\ref{fig_subarray_dram}d) to latch data from an activated row, and
\liv~\emph{helper flip-flops} (HFFs) that drive a portion (e.g., 4-bit) of the data in the local
row buffer to the global bitlines.
A DRAM cell (\omcri{\circlediv{2}}) consists of a \emph{cell capacitor} and an \emph{access transistor}, which connects the cell capacitor with a \emph{local bitline} shared by all cells in the same column.
Modern DRAM employs an \emph{open bitline architecture}~\omcri{\cite{lim20121,takahashi2001multigigabit, chang2016low}}, fitting only enough SAs in one local row buffer to latch half a row of cells\revdelm{. 
To latch the entire row, a subarray connects to \emph{two} local row buffers, one above the cell array and one below} (\omcri{\circlediv{3}}).}

\sgdel{\gf{A \emph{DRAM subarray} (Fig.~\ref{fig_subarray_dram}d) organizes DRAM cells into multiple \emph{rows} (e.g., 512--1024) and multiple \emph{columns} (e.g., 2--8~kB~\cite{kim2018solar, lee2017design, kim2002adaptive}\gfisca{)} and contains
\li~a \emph{local row decoder} that drives the \emph{local wordlines} to the appropriate voltage levels to activate a row; and  
\lii~a row of sense amplifiers (also called a \emph{local row buffer}) that senses and latches data from the activated row.
A \emph{sense amplifier} \sg{(SA)} comprises of two back-to-back inverters (\circled{1} in Fig.~\ref{fig_subarray_dram}d).
A \emph{DRAM cell} (\circled{2}) consists of an \emph{access transistor} and a \emph{storage capacitor}. 
The source nodes of the access transistors of all the DRAM cells in the same column connect the cells' storage capacitors to the same \emph{local bitline}. The gate nodes of the access transistors of all the DRAM cells in the same row connect the cells' access transistors to the same \emph{local wordline}. 
To achieve high density, modern DRAM designs employ an \emph{open bitline architecture}~\cite{lim20121,takahashi2001multigigabit}, fitting only enough sense amplifiers in a local row buffer to sense half a row of cells. To sense the entire row of cells, each subarray has local bitlines connecting to two rows of local sense amplifiers -- one above and one below the cell array (\circled{3}).}}

\paratitle{DRAM Operation \& Data Organization}
\sgi{The memory controller issues three commands to service a DRAM request.
\revdel{Initially, the local bitlines are set at a reference voltage.}
The first command, \texttt{ACTIVATE} (\texttt{ACT}),\revdel{ selects a specified DRAM row.
This} connects \omcri{each} DRAM \omcri{cell} in \gfmicro{a} row to its local bitline,
and \omcri{each} cell's transistor shares its charge with the bitline to
shift the bitline voltage higher (or lower) if the cell stores a `1' (`0').
The local \omcri{sense amplifier} amplifies the \omcri{shift} to CMOS-readable values (simultaneously restoring charge to the DRAM cell).
The latency from the start of activation until charge restoration is called $t_{RAS}$.
The second command, \texttt{READ} (\texttt{RD}), returns a cache line of data from the \gfmicro{local \omcri{sense amplifier}}.
The third command, \texttt{PRECHARGE} (\texttt{PRE}), disconnects DRAM cells from the bitlines, and returns the bitlines to their reference voltage.
The precharge latency is called $t_{RP}$.}
\sgdel{\gf{Three major steps are involved in serving a main memory request. First, to select a DRAM row, the memory controller issues an \texttt{ACTIVATION} (\texttt{ACT}) command with the row address. On receiving this command, DRAM transfers all the data in the row to the corresponding local row buffers. \sgdel{(i.e., the row buffer at the top of the subarray and the one at the bottom).}
The two-terminal sense amplifier in the local row buffers senses the voltage difference between the local bitline and a reference voltage and amplifies it
to a CMOS-readable value \gfisca{until the cell charge is restored}. 
\gfisca{The latency from the start of row activation until the completion of the DRAM cell's charge restoration is called \emph{charge restoration latency} ($t_{RAS}$).}
Second, to access a cache line from the activated row, the memory controller issues a \texttt{READ} (\texttt{RD}) command with the column address of the request. 
Third, to enable the access of another DRAM row in the same bank, the memory controller issues a  \texttt{PRECHARGE} (\texttt{PRE}) command. 
This command disconnects the local bitline and restores the local bitline voltage to its quiescent state. \gfisca{The latency between issuing a \texttt{PRE} and when the DRAM bank is ready for a new row activation is called \emph{precharge latency} ($t_{RP}$)}.}} \revdelmrev{A DRAM data transfer proceeds in three steps, which Fig.~\ref{fig_dram_bust} illustrates\revdelm{ with an example cache line mapping consistent with common DRAM data interleaving (we use a \texttt{WRITE} transfer as an example; a \texttt{READ} transfer proceeds analogously)}.
First, the memory controller drives 64 DQ (data) signals, transferring one 64-bit \gfcri{of} data per \omcri{\gls{DDR} interface} cycle \gfcri{(i.e., a \emph{data beat})} over an 8-beat burst to \omcri{supply} an entire \SI{64}{\byte} cache line.
Second, each $\times$16 DRAM chip receives 16 bits per beat and accumulates 128~bits in its prefetch buffer \omcri{(Fig.~\ref{fig_dram_bust})} during the burst.
Third, the 128~bits buffered in each DRAM chip are written into the DRAM mats. 
\revdelm{In our example, only 4~bits are transferred
in a burst to (from) a DRAM mat with a \texttt{WRITE} (\texttt{READ}) DRAM command, with each scalar element striped across multiple chips and mats to maximize memory bandwidth.}

\begin{figure}[!ht]
    \centering
    \includegraphics[width=\linewidth]{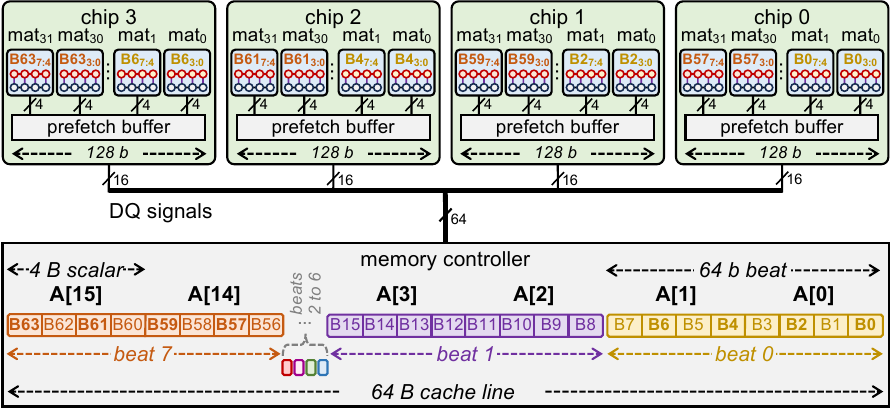}
   \caption{Example cache line placement.\revdelm{ A \SI{64}{\byte} cache line is transferred over eight 64-bit beats. We assume a DRAM rank comprising four $\times$16 DRAM chips, each containing 32 DRAM mats.} ``B'' denotes byte. $B_{j:i}$ denotes bits $\{\mathrm{bit}_j, \ldots, \mathrm{bit}_i\}$, where $i < j$.}
    \label{fig_dram_bust}
\end{figure}
}

\subsection{Processing-\sgi{Using}-DRAM} 
\label{sec:background:PUD}

\paratitle{In-DRAM Row Copy} RowClone~\cite{seshadri2013rowclone}  enables copying a row~$A$ to a row~$B$ in the \emph{same} subarray by issuing two consecutive \texttt{ACT} commands to these two rows, followed by a \texttt{PRE} command. This command sequence is called \texttt{AAP}.
LISA~\cite{chang2016low}\revdelm{ expands RowClone functionally to}  enables the execution of in-DRAM row copy operations across DRAM rows in \emph{different} subarrays of a DRAM \omcri{bank} by connecting local row buffers of neighbor subarrays using isolation transistors, providing \emph{high-throughput inter-subarray data copy}.
To do so, LISA exposes a new command to the memory controller called \texttt{LISA-RISC} (\underline{r}apid \underline{i}nter-\underline{s}ubarray bulk data \underline{c}opying), which moves an entire DRAM row between neighboring DRAM subarrays with a latency of $3 \times t_{RAS} + 2 \times t_{RP} + 2 
\times t_{RBM}$.\footnote{The row buffer movement (RBM) command asserts the isolation transistors between neighboring DRAM subarrays, which takes. $t_{RBM}$~ns.} 
\omcri{FIGARO}~\cite{wang2020figaro} allows copying one column from the local row buffer of one subarray to the local row buffer of another subarray within the same bank using a new DRAM command called \texttt{RELOC}.
\revdelm{Prior works~\cite{olgun2022pidram,gao2019computedram,yuksel2024simultaneous,mcciede2024} demonstrate the feasibility of performing in-DRAM row copy operations in commodity off-the-shelf (COTS) DRAM chips.}

\paratitle{In-DRAM Bitwise Operations} 
%
\sgi{Ambit~\omcri{\cite{seshadri2017ambit,seshadri2015fast}} shows that a simultaneous \emph{{\gls{TRA}}} can perform \emph{in-DRAM} bitwise AND/OR\omcri{/MAJ} operations.}
\revdelm{When activating three rows, three cells connected to each local bitline share charge simultaneously and contribute to the perturbation of the local bitline.}
Upon sensing the perturbation of the three simultaneously activated rows, the sense amplifier amplifies the local bitline voltage to $V_{DD}$ or 0 if at least two of the capacitors of the three DRAM cells are charged or discharged, respectively.
As such, a \gls{TRA} results in a Boolean majority operation (MAJ).
\sgi{Ambit implements \gls{TRA} using a custom row decoder, and introduces a new command called \texttt{AP} that issues a \gls{TRA} followed by a \texttt{PRE}.
Since TRA operations are destructive, Ambit divides DRAM rows into three groups\revdelm{ for \gls{PuD} computing}: 
\li~the \textbf{D}-group, which contains regular data rows; 
\lii~the \textbf{C}-group,
which consists of two rows (\texttt{C0} and \texttt{C1}) with all-`0' and all-`1'
values; and 
\liii~the \textbf{B}-group, which contains six rows designated for computation (four regular rows, \texttt{T0}, \texttt{T1}, \texttt{T2}, \texttt{T3};
and two rows, \texttt{DCC0} and \texttt{DCC1}, of dual-contact cells for \omcri{enabling the} NOT \omcri{operation}). \gfcr{Ambit implements an in-DRAM NOT operation by simply forwarding the complement of the sensed value as part of the activation process ($\overline{\mbox{bitline}}$ in Fig.~\ref{fig_subarray_dram}) to a special DRAM row in the subarray that consists of DRAM cells with two access transistors, called \emph{dual-contact cells} (DCCs). Each access transistor is connected to one side of the sense amplifier and is controlled by a separate wordline (\emph{d-wordline} or \emph{n-wordline}). By activating either the \emph{d-wordline} or the \emph{n-wordline}, the row of DCCs can provide the true or negated value stored in the row's cells, respectively.}
SIMDRAM~\cite{hajinazarsimdram}\revdel{ is a framework that} builds on Ambit to implement and expose high-level \omcri{instructions executed} in DRAM (\emph{{\uprog}s}).
A \uprog consists of a sequence of \emph{\gls{PuD} primitives}, i.e., \texttt{AAP}s (row copies) and \texttt{AP}s ($MAJ$), that are generated offline, and exposed to the programmer as \emph{bbop} instructions.
\revdel{During program execution, SIMDRAM
\li~decodes a \emph{bbop} instruction into its respective \uprog and
\lii~dispatches the \aaps in the \uprog to DRAM.}
To implement carry propagation, SIMDRAM (\omcri{as well as} other \gls{PuD} architectures~\omcri{\cite{peng2023chopper, zhou2022transpim,angizi2019graphide,ali2019memory,mimdramextended,oliveira2025proteus,tokuda2026clutch,liu2025optipim}}) employs a \emph{vertical} data layout, where all bits of a data word (e.g., a 32-bit integer) are stored in a single DRAM column (or bitline), and executes \omcri{each in-DRAM operation} bit-serially (i.e., one row at a time) in a \gls{SIMD} manner.
\omcri{\emph{Proteus}~\cite{oliveira2025proteus} further extends SIMDRAM's bit-serial execution model by implementing \emph{bit-parallel} in-DRAM operations, i.e., variants of SIMDRAM's \uprogs that leverage carry-lookahead logic to decouple the calculation of the carry bits and arithmetic logic.}}

\sgdel{SIMDRAM~\cite{hajinazarsimdram} builds on top of Ambit by proposing a three-step framework that translates an operation into its in-DRAM representation called \emph{\uprog}. A \uprog consists of a sequence of \texttt{AAP}s (row copies) and \texttt{AP}s (MAJ/\gls{TRA}s) that implements an operation (e.g., addition) in DRAM. 
In SIMDRAM, \uprogs are generated offline, stored in DRAM for future use, and exposed to the programmer as \emph{bbop} instruction. 
To implement complex arithmetic operations that require carry propagation (e.g., addition and multiplication), SIMDRAM employs a \emph{vertical} data layout, where all bits of a data word (e.g., a 32-bit integer) are stored in a single DRAM column (or bitline). 
In this way, SIMDRAM implements \emph{bit-serial} computation. }

\begin{figure}[!ht]
    \centering
    \includegraphics[width=\linewidth]{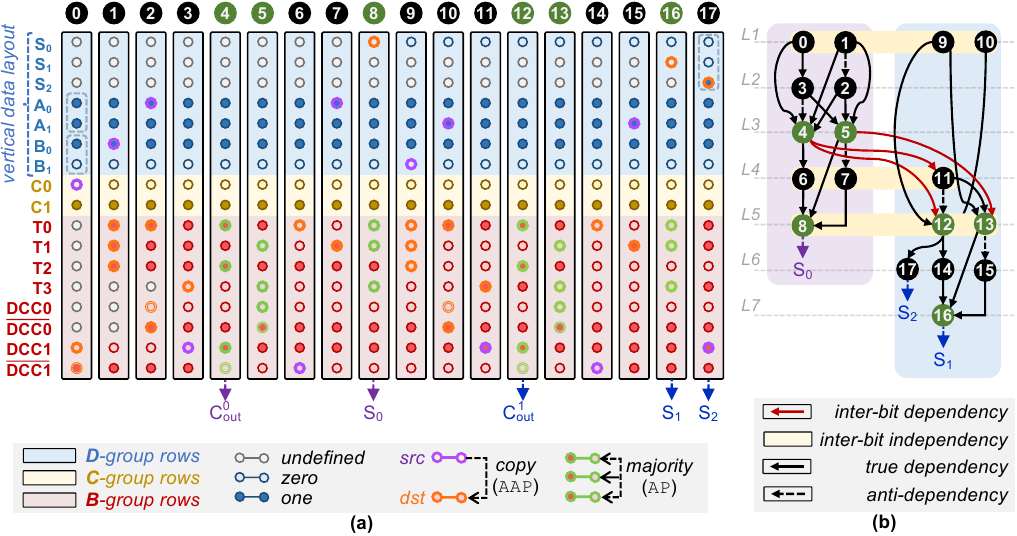}
    \caption{2-bit \gls{PuD} addition with vertical data layout.
    (a)~Step-by-step \gls{PuD} primitive execution. (b)~Corresponding DDG.
    \revdelm{2-bit in-DRAM bit-serial addition. (a)~Step-by-step execution of each \gls{PuD} primitive (i.e., $AAP$s and $AP$s) required to perform the bit-serial addition within a single DRAM subarray, where the input data $A$ and $B$ are stored using a vertical data layout.
    (b)~The corresponding DDG\revdelm{, highlighting true dependencies (intra-bit), anti-dependencies (stemming from the limited number of B-group rows), and inter-bit (true) dependencies}.}}
    \label{fig_bit_serial_example_new}
\end{figure}

Fig.~\ref{fig_bit_serial_example_new} illustrates the execution of a 2-bit in-DRAM bit-serial addition\gfcr{, which computes $C_{\mathrm{out},i} = \mathrm{MAJ}(A_i, B_i, C_{\mathrm{in},i})$ and
$S_i = \mathrm{MAJ}(A_i, \overline{C_{\mathrm{out},i}}, \mathrm{MAJ}(\overline{A_i}, B_i, C_{\mathrm{in},i}))$} using a sequence of \texttt{AAP} (row copy) and \texttt{AP} (MAJ) primitives within a single DRAM subarray. The inputs $A$ and $B$ are stored using a vertical data layout. \gf{Each number \omcri{at} the top of Fig.~\ref{fig_bit_serial_example_new}a indicates the execution of a \gls{PuD} primitive}. We make two observations.
First, data copy dominates computation due to the destructive nature of TRA. 
Copy operations are required either to preserve input operands (e.g., \omcri{\circled{0}, \circled{1}, \circled{2}, \circled{7}, \circled{9}, \circled{10}, \circled{14}}) or to preserve intermediate values consumed by multiple TRA operations (e.g., \omcri{\circled{3}, \circled{6}, \circled{14}}). 
Second, the data-dependency graph (DDG) (Fig.~\ref{fig_bit_serial_example_new}b) \gf{of the \gls{PuD} operation} reveals available \emph{bit-level parallelism} within the operation, since 
\li~several \gls{PuD} primitives are independent of carry propagation (e.g., initialization operations \omcri{\circled{0}, \circled{1}, \circled{9}, \circled{10}}) and 
\lii~others become independent once the carry resolves (e.g., \omcri{\circled[green!60!black]{8}, \circled[green!60!black]{12}, \circled[green!60!black]{13}}). 
The \gls{PuD} substrate can exploit this bit-level parallelism to reduce overall latency by enabling the concurrent execution of independent \gls{PuD} primitives within a DRAM subarray or bank. 
Since sense amplifiers are the primary \omcri{computation} drivers of in-DRAM operations, achieving such parallelism requires mechanisms that allow subsets of sense amplifiers to operate independently rather than in a fully lockstep manner (e.g., via fine-grained DRAM~\omcri{\cite{cooper2010fine,udipi2010rethinking,zhang2014half,ha2016improving,lee2017partial,olgun2022sectored,o2021energy,oconnor2017fine, olgun2024sectored}} or \gls{SLP}~\omcri{\cite{kim2012case,oliveira2025proteus,yauglikcci2022hira}}). 
\revdelm{\emph{Proteus}~\cite{oliveira2025proteus} exploits this bit-level parallelism by distributing the bit positions of a word across multiple DRAM subarrays within a bank and leveraging LISA~\cite{chang2016low} to propagate intermediate (e.g., carry) values between subarrays, enabling bit-parallel \gls{PuD} arithmetic.  
However, \emph{Proteus} requires simultaneously operating multiple subarrays within a bank (constrained by the DRAM power-delivery limits, e.g., $t_{FAW}$) and still relies on transposing data into a \gls{PuD}-friendly vertical layout.} 

\paratitle{Fine-Grained In-DRAM Computing} MIMDRAM~\cite{mimdramextended}\revdelm{ further enhances the bit-serial \gls{PuD} execution model by} introduces the ability to allocate only the required resources for a given \gls{PuD} operation via fine-grained DRAM access and activation~\omcri{\cite{cooper2010fine,udipi2010rethinking,zhang2014half,ha2016improving,lee2017partial,olgun2022sectored,o2021energy,oconnor2017fine, olgun2024sectored}}. 
\revdelm{Fig.~\ref{fig_mimdram_subarray} illustrates MIMDRAM's subarray organization.}
Concretely, MIMDRAM modifies
\li~the DRAM subarray circuitry with \emph{mat isolation transistors},  \emph{row decoder latches}, and a \emph{mat selector} logic to allow individual DRAM mats to be activated during a \gls{PuD} operation; and
\lii~the global and local sense amplification logic within a DRAM bank and DRAM mat to implement \emph{low-cost inter-/intra-mat interconnects}, respectively, to allow portions of an activated DRAM row to be moved across \omcri{mats} (via a new DRAM command called \texttt{GB\_MOV}) and within \omcri{a mat} (via a new DRAM command called \texttt{LC\_MOV}).
After row activation, moving (\#HFFs)-bits inter-/intra-\omcri{mat takes} $t_{RELOC}$~ns. 
MIMDRAM provides hardware\revdelm{ (through a specialized control unit placed in the memory controller)} and software support\revdelm{ (through compiler passes)} to map, schedule, and orchestrate the execution of data-independent \gls{PuD} operations across different DRAM mats within a DRAM subarray, in a \emph{\gls{MIMD}} manner. 

\revdelm{\begin{figure}[!ht]
    \centering
    \includegraphics[width=\linewidth]{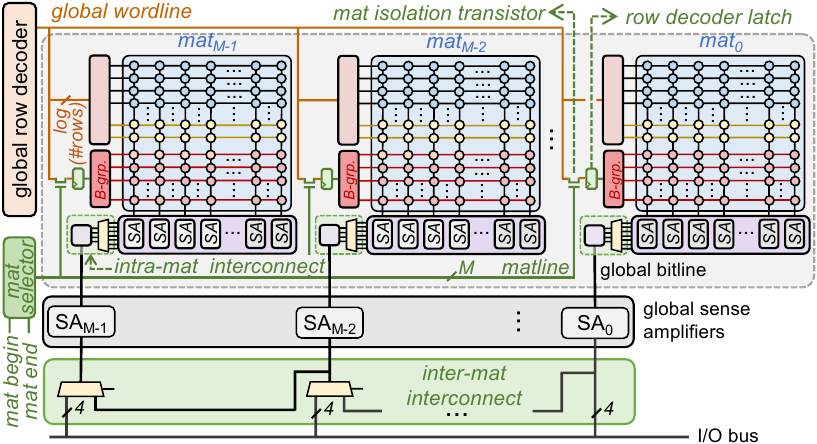}
    \caption{MIMDRAM subarray organization. Adapted from~\cite{mimdramextended}.}
    \label{fig_mimdram_subarray}
\end{figure}}

\paratitle{\revA{The \gls{PuD} Landscape}} \revA{\changeA{\#A2} Prior \gls{PuD} substrates~\cite{hajinazarsimdram, mimdramextended, oliveira2025proteus} differ along the axes summarized in Table~\ref{table:pudsummary}.
\li~\emph{Parallelism mode} is the \omcri{type of} concurrency a substrate exploits within a DRAM subarray: data-level (one \gls{PuD} operation across many data-independent elements), instruction-level (multiple independent \gls{PuD} operations at once), or bit-level (the bit-independent \gls{PuD} primitives of one \gls{PuD} operation).
\lii~\emph{Execution mode} is how that concurrency is realized, as \omcri{\emph{\gls{SIMD}}}, \emph{MIMD}, or \emph{pipelined} execution. 
\liii~\emph{Computing granularity} is the smallest unit a substrate can activate, either fine-grained over an individual DRAM mat (512 columns) or coarse-grained over the full row (65{,}536 columns).
\liv~\emph{Peak SIMD width} is the number of columns one single \gls{PuD} operation \omcri{within a bank} computes on.
\lv~\emph{Data transposition} \omcri{indicates} whether the substrate must first transpose data from the conventional horizontal layout to a vertical one.
SIMDRAM~\cite{hajinazarsimdram} computes a single \gls{SIMD} \gls{PuD} operation at coarse, full-row granularity. 
MIMDRAM~\cite{mimdramextended} adds fine-grained, mat-level computation and instruction-level parallelism, running different \gls{PuD} operations on different mats (\gls{MIMD}) and \omcri{supporting \gls{SIMD} widths between} 512  (one mat) to 65{,}536 lanes (all mats). 
\emph{Proteus}~\cite{oliveira2025proteus} adds bit-level parallelism, pipelining bit-independent \gls{PuD} primitives across the subarrays of a bank. 
These substrates differ in parallelism and granularity, yet they share one requirement: regardless of mode or granularity, \emph{all} have to transpose the entire input from the horizontal layout the processor produces to a vertical layout before computation. 
\prop, the substrate we present in \cref{sec:overview}, is the first \gls{PuD} substrate to \emph{fully} remove this \omcri{taxing} data transposition requirement while enabling all three data-parallelism  \omcri{modes} and \omcri{all three} execution \omcri{modes} to be exploited \omcri{within} a \emph{single} \gls{PuD} substrate.}

\begin{table}[ht]
\centering
\renewcommand{\arraystretch}{1.1}
\setlength{\tabcolsep}{4pt}
\caption{\revA{Summary of the \gls{PuD} landscape.}}
\label{table:pudsummary}
\revA{
\resizebox{\columnwidth}{!}{%
\begin{tabular}{@{}l|| ccc | ccc | c | c | c@{}}
\toprule
\multicolumn{1}{@{}l||}{}
& \multicolumn{3}{c|}{Parallelism Mode}
& \multicolumn{3}{c|}{Execution Mode}
& \multicolumn{1}{c|}{}
& \multicolumn{1}{c|}{}
& \multicolumn{1}{c@{}}{} \\
\cmidrule(lr){2-4}\cmidrule(lr){5-7}
\makecell[c]{System}
& \makecell[c]{Data} & \makecell[c]{Instr.} & \makecell[c]{Bit}
& \makecell[c]{SIMD} & \makecell[c]{MIMD} & \makecell[c]{Pipe.}
& \makecell[c]{Fine-\\Grained}
& \makecell[c]{Peak SIMD\\Width \\ \omcri{(per-bank)}}
& \makecell[c]{No\\Transp.} \\
\midrule\midrule
SIMDRAM~\cite{hajinazarsimdram}      & \cmark &        &        & \cmark &        &        &        & 65{,}536              &        \\
MIMDRAM~\cite{mimdramextended}       & \cmark & \cmark &        & \cmark & \cmark &        & \cmark & 512$\times$\#mats     &        \\
\emph{Proteus}~\cite{oliveira2025proteus}   & \cmark &        & \cmark & \cmark &        & \cmark &        & 65{,}536              &        \\
\midrule
\rowcolor[gray]{0.9}
\textbf{\prop (ours)} & \cmark & \cmark & \cmark & \cmark & \cmark & \cmark & \cmark & 2{,}048$\times$\#\omcri{bit-position} & \cmark \\
\bottomrule
\end{tabular}%
}
}
\end{table}

\section{Motivation}
\label{sec_motivation}

\gfcut{\gfcr{In this section, we discuss the three limitations imposed by requiring a vertical data layout for bit-serial \gls{PuD} execution: 
\li~costly runtime data layout transformation, 
\lii~amplified data movement and cache interference, and 
\liii~a dual data-layout view of memory that complicates system integration.}
}

\begin{figure}[ht]
    \centering
    \includegraphics[width=0.85\linewidth]{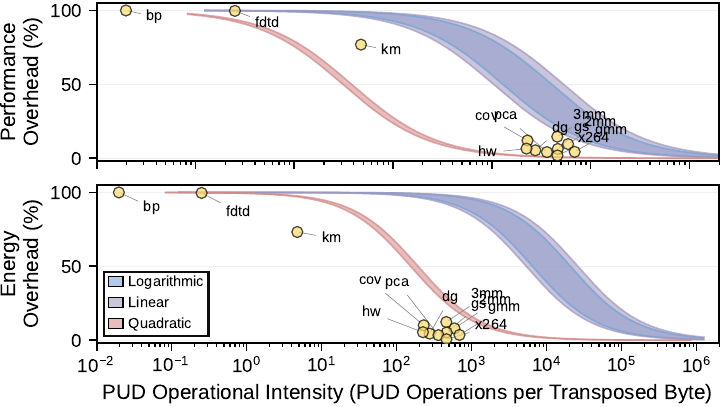}
    \caption{Data transposition overheads \omcri{on 16 \gls{PuD} operations (shaded bands) and 12 real-world applications (scatter points).}}
    \label{fig:motivation:transoverhead}
\end{figure}

\paratitle{Costly Runtime Data Layout Transformation} 
\revdelmrev{Requiring a vertical data layout for bit-serial \gls{PuD} execution necessitates frequent and \emph{costly runtime data layout transformation} between the system's native \omcri{CPU-friendly} horizontal format and the \gls{PuD}-friendly vertical format.
\revdelm{Fig.~\ref{fig_data_transposition_example} illustrates this transformation process in a representative system.}
In conventional \omcri{processor-centric} systems, cache lines are organized horizontally across DRAM chips and columns to preserve the cache line abstraction and maximize memory throughput. 
Each cache line access is serviced by a single sequence of DRAM commands (\texttt{ACT}-\texttt{RD}/\texttt{WR}-\texttt{PRE}), since all bits of the cache line reside within the same DRAM row across chips.
In contrast, under vertical-layout \gls{PuD} execution, the bits of an $N$-bit data element are distributed across $N$ different DRAM rows, with each row corresponding to a distinct bit-position. 
Consequently, reading \omcri{or writing} a vertically organized cache line requires $N$ separate DRAM \texttt{ACT}-\omcri{\texttt{RD}}/\texttt{WR}-\omcri{\texttt{PRE}} execution, one per row. 
\revdelm{Importantly, this overhead is incurred \emph{dynamically} during program execution. 
Every time the host processor reads, writes, evicts, or flushes a cache line belonging to a vertically organized \gls{PuD} memory object, the memory controller must perform such a data layout transformation. 
Unlike one-time data pre-processing or bulk data copying in other accelerator-based systems (e.g., data transposition in loosely-coupled \gls{PnM} systems~\cite{upmem,gomez2022benchmarking}), vertical-layout \gls{PuD} requires \emph{repeated runtime transformation throughout execution}.}
Since DRAM row activations dominate both latency and energy consumption, this $N$-fold amplification of DRAM traffic substantially increases memory access cost and undermines the efficiency benefits of in-DRAM computation.

}We quantify, in Fig.~\ref{fig:motivation:transoverhead}, the impact of data transposition on the performance (top) and energy efficiency (bottom) of \gls{PuD} operations by analyzing the \emph{operational intensity} (i.e., the ratio of \gls{PuD} operations to transposed bytes) across 16 bit-serial \gls{PuD} operations and 12 real-world applications using SIMDRAM~\cite{hajinazarsimdram} as the base \gls{PuD} architecture.\footnote{See~\cref{sec:methodology} for details of our experimental methodology.}
The shaded bands represent the data transposition overhead envelope for three \gls{PuD} complexity classes (logarithmic, linear, and quadratic\omcri{; see SIMDRAM~\cite{hajinazarsimdram}}), while the scatter points indicate the overhead \omcri{incurred by} the 12 real-world applications.
We make three observations.
First, three applications (\texttt{bp}, \texttt{fdtd}, and \texttt{km}) exhibit low operational intensity (below 10~Op/B), causing data transposition to account for most of the total execution time (77\%--100\%) and energy consumption (73\%--100\%)\revdelm{, effectively making \gls{PuD} computation negligible compared to the data layout transformation cost}.
Second, even for the nine compute-intensive applications with higher operational intensity, \omcri{data} transposition overhead still accounts for up to 14.6\% of execution time and 12.3\% of energy consumption.
Third, the operational intensity required to amortize data transposition overhead varies \emph{significantly} across operation complexity classes: quadratic operations\revdelmrev{ (e.g., multiplication, division)} reach low overhead at much lower operational intensities than logarithmic operations (e.g., reductions), because their higher per-element compute cost yields a more favorable compute-to-transposition ratio. 
\revdelm{This means that applications dominated by simple operations are disproportionately impacted by data transposition.
We conclude that data layout transformation significantly hinders the performance and energy efficiency of bit-serial \gls{PuD} architectures across a wide range of operational intensities.}

\paratitle{Amplified Data Movement \& Cache Interference}
Vertical-layout \gls{PuD} incurs \emph{amplified data movement and cache interference} at cache line granularity, by tightly coupling \gls{PuD} execution with cache hierarchy behavior.
\revdelm{Fig.~\ref{fig_data_transposition_example}b illustrates the data transposition process.} 
When a cache line $A$ belonging to a \gls{PuD} memory object is evicted from the \gls{LLC}, the data transposition unit (placed in the memory controller) intercepts the request (see~\cite{hajinazarsimdram}). Since a vertically-organized \gls{PuD} memory object distributes the $N$ bit-positions of each data word across 
$N$ DRAM rows, writing such data at high throughput requires assembling a full vertical memory slice that spans $N$ cache lines. 
\revdelm{The bits that populate a single vertically written row originate from different horizontal cache lines.
To construct this vertical slice, the data transposition unit~\cite{hajinazarsimdram, mimdramextended,oliveira2025proteus} 
\li~issues invalidation or writeback requests for the remaining $N-1$ cache lines of the same \gls{PuD} memory object, 
\lii~gathers their data into a transpose buffer, and 
\liii~reorganizes the bits into vertically laid-out cache lines.} 
Only after collecting all $N$ cache lines, the memory controller can issue $N$ DRAM write operations, each targeting a distinct DRAM row.
\revdelm{As a result, a single cache line eviction may trigger data transfers equivalent to $N$ cache lines between the cache hierarchy and main memory. }

\revdelm{\begin{figure}[!ht]
    \centering
    \includegraphics[width=\linewidth]{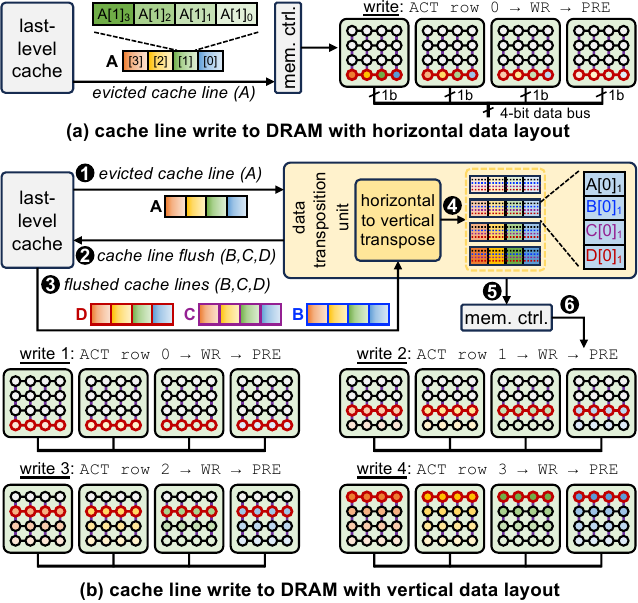}
    \caption{\revdelm{Overview of the data transposition process in bit-serial \gls{PuD} architectures~\cite{hajinazarsimdram,mimdramextended,oliveira2025proteus}. In this example, we assume a cache line size of \SI{16}{\byte} and a DRAM rank comprising four DRAM chips, each with a 1-bit DQ pin (i.e., a 4-bit-wide data bus). 
    (a)~Cache line write to DRAM using the conventional horizontal data layout. 
    (b)~Cache line write using the \gls{PuD}-friendly vertical data layout, where data are transposed before being written to DRAM. Different colors denote different 4-bit scalar elements within the cache line, and different shades denote different bit positions within each 4-bit scalar.}
    Data transposition in \gls{PuD} architectures~\cite{hajinazarsimdram,mimdramextended,oliveira2025proteus}
    (\SI{16}{\byte} cache line; four 1-bit-DQ chips). (a)~Horizontal layout write.
    (b)~Vertical layout write. Colors denote distinct 4-bit scalars;
    shades denote bit positions within each scalar.}
    \label{fig_data_transposition_example}
\end{figure}}

\paratitle{Dual Data-Layout View of Memory} 
\gls{PuD} architectures that rely on a vertical data layout impose a \emph{dual data-layout view of memory address space}, since the system must maintain two incompatible physical organizations, i.e., the native horizontal layout for host execution and the vertical layout required for in-DRAM computation. 
This duality breaks the uniform memory abstraction expected by software and complicates system integration, as the programmer (or compiler) must \emph{explicitly} distinguish between horizontally and vertically organized memory objects and determine when \omcri{a} data layout transformation should occur. 
This decision is inherently fragile: if \omcri{a} transformation occurs \emph{too late}, it lies on the critical path of the \gls{PuD} operation; if it occurs \emph{too early}, subsequent host accesses may consume data already reorganized into the vertical format, forcing additional layout conversions. 
As a result, performance depends on anticipating dynamic cache residency and future access behavior.
\revdelm{Second, the operating system and runtime must track and manage data layout as an explicit property of memory objects, since the same logical data may exist in different physical formats at different points during execution. 
This tight coupling between data format management and \gls{PuD} execution complicates programming models, increases system-level coordination overhead, and reduces the transparency of in-DRAM computation.}

\paratitle{Goal} Our \emph{goal} in this work is to eliminate the fundamental limitations imposed by vertical data layout in bit-serial \gls{PuD} architectures. 
We aim to remove the root cause of these inefficiencies by enabling high-throughput in-DRAM computation \emph{directly} over horizontally laid-out data\revdelmrev{, thereby eliminating runtime data transposition and avoiding dual physical data layouts while remaining compatible with existing \omcri{conventional} memory systems.
\revdel{To this end, we combine fine-grained DRAM access with hardware and software pipelining techniques to support efficient bit-serial/bit-parallel \gls{PuD} execution without disrupting the system-level horizontal data layout and cache line abstraction of modern memory systems}}  

\section{\prop Overview}
\label{sec:overview}

\prop is a \gls{PuD} architecture that enables bit-serial/bit-parallel execution directly over horizontally laid-out data\revdelm{, eliminating the need for runtime data layout transformation}. 
\prop is built on two key ideas. 
First, it \emph{deterministically} reorganizes bits within each cache line to construct a \gls{PuD}-friendly data placement inside DRAM while preserving the system-level horizontal data layout and cache line abstraction. 
Second, it combines fine-grained DRAM access with hardware/software pipelining to exploit bit-level parallelism across DRAM mats during \gls{PuD} execution.
\gfcr{\prop consists of three main components. 
First, a \emph{mat-aware bit mapping (MABM) mechanism} (\cref{sec:mabm}), implemented in the memory controller, that deterministically redistributes bits within each cache line to enable mat-level data interleaving without requiring runtime data layout transformation. 
Second, a \emph{software-assisted \omcri{pipelined} execution model} (\cref{sec:pipeline}) that statically schedules \gls{PuD} primitives across DRAM mats and subarrays to sustain high-throughput bit-serial execution.
Third, \prop implements a \emph{hierarchical in-DRAM data copy scheme} (\cref{sec:indramcopy}) that accelerates carry propagation and removes \omcri{inter-bit dependency} communication from the critical path of the pipeline.}

\subsection{Subarray Organization}

\begin{figure}[!ht]
    \centering
    \includegraphics[width=\linewidth]{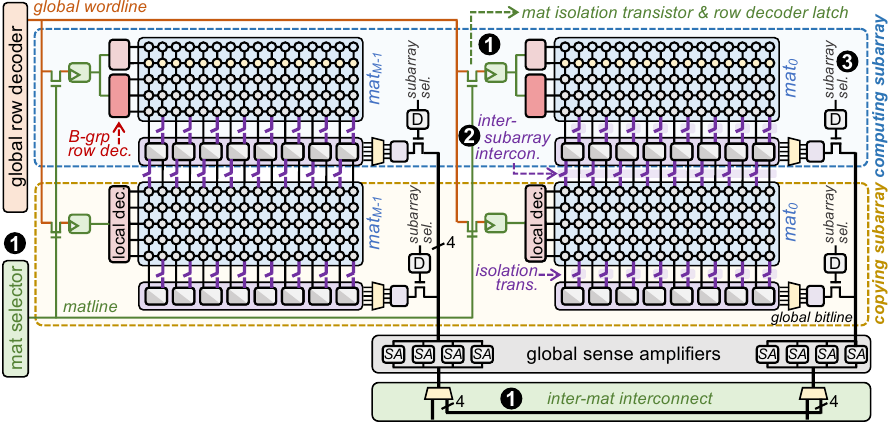}
    \caption{\prop subarray organization\omcri{, building on MIMDRAM~\cite{mimdramextended}, LISA~\cite{chang2016low}, and SALP~\cite{kim2012case}}.}
    \label{fig_pipedram_subarray}
\end{figure}

Fig.~\ref{fig_pipedram_subarray} shows the subarray organization of \prop\revdelm{, which builds on top of three existing DRAM mechanisms}. 
First, \prop leverages the fine-grained mat access substrate of MIMDRAM~\cite{mimdramextended} (\circled{1} in Fig.~\ref{fig_pipedram_subarray})\revdelm{, including the mat isolation transistors, row decoder latches, mat selector, and the inter-mat interconnect, to enable selective mat activation and low-cost data movement across mats within the same subarray}. 
Second, \prop leverages the isolation transistors of LISA~\cite{chang2016low} (\circled{2}) to provide high-throughput \omcri{connections} between neighboring subarrays\revdelm{, enabling fast transfer of data between local row buffers in adjacent subarrays}. 
Third, \prop incorporates the SALP~\cite{kim2012case} row-decoder latch (\circled{3}) to
enable subarray-level parallelism\revdelm{ during \gls{PuD} execution}.
\omcri{In addition} to these works, \prop introduces \emph{no} additional \omcri{circuitry} substrate; it only logically designates neighboring subarrays as \emph{copying subarrays}, \omcri{as} we describe in~\cref{sec:indramcopy}.

\subsection{Mat-Aware Bit Mapping (MABM)}
\label{sec:mabm}

\prop's MABM mechanism eliminates the need for runtime data layout transformation by redefining how bits within a cache line are mapped to DRAM chips and mats while addressing the \emph{data locality} problem (\cref{sec:introduction}) imposed by conventional DRAM interleaving. 
\revdelm{In conventional systems, data is interleaved across DRAM chips and mats to maximize memory bandwidth. While this organization is well-suited for host execution, it is poorly matched to in-DRAM arithmetic for two reasons. 
First, the bits of a scalar value are distributed across multiple DRAM chips, so no single chip contains the full operand required to perform the operation. 
Second, successive bit positions of a scalar do \emph{not} have a low-cost communication path for dependency-critical intermediate values, such as carry bits, since conventional DRAM does \emph{not} provide direct connectivity across bitlines or columns; adding such support would require costly additional in-DRAM communication hardware (prior work~\cite{mimdramextended} reports a 21\% DRAM area overhead for the implementation of an inter-lane shifting network with a DRAM subarray~\cite{li2017drisa}).}
MABM simultaneously
\li~localizes each scalar value within a single DRAM chip; and
\lii~interleaves the bit-positions of that scalar \omcri{value} across multiple DRAM mats using a \emph{deterministic mapping function}.
This data placement preserves the host's horizontal data layout and cache line abstraction, while creating the bit distribution that \prop requires for pipelined bit-serial execution. 
Each mat then holds one operand bit-position for many data elements, enabling successive mats to operate as pipeline stages and allowing dependency-critical values to be propagated through the data movement substrate described in \cref{sec:indramcopy}, without runtime transposition or costly cross-bitline communication hardware.

\paratitle{Implementation} Formally, let $A[i]_j$ denote bit-position $j$ of scalar $A[i]$, where $N$ is the scalar bit-width.
Let $N_{chips}$ denote the number of DRAM chips in a rank, and let $N_{mats}$ denote the number of DRAM mats participating in \gls{PuD} execution within a chip.
MABM defines a deterministic mapping $\Phi(A[i]_j) = (c,m,\ell)$, where $c = i \bmod N_{\text{chips}}$, $\ell = \lfloor i / N_{\text{chips}} \rfloor$, and $m = j \bmod N_{\text{mats}}$. 
Here, $c$ denotes the target DRAM chip, $m$ denotes the target DRAM mat within that chip, and $\ell$ denotes the column group index within the mat, derived from the scalar index $i$; higher-order bit positions are mapped to successive mat groups.\footnote{For clarity, the expression $m = j$ assumes that the number of participating mats equals the scalar bit-width ($N_{mats} = N$). More generally, with $N_{mats}$ participating mats, $m = j \bmod N_{mats}$. \omcri{For a bit-precision wider than $N_{mats}$ bits, bit-positions $j$ and $j + N_{mats}$ share one mat and
occupy distinct column positions within it. \omcrii{As we discuss in~\cref{sec:discussion}, implementing MABM requires \emph{no} information of the bit-width of individual DRAM requests.}}} 
This mapping ensures that:
\li~all bits of scalar $A[i]$ reside within a single DRAM chip, and  
\lii~distinct bit-positions of the scalar are distributed across different DRAM mats, enabling mat-level parallel execution of \gls{PuD} primitives.
When retrieving a cache line from DRAM during a read operation, the MABM mechanism applies the inverse permutation $\Phi^{-1}$ to the returned data before injecting it into the cache subsystem.
Because $\Phi$ is bijective over the bits of a single cache line, the original horizontal cache line organization is reconstructed by applying $\Phi^{-1}$ to each bit read from DRAM.
Specifically, a bit read from physical location $(c,m,\ell)$ is restored to $\Phi^{-1}(c,m,\ell) = A[\ell \cdot N_{chips} + c]_m$. Equivalently, the scalar index is recovered as $i = \ell \cdot N_{chips} + c$, and the bit index is recovered as $j = m$. 
\revdelm{Thus, writes apply $\Phi$ to materialize the MABM layout in DRAM, while reads apply $\Phi^{-1}$ across the returning DQ beats to recompose the cache line before it is delivered to the cache hierarchy.
Since both $\Phi$ and $\Phi^{-1}$ are fixed permutations confined to a single cache line, reconstruction requires neither a transpose buffer, multi-row assembly, nor additional DRAM commands.}

\begin{figure}[!ht]
    \centering
    \includegraphics[width=\linewidth]{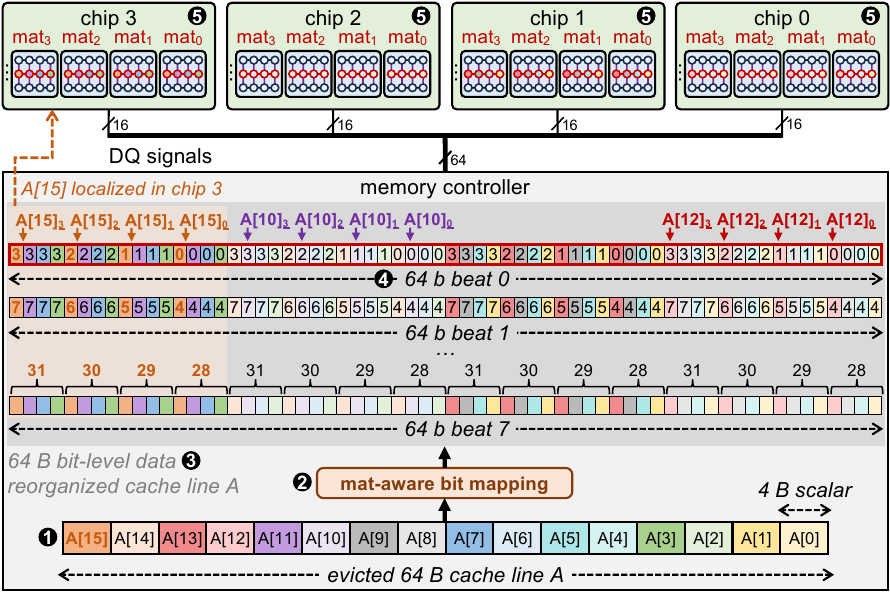}
    \caption{Mat-aware bit mapping (MABM) mechanism placed at the memory controller.
    A \SI{4}{\byte} scalar data element is denoted $A[i]$, and its bit $j$ is denoted $A[i]_j$.}
    \label{fig_mabm}
\end{figure}

Fig.~\ref{fig_mabm} illustrates the operation of the MABM mechanism during a cache line eviction to DRAM (a cache line retrieval, i.e., a DRAM read, is analogous). 
The MABM mechanism operates in three steps.
In the first step, the memory controller receives an evicted cache line (e.g., \SI{64}{\byte}) containing 16 scalar elements $A[i]$, where each scalar is \SI{4}{\byte} (32-bit) in this example (\circled{1} in Fig.~\ref{fig_mabm}). 
The cache line is organized in the conventional horizontal data layout.
In the second step, upon receiving the cache line and prior to injecting its data into the memory controller's write queue (not shown), the MABM mechanism applies $\Phi$ independently to each bit $A[i]_j$ (\circled{2}). 
The cache line data is logically permuted such that each scalar element is confined to a single DRAM chip (determined by $c = i \bmod N_{chips}$) and its bits are distributed across that chip's DRAM mats according to $m = j \bmod N_{mats}$. 
The output of this step is a bit-level reorganized version of cache line $A$ (\circled{3}), which preserves the cache line abstraction while enabling mat-level bit distribution.
In the third step, when the memory controller issues a DRAM write request, the reordered cache line is transferred over the standard DQ interface using the same beat-sized burst structure as the baseline system (\circled{4}). 
For example, during \emph{beat 0} (\circled{5}), the last 16 DQ signals deliver bit-slices \omcri{\texttt{\{3,2,1,0\}}} from scalar elements $\{A[15], A[11], A[7], A[3]\}$ to \emph{chip 0}.\revdelm{, such that
\[
\begin{alignedat}{1}
\mathrm{mat}_0 &\leftarrow \{A[15]_0, A[11]_0, A[7]_0, A[3]_0\} \\
\mathrm{mat}_1 &\leftarrow \{A[15]_1, A[11]_1, A[7]_1, A[3]_1\} \\
\mathrm{mat}_2 &\leftarrow \{A[15]_2, A[11]_2, A[7]_2, A[3]_2\} \\
\mathrm{mat}_3 &\leftarrow \{A[15]_3, A[11]_3, A[7]_3, A[3]_3\}.
\end{alignedat}
\]}
\noindent During \emph{beat 1}, the last 16 DQ signals deliver bit-slices \omcri{\texttt{\{7,6,5,4\}}} from the same scalar elements, which are written to the next set of mats (i.e., $\mathrm{mat}_4$ to $\mathrm{mat}_7$). The remaining beats proceed analogously.
After all beats in the burst are transferred to DRAM, each scalar element remains interleaved across DRAM chips and columns at cache line granularity, thereby preserving the horizontal data layout. 
However, within a given chip, all bits of a scalar are localized and distributed across that chip's mats, enabling efficient bit-serial \gls{PuD} execution without runtime data layout transformation.

\begin{figure*}[!ht]
    \centering
    \includegraphics[width=\linewidth]{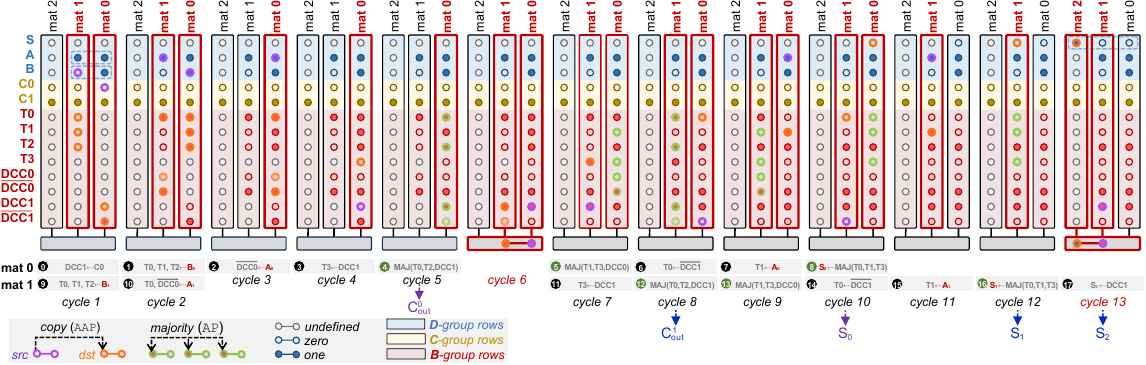}
\caption{Example of a 2-bit in-DRAM bit-serial addition,
\texttt{A (2'b11) + B (2'b01) = S (3'b100)}, in \prop, \omcri{where bit~$\boldsymbol{i}$
of both operands, $\boldsymbol{A_i}$ and $\boldsymbol{B_i}$, resides in
$\boldsymbol{\mathrm{mat}_i}$}. Each cycle executes \gls{PuD} primitives
according to the DDG in Fig.~\ref{fig_bit_serial_example_new}\omcrii{, whose
primitives compute
$\boldsymbol{C_{out}^{i} = \mathrm{MAJ}(A_i, B_i, C_{in}^{i})}$ and
$\boldsymbol{S_i = \mathrm{MAJ}(B_i, \overline{C_{out}^{i}},
\mathrm{MAJ}(A_i, C_{in}^{i}, \overline{C_{out}^{i}}))}$. A bold outline
marks the DRAM mats active in a cycle.}}
    \label{fig_bit_serial_pipeline}
\end{figure*}



\subsection{Software-Assisted \omcri{Pipelined} \gls{PuD} Execution}
\label{sec:pipeline}

\subsubsection{\omcri{Pipelined} Execution Flow}

\revdelmrev{\omcri{\prop} exploits the horizontal data layout produced by MABM (\cref{sec:mabm}) to pipeline the execution of the several \gls{PuD} primitives that compose a single bit-serial \gls{PuD} operation across the DRAM mats within a DRAM chip. Instead of executing all \gls{PuD} primitives of bit-position $i$ before starting the execution of bit-position $i+1$, \prop assigns successive bit-positions to neighboring mats and overlaps their execution whenever bit-dependencies permit. 
In this way, the mats within a chip act as \emph{pipeline stages} that process different bit-positions of the same operation, while intermediate values propagate between mats according to the inter-bit dependencies of the \gls{PuD} operation.
\revdelm{This pipeline execution model is enabled by two observations from the data-dependency graph (DDG) of a \gls{PuD} operation (\cref{sec:background:PUD}). 
First, only a subset of \gls{PuD} primitives lie on the inter-bit dependency chain, such as the carry propagation in a bit-serial addition. 
Second, many \gls{PuD} primitives are bit-independent and can execute either before the carry of the current bit-position is produced or after the carry dependency is resolved. 
Thereby, \prop pipelines these bit-independent \gls{PuD} primitives across neighboring mats while advancing the bit-dependent \gls{PuD} primitives respecting the dependency order.} 
As a result, multiple bit-positions of the same \gls{PuD} operation can be in flight \emph{simultaneously} across a single DRAM subarray, converting mat-level parallelism into pipeline parallelism.

}Fig.~\ref{fig_bit_serial_pipeline} illustrates \prop's \omcri{pipelined} execution flow using the same 2-bit in-DRAM bit-serial addition shown in the DDG of Fig.~\ref{fig_bit_serial_example_new}. 
The figure maps the two operand bit-positions to different DRAM mats and shows, cycle by cycle, which
\gls{PuD} primitives are issued to each mat. 
In this example, $\mathrm{mat}_0$ executes the \gls{PuD} primitives of the least-significant bit-position, while $\mathrm{mat}_1$ executes the \gls{PuD} primitives of the next bit-position once the carry dependency becomes available. 
The execution proceeds in five intervals: \emph{initialization}, \emph{carry generation}, \emph{carry propagation}, \emph{steady-state}, and \emph{pipeline drain}. 
\li~During initialization (\emph{cycles 1--2}), \prop concurrently executes \gls{PuD} primitives in $mat_0$ and $mat_1$ to copy input operand data into B-group rows (\cref{sec:background:PUD}). 
\lii~During carry generation (\emph{cycles 3--5}), \prop issues \gls{PuD} primitives only to $mat_0$ to produce the first carry-out value, $C_{out}^{0}$. 
\liii~In \emph{cycle 6}, \prop propagates $C_{out}^{0}$ from $mat_0$ to $mat_1$ via the inter-mat interconnect (\cref{sec:background:PUD}). \liv~Once \emph{cycle 6} completes, the inter-bit dependency is resolved and \prop proceeds by concurrently executing \gls{PuD} primitives over $mat_0$ and $mat_1$ during \emph{cycles 7--10}, reaching a steady-state pipeline execution. 
\lv~After \emph{cycle 10}, $mat_0$ completes all \gls{PuD} primitives associated with the least-significant bit-position, producing $S_0$. 
The remaining cycles (\emph{cycles 11--13}) drain the pipeline by executing the remaining \gls{PuD} primitives in $mat_1$ for the next
bit-position, producing $S_1$ and $C_{out}^{1}$. 
Finally, in \emph{cycle 13}, \prop forwards the carry produced in $mat_1$ to the next pipeline stage, $mat_2$, which corresponds to the most-significant output bit, $S_2$.
Compared to the serialized execution of Fig.~\ref{fig_bit_serial_example_new}, \prop reduces the execution of the operation from 18 cycles to 13 cycles by overlapping \gls{PuD} primitives from different bit-positions across mats. 
\revdelm{More generally, as operand precision increases, additional bit-positions occupy additional mats and the pipeline sustains high throughput by continuously overlapping bit-dependent and bit-independent \gls{PuD} primitives across DRAM mats.}

\subsubsection{Software-Assisted Instruction Scheduling}

\prop employs a software-assisted static scheduler to construct the \omcri{pipelined} execution schedule of a \gls{PuD} operation across DRAM mats.
\revdelmrev{Unlike conventional processors, \gls{PuD} execution is fully deterministic: each \gls{PuD} operation corresponds to a fixed sequence of \gls{PuD} primitives defined by a \uprog, with no conditional branches, dynamic memory accesses, or runtime control flow. 
Consequently, all data-/bit-dependencies between \gls{PuD} primitives are known prior to execution. 
This property enables \prop to compute the instruction schedule \emph{offline} and provide the memory controller with a static schedule that maps each \gls{PuD} primitive to a pipeline stage (i.e., DRAM mat) and execution cycle.
To do so, \prop adopts a scheduling strategy inspired by \emph{iterative modulo scheduling}~\cite{ramakrishna1994ims}, widely used in \omcri{\gls{VLIW}} compilers~\omcrii{\cite{mei2004design,fisher2005embedded,faraboschi2000lx, fisher1983very, ellis1985bulldog}} to construct \omcri{static} execution schedules.

}Algorithm~\ref{alg:pipedram_scheduler} shows the scheduling procedure (inspired by \emph{iterative modulo scheduling}~\cite{ramakrishna1994ims}), which works in four main steps.
Given a \uprog $U$ describing a \gls{PuD} operation and a target operand bit-position $b$, the scheduler determines when each \gls{PuD} primitive should execute and on which DRAM mat it should be issued. 
The scheduler constructs a pipeline schedule that respects both
\li~data-/bit-dependencies between \gls{PuD} primitives and 
\lii~resource constraints imposed by the limited number of DRAM mats that can execute \gls{PuD} primitives concurrently. 
First, the scheduler constructs the DDG of the \gls{PuD} primitives in the input \uprog (line~\ref{alg:line:ddg}). 
Second, it computes the minimum feasible initiation interval $II$, which defines the rate at which new operand bit-positions may enter the pipeline. 
This interval is constrained by both data-dependencies and hardware resources. 
The \emph{recurrence-constrained initiation
interval} captures the dependency-critical paths in the \uprog, such as carry propagation, and ensures that a bit-position is \emph{not} issued before the intermediate values it depends on become available. The \emph{resource-constrained initiation interval} captures the limited number of DRAM mats available for execution and ensures that the schedule does \emph{not} oversubscribe mat-level resources in any cycle. 
The
scheduler therefore sets $II$ to the maximum of these two bounds (lines~\ref{alg:line:ii_rec}--\ref{alg:line:ii_init}), since any smaller
value would yield an illegal schedule.
Third, given this $II$, the scheduler iteratively attempts to construct a legal pipeline schedule.
For each \gls{PuD} primitive, it computes the earliest legal cycle that satisfies the dependency constraints (line~\ref{alg:line:earliest}), derives the pipeline stage as $t \bmod II$ (line~\ref{alg:line:stage}), and maps the \gls{PuD} primitive to the DRAM mat corresponding to the target operand bit-position and pipeline stage (line~\ref{alg:line:mat}).
If the assignment violates either a data-/bit-dependency or a mat-level resource constraint, the current attempt fails, and the scheduler increases $II$ (lines~\ref{alg:line:conflict}--\ref{alg:line:ii_inc}).
Fourth, once a legal schedule is found, the resulting mapping assigns each \gls{PuD} primitive to a specific DRAM mat, execution cycle, and pipeline stage.

\begin{algorithm}[t]
\caption{\prop Static Instruction Scheduler}
\label{alg:pipedram_scheduler}
\KwIn{\uprog $U$, target bit-position $b$}
\KwOut{Static schedule $S$, where $S : p \mapsto (m,t,s)$, \omcri{where $m$ is the DRAM mat that executes $p$, $t$ is the cycle at which $p$ issues, and $s$ is the pipeline stage of $p$} }
Construct DDG $G$ from \gls{PuD} primitives in $U$\; \nllabel{alg:line:ddg}
$II_{rec} \leftarrow$ recurrence-constrained II from inter-primitive
    dependencies in $G$\; \nllabel{alg:line:ii_rec}
$II_{res} \leftarrow$ resource-constrained II from the number of
    available DRAM mats\; \nllabel{alg:line:ii_res}
$II \leftarrow \max(II_{rec}, II_{res})$\; \nllabel{alg:line:ii_init}
\Repeat{a legal schedule is found}{
    Clear schedule $S$\; $failed \leftarrow \textbf{false}$\;
    \ForEach{\gls{PuD} primitive $p \in U$ in dependency order}{
        $t \leftarrow$ earliest legal cycle satisfying all dependencies
            of $p$ in $G$\; \nllabel{alg:line:earliest}
        $s \leftarrow t \bmod II$\; \nllabel{alg:line:stage}
        $m \leftarrow$ DRAM mat for bit-position $b$ and stage $s$\; \nllabel{alg:line:mat}
        \lIf{$(m,t,s)$ violates a dependency or resource constraint}
            {$failed \leftarrow \textbf{true}$; \textbf{break}} \nllabel{alg:line:conflict}
        Insert $S[p] \leftarrow (m,t,s)$\;
    }
    \lIf{$failed$}{$II \leftarrow II + 1$} \nllabel{alg:line:ii_inc}
}
\Return{$S$}\;
\end{algorithm}

\revdelm{Because the dependency structure of a \uprog is fixed, the computed schedule is independent of input data values and can be reused across all data elements processed by the pipeline. 
During execution, the memory controller simply issues \gls{PuD} primitives to the corresponding DRAM mats according to this precomputed schedule, eliminating the need for dynamic instruction scheduling or hazard-detection hardware while sustaining high-throughput pipeline execution.}

\subsection{Hierarchical In-DRAM Data Copy Scheme}
\label{sec:indramcopy}
\revdelm{\prop uses two existing in-DRAM data movement substrates to propagate intermediate values between pipeline stages. 
First, it inherits the \emph{inter-mat interconnect} proposed by MIMDRAM~\cite{mimdramextended}, which enables low-cost data movement across DRAM mats through the peripheral logic of a DRAM bank.
Second, it leverages the \emph{inter-subarray interconnect} proposed by LISA~\cite{chang2016low}
to enable high-throughput data transfer across neighboring subarrays.
\prop combines these two mechanisms into a \emph{hierarchical in-DRAM data copy scheme} that separates latency-critical communication from non-critical intermediate data propagation.}

During \prop's pipelined execution of a bit-serial \gls{PuD} operation across DRAM mats, bit-dependent intermediate values (e.g., carry values) must be propagated between neighboring pipeline stages (i.e., DRAM mats). 
However, if such values are propagated only through the low-cost inter-mat interconnect, communication across pipeline stages can become a throughput bottleneck.
The key reason is that the inter-mat interconnect is \emph{fundamentally} limited by the width of the helper flip-flops (HFFs) in the DRAM bank periphery.
An inter-mat transfer can move only as many bits as the number of available HFFs, e.g., 4--8 bits in a typical DRAM organization~\omcri{\cite{mimdramextended,olgun2022sectored,zhang2014half,oconnor2017fine,ha2016improving,olgun2024sectored}}.
Thus, propagating an intermediate value spanning an entire mat width requires repeating the transfer operation $\frac{\mathit{mat\_width}}{\#HFFs}$ times.
\revdelmrev{For example, for a DRAM mat with 512 DRAM columns and 4 HFFs, the transfer must be repeated $\frac{512}{4}=128$ times, where each iteration takes $t_{\mathrm{RELOC}}$~(\si{\nano\second}). 
This serialized transfer cost is acceptable for non-critical inter-mat data transfers, but it can become too slow for bit-dependency-critical data values. 
A naive way to reduce the latency of inter-mat data transfers would be to increase the width of the inter-mat interconnect by adding more HFFs per DRAM mat. 
However, such a solution is prohibitively expensive, since wide HFFs incur area overhead comparable to duplicating portions of the local row buffer.\footnote{Making the HFF structure as wide as the mat width would incur an area overhead of 20–-26\% to a \SIrange{4}{8}{\giga\bit} DDR4 DRAM chip, according to our CACTI~\cite{cacti} analysis.}}

To mitigate this issue, \prop adopts a \emph{hierarchical in-DRAM data copy scheme} (\circled{3} in Fig.~\ref{fig_pipedram_subarray}) that separates latency-critical communication from non-critical data propagation.
The scheme combines two complementary data transfer paths: 
\li~a high-throughput inter-subarray copy path \omcri{(building in LISA~\cite{chang2016low}, as Fig.~\ref{fig_pipedram_subarray} illustrates)} for dependency-critical values, and
\lii~a low-cost inter-mat copy path for asynchronous propagation within a DRAM bank. 
By assigning each path to the communication it serves best, \prop removes carry propagation from the critical path of pipelined \gls{PuD} execution while preserving low hardware cost.
\prop logically partitions neighboring \revCommon{(i.e., physically adjacent, sharing a local row buffer)} subarrays into two roles.
A \emph{computing subarray} executes \gls{PuD} primitives over the active pipeline stages, while a neighboring \emph{copying subarray} serves as a temporary staging area for intermediate values that must be forwarded to subsequent pipeline stages. 
\omcri{This organization allows \prop to
remove carry-dependent communication from the critical path of computation: latency-critical values are first transferred quickly to a copying subarray using the inter-subarray interconnect, and are then propagated asynchronously to the appropriate destination mats using the low-cost inter-mat interconnect.} 
\revCommon{\changeCM{\#CQ2}It is important to point out that the adoption of a copying subarray during \gls{PuD} execution does \emph{not} reduce the effective memory capacity of the DRAM subsystem, since the copying subarray is used as a staging buffer that only occupies the copying subarray's local row buffer and does \emph{not} hold any state during PUD operation, i.e., user data stored in the copying DRAM subarray is preserved. 
In case the \gls{PuD} system enables \gls{PuD} execution across all DRAM subarrays in a DRAM bank, a copying subarray can be time multiplexed between copying and computing across \gls{PuD} execution, since enabling a copying subarray does \emph{not} involve any static resource partitioning and \omcri{requires only that \prop appropriately schedules} the inter-subarray/-mat DRAM commands (discussed below) during \gls{PuD} execution.}

\subsubsection{Asynchronous Carry Propagation}

\revdelm{\prop leverages the hierarchical in-DRAM data-copy scheme to asynchronously propagate dependency-critical intermediate values during \gls{PuD} execution. 
The mechanism operates in two steps. 
In the first step, once a critical data value is produced in the computing subarray, \prop transfers it to the neighboring copying subarray using the high-throughput inter-subarray interconnect. 
This step quickly removes the data value from the critical path of the originating pipeline stage. 
In the second step, the data value is propagated within the copying subarray to the destination mat using the low-cost inter-mat interconnect while useful computation continues in the original computing subarray. 
Once the data value reaches the target  destination DRAM mat location in the copying subarray, it is copied back to the computing subarray so that execution of the dependent
\gls{PuD} primitives in the next pipeline stage in the computing subarray can proceed.}

\begin{figure}[!ht]
    \centering
    \includegraphics[width=\linewidth]{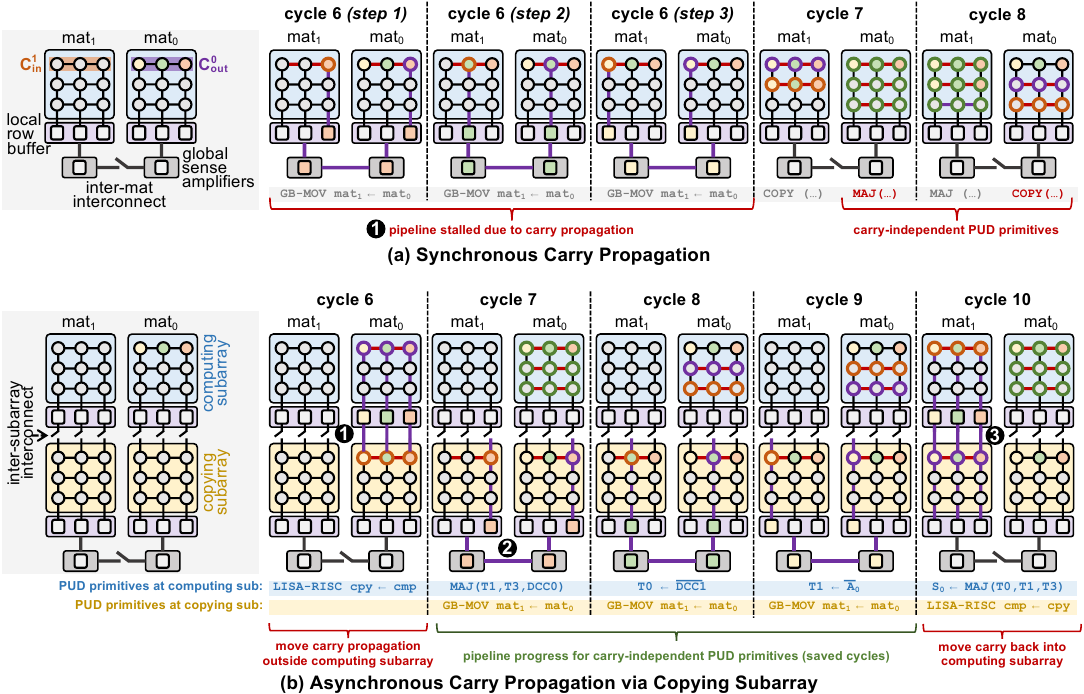}
    \caption{Carry propagation in \prop (2-bit addition of \omcri{Fig.~\ref{fig_bit_serial_pipeline}}). (a)~Baseline: carry propagation blocks computation. (b)~\prop's hierarchical in-DRAM data copy scheme.}
    \label{fig_execution_data_copy}
\end{figure}

Fig.~\ref{fig_execution_data_copy} illustrates our hierarchical in-DRAM data copy scheme using the same 2-bit bit-serial addition as in \omcri{Fig.~\ref{fig_bit_serial_pipeline}}. 
In the baseline execution (Fig.~\ref{fig_execution_data_copy}(a)), the carry data produced in $mat_0$ by \gls{PuD} primitive \texttt{\#4}, i.e., $C_{out}^0$, is propagated to the consumer pipeline stage in $mat_1$ using the \texttt{GB-MOV} DRAM command (\circled{1} in Fig.~\ref{fig_execution_data_copy}(a)).
This inter-mat data copy blocks the pipeline execution to process the following bit-independent \gls{PuD} primitives in $mat_0$ (i.e., \gls{PuD} primitives \texttt{\#5--\#8}) by $t_{RELOC} \times \frac{mat_{width}}{\#HFFs}$~ns. 
In contrast, Fig.~\ref{fig_execution_data_copy}(b) shows \prop's asynchronous carry propagation in three main steps. 
First, once the carry $C_{out}^0$ is produced in $mat_0$ by the computing subarray, \prop immediately moves it to the neighboring $mat_0$ in copying subarray using the high-throughput inter-subarray interconnect by issuing a \texttt{\omcri{LISA}-RISC} DRAM command (\circled{1} in Fig.~\ref{fig_execution_data_copy}(b)).
This allows the carry originating mat ($mat_0$) to continue executing carry-independent \gls{PuD} primitives or begin processing a new subset of data elements after $t_{RBM}$~ns. 
Second, the memory controller issues a \texttt{GB-MOV} command to the copying subarray to asynchronously move $C_{out}^0$ from $mat_0$ to $mat_1$ using the inter-mat interconnect in \omcri{the copying subarray} \circled{2}).
Third, once the \texttt{GB-MOV} command finishes executing, the memory controller issues a \texttt{\omcri{LISA}-RISC} command to copy $C_{out}^0$ \omcrii{from} $mat_1$ \omcrii{in the copying subarray} to $mat_1$ in the computing \omcri{subarray} (\circled{3}), \omcri{thereby} enabling the bit-dependent \gls{PuD} primitives of the next pipeline stage to execute.

\subsection{Discussion}
\label{sec:discussion}

\paratitle{\revA{Execution Model}} 
\revA{\changeA{\#A2}\prop converts mat-level parallelism within a chip into pipeline parallelism for a single \gls{PuD} operation, while preserving chip-level MIMD parallelism across the logical subarray.
This execution model differs from prior fine-grained \gls{PuD} architectures, such as MIMDRAM~\cite{mimdramextended}: while they exploit \emph{data-level parallelism} across multiple \gls{PuD} operations, \prop exploits \emph{bit-level parallelism} within a \emph{single} \gls{PuD} operation. 
However, it is important to point out that \prop can also exploit data-level parallelism, since operand bits are interleaved \omcri{\emph{across}} DRAM chips \omcri{via} our MABM scheme, enabling different DRAM chips within the same logical DRAM subarray to execute independent \gls{PuD} operations \emph{concurrently}, and thereby support MIMD-style execution.}

\paratitle{Programming Model, ISA \& System Support} \prop leverages the \emph{same} ISA extensions, compiler support, control unit, and coherency mechanism as prior \gls{PuD} systems~\cite{hajinazarsimdram,mimdramextended,oliveira2025proteus}.
\revdelm{\prop leverages the \emph{same} system integration solutions as in prior \gls{PuD} systems~\cite{hajinazarsimdram,mimdramextended,oliveira2025proteus}, including: 
\li~ISA extensions to the host CPU ISA that the programmer/compiler uses to launch \bbop instructions;
\lii~compiler support to identify \gls{PuD}-friendly loops in applications; 
\liii~a hardware control unit, alongside the memory controller, to control the execution of \uprogs; and
\liv~a coherency mechanism based on \emph{cache line flushing} to guarantee data coherence between host and \gls{PuD} execution. }
\omcri{It is important to point out that our MABM mechanism (\cref{sec:mabm}) requires \emph{no} knowledge of the data type or bit-precision of the cache lines the memory controller writes/reads to/from DRAM. 
MABM's deterministic mapping $\Phi$ is defined over fixed-size $M$-bit words ($M = 64$~bits in our design) and is applied identically to \emph{every} cache line, so a DRAM read returns exactly the cache line the preceding write stored, regardless of how software interprets its bytes. 
The bit-precision $M$ is needed \emph{only} during the execution of a \gls{PuD} operation, and such information is present in the ISA extensions we inherit from SIMDRAM~\cite{hajinazarsimdram} and MIMDRAM~\cite{mimdramextended}.}

\paratitle{Execution Control} \prop leverages MIMDRAM's control unit~\cite{mimdramextended}, placed at the memory controller, to fetch, decode, and dispatch \bbop instructions to DRAM. 
When the memory controller receives a \bbop instruction, the control unit decodes the instruction to identify the corresponding \uprog and retrieves its precomputed pipeline schedule from a dedicated \emph{pipeline schedule table}. This table stores the static schedule produced by Algorithm~\ref{alg:pipedram_scheduler} for each \uprog, mapping every \gls{PuD} primitive to a (mat, cycle, stage) tuple.
\revdelm{Because the schedule depends solely on the \uprog's dependency structure and the number of DRAM mats (both known at compile time), it is computed entirely offline in software, requiring \emph{no} runtime scheduling logic in the memory controller.} 
The control unit streams the retrieved schedule entries to the \emph{\uprog processing engine}, which issues \gls{PuD} primitives to the corresponding mats according to the precomputed order.




\section{Methodology}
\label{sec:methodology}

We implement \prop using an in-house cycle-level simulator \omcri{(which we open-source at~\cite{pipedramgit})} and compare it to prior simulated state-of-the-art \gls{PuD} frameworks, i.e., \li~SIMDRAM~\cite{hajinazarsimdram}, a \emph{coarse-grained bit-serial} \gls{PuD} architecture;
\lii~MIMDRAM~\cite{mimdramextended}, a \emph{fine-grained bit-serial} \gls{PuD} architecture; and \liii~\emph{Proteus}~\cite{oliveira2025proteus}, a \emph{coarse-grained bit-parallel} \gls{PuD} architecture.
Our in-house simulator was rigorously validated against the gem5~\cite{gem5} implementations of SIMDRAM and MIMDRAM~\cite{mimdramgit} and accounts for the additional latency imposed by SALP\omcri{-MASA (multitude of activated subarrays)}~\cite{kim2012case} on DRAM \texttt{ACT} commands, i.e., the extra circuitry required to support SALP incurs an extra latency of \SI{0.028}{\nano\second} to an \texttt{ACT}~\cite{hassan2022case} (which is less than  0.11\% extra latency for an \texttt{AAP}).\changeE{\#E2}\footnote{\revE{The public gem5 \omcri{MIMDRAM} simulator~\cite{mimdramgit} executes \omcri{a} vertical-layout bit-serial \uprog at full-row granularity; it \emph{cannot} represent \prop's mechanisms, which require an explicit per-primitive DDG and a pipelined per-mat engine. Thus, we built a cycle-level, per-\gls{PuD} primitive simulator that reuses the identical DRAM timing and energy parameters as SIMDRAM/MIMDRAM/\emph{Proteus}, so the only difference from prior work is the execution model, keeping comparisons apples-to-apples.}} 
\revdelm{To verify the functional correctness of our target applications, our simulation infrastructure considers the application's data when performing \gls{PuD} operations.
We did \emph{not} observe any deviation from the expected outputs.}
We also compare \prop against processor-centric systems using publicly available performance and energy measurements for a real multicore CPU~\cite{intelskylake} and a real high-end GPU~\cite{a100} on our evaluated workloads. 
These CPU \omcri{and} GPU measurements are \emph{not} collected in this work; instead, we obtain them from the publicly available open-source artifact of prior work~\cite{proteusgit}. 
To ensure a fair comparison, we use the same workloads, compilation settings, and input datasets as those used to obtain these CPU \omcri{and} GPU measurements.
Table~\ref{table_parameters} shows the system parameters we use in our evaluations. 
We use CACTI~\cite{cacti} to evaluate \prop and SIMDRAM/MIMDRAM/\emph{Proteus} energy consumption, where we take into account that each additional simultaneous row activation increases energy consumption by 22\%~\omcri{\cite{seshadri2017ambit, hajinazarsimdram, mimdramextended, oliveira2025proteus,yuksel2024simultaneous,missingnot}}. All evaluated \gls{PuD} architectures use 32 DRAM subarrays in a single DRAM bank for \gls{PuD} execution if not otherwise stated.
We evaluate three \prop configurations:
\li~\prop-Seq, a baseline \prop implementation that performs a simple first-in-first-out scheduling algorithm to schedule \gls{PuD} primitives across DRAM mats;
\lii~\prop-Pipe, a \prop implementation that implements our scheduling algorithm (from Algorithm~\ref{alg:pipedram_scheduler}); and
\liii~\prop-Hier, which adds our \emph{hierarchical in-DRAM copy scheme} (\cref{sec:indramcopy}) on top of \prop-Pipe.

\begin{table}[ht]
   \caption{Evaluated system configurations.}
   \centering
   \footnotesize
   \tempcommand{1}
   \renewcommand{\arraystretch}{0.7}
   \resizebox{\columnwidth}{!}{
   \begin{tabular}{@{} c l @{}}
   \toprule
   \multirow{5}{*}{\shortstack{\textbf{Multicore CPU~\cite{intelcometlake}}\\ \textbf{(\omcri{From} Artifact Results~\cite{proteusgit})}}} & x86~\cite{guide2016intel}, 16~cores, 8-wide, out-of-order, 3.8~GHz;  \\
                                                                           & \emph{L1 Data + Inst. Private Cache:} 256~kB, 8-way, 64~B line; \\
                                                                           & \emph{L2 Private Cache:} 2~MB, 4-way, 64~B line; \\
                                                                           & \emph{L3 Shared Cache:} 16~MB, 16-way, 64~B line; \\
                                                                           & \emph{Main Memory:} 64~GB DDR4-2133, 4~channels, 4~ranks \\
   \midrule
      \multirow{3}{*}{\shortstack{\textbf{High-End GPU~\cite{a100}}\\ \textbf{(\omcri{From} Artifact Results~\cite{proteusgit})}}} &  7~nm technology node; 826~mm$^2$ die area~\cite{a100};\\ 
                                                                            & 108 streaming multiprocessors, 1.4~GHz base clock; \\
                                                                            & \emph{L2 Cache:} 40~MB L2 Cache; \emph{Main Memory:} 40~GB HBM2~\mbox{\cite{HBM,lee2016simultaneous}} \\
   \midrule

   \multirow{7}{*}{\shortstack{\textbf{SIMDRAM~\cite{hajinazarsimdram}}\\ 
   \textbf{MIMDRAM~\cite{mimdramextended}}\\ \textbf{\emph{Proteus}~\cite{oliveira2025proteus}}\\  \textbf{\& \prop}}} &  gem5-based in-house simulator;  x86~\cite{guide2016intel};  \\ 
                                                            & \omcri{one} \gfmicro{out-of-order core @ 4~GHz (\emph{only} for instruction offloading});\\
                                                                             & \emph{L1 Data + Inst. Cache:} 32~kB, 8-way, 64~B line;\\
                                                                             & \emph{L2 Cache:} 256~kB, 4-way, 64~B line; \\
                                                                             & \emph{Memory Controller:}  8~kB row size, FR-FCFS~\cite{mutlu2007stall,zuravleff1997controller}\\
                                                                             & \emph{Main Memory:}  \gfmicro{DDR5-5200}, 1~channel, 1~rank, 16~banks \\ 
                                                                             & 32 DRAM subarrays \omcri{in a single bank} for \gls{PuD} execution \\
                                      
   \bottomrule
   \end{tabular}
   }
   \label{table_parameters}
\end{table}

\paratitle{Real-World Applications} We select twelve workloads from four popular benchmark suites in our real-workload analysis \changeD{\#D3}\revD{(as Table~\ref{table:workload:properties} describes)}, including 
\li~525.x264\_r (\texttt{x264}) from SPEC 2017~\cite{spec2017};
\lii~\texttt{pca} from Phoenix~\cite{yoo_iiswc2009};
\liii~\texttt{2mm}, 
\texttt{3mm}, 
convolution (\texttt{cov}),
doitgen (\texttt{dg}), 
fdtd-apml (\texttt{fdtd}),
gemm  (\texttt{gmm}), and
gramschmidt (\texttt{gs}) from Polybench~\cite{pouchet2012polybench};
and
\liv~heartwall (\texttt{hw}), kmeans (\texttt{km}), and backprop (\texttt{bp}) from Rodinia~\cite{che_iiswc2009}.
These workloads were identified as \gls{PuD}-friendly by prior work~\cite{mimdramextended} through an extensive analysis of 117 applications, and have been used to evaluate MIMDRAM~\cite{mimdramextended} and \emph{Proteus}~\cite{oliveira2025proteus}. 
\revdelmrev{We use the LLVM-based compilation infrastructure of MIMDRAM~\cite{mimdramgit} to identify memory-bound loops that can exploit SIMD parallelism and generate the corresponding \bbop instructions.}
We use the largest input dataset available in all evaluations \revD{\changeD{\#D3}and native data precisions (32-bit for all workloads other than \texttt{x264}, which natively employs 8-bit precision).}

\begin{table}[ht]
   \caption{\revD{Evaluated applications and their characteristics.}}
   \tempcommand{0.8}
   \centering
\revD{
   \resizebox{\columnwidth}{!}{%
    \begin{tabular}{|c|c||c|c|}
\hline
\textbf{\begin{tabular}[c]{@{}c@{}}Benchmark\\ Suite\end{tabular}} &
\textbf{\begin{tabular}[c]{@{}c@{}}Application\\ (Short Name)\end{tabular}} &
\textbf{\begin{tabular}[c]{@{}c@{}}Dataset\\ Size\end{tabular}} &
\textbf{\begin{tabular}[c]{@{}c@{}}PUD\\ Ops.$^\dag$\end{tabular}} \\ \hline\hline
Phoenix~\cite{yoo_iiswc2009} & pca (\texttt{pca}) & R = C = 16000 & D, S, M, R \\ \hline
\multirow{7}{*}{\begin{tabular}[c]{@{}c@{}}Polybench\\ \cite{pouchet2012polybench}\end{tabular}}
 & 2mm (\texttt{2mm}) & NI = NJ = NK = NL = 32000 & M, R \\ \cline{2-4}
 & 3mm (\texttt{3mm}) & NI = NJ = NK = NL = NM = 32000 & M, R \\ \cline{2-4}
 & covariance (\texttt{cov}) & N = M = 32000 & D, S, R \\ \cline{2-4}
 & doitgen (\texttt{dg}) & NQ = NR = 100, NP = 10000 & M, C, R \\ \cline{2-4}
 & fdtd-apml (\texttt{fdtd}) & CZ = CYM = 400, CXM = 10000 & D, M, S, A \\ \cline{2-4}
 & gemm (\texttt{gmm}) & NI = NJ = NK = 32000 & M, R \\ \cline{2-4}
 & gramschmidt (\texttt{gs}) & NI = NJ = 32000 & M, D, R \\ \hline
\multirow{3}{*}{\begin{tabular}[c]{@{}c@{}}Rodinia\\ \cite{che_iiswc2009}\end{tabular}}
 & backprop (\texttt{bp}) & 134217728 input elm. & M, R \\ \cline{2-4}
 & heartwall (\texttt{hw}) & test\_4k.avi, 20 frames & M, R \\ \cline{2-4}
 & kmeans (\texttt{km}) & 16384 data points & S, M, R \\ \hline
\begin{tabular}[c]{@{}c@{}}SPEC 2017\\ \cite{spec2017}\end{tabular}
 & 525.x264\_r (\texttt{x264}) & 1280$\times$720, 1000 frames & A, R \\ \hline
\end{tabular}%
}
\\[2pt]
{\scriptsize $^\dag$: D = division, S = subtraction, M = multiplication, A = \omcrii{add}, R = reduction, C = copy}
}
\label{table:workload:properties}
\end{table}

\section{Evaluation}
\label{sec:eval}

\gfcriii{We demonstrate the advantages of \prop by evaluating 
\li~\prop's throughput for sixteen bit-serial and four bit-parallel \gls{PuD} operations at 4- to 32-bit precision, in comparison to three state-of-the-art \gls{PuD} systems, i.e., SIMDRAM~\cite{hajinazarsimdram}, MIMDRAM~\cite{mimdramextended}, and 
\emph{Proteus}~\cite{oliveira2025proteus} (\cref{sec:eval:throughput}); 
\lii~\prop's performance, energy consumption, and energy efficiency for twelve real-world applications, in comparison to the three \gls{PuD} systems, a multicore CPU, and a high-end GPU (\cref{sec:eval:real}); 
\liii~\prop's load balance and pipeline utilization across DRAM mats and DRAM chips (\cref{sec:eval:utilization}); and 
\liv~the \gls{SIMD}-width trade-off between \prop's pipelined execution and SIMDRAM's full-row execution for varying bit-precision and number of data elements (\cref{sec:eval:simdwidth}). 
Finally, we evaluate two key overheads in \prop: exposed carry-propagation latency (\cref{sec:eval:carry}) and area cost on top of a DRAM chip and CPU die (\cref{sec:eval:area}).}

\subsection{Throughput Analysis}
\label{sec:eval:throughput}

\begin{figure}[ht]
    \centering
    \includegraphics[width=\linewidth]{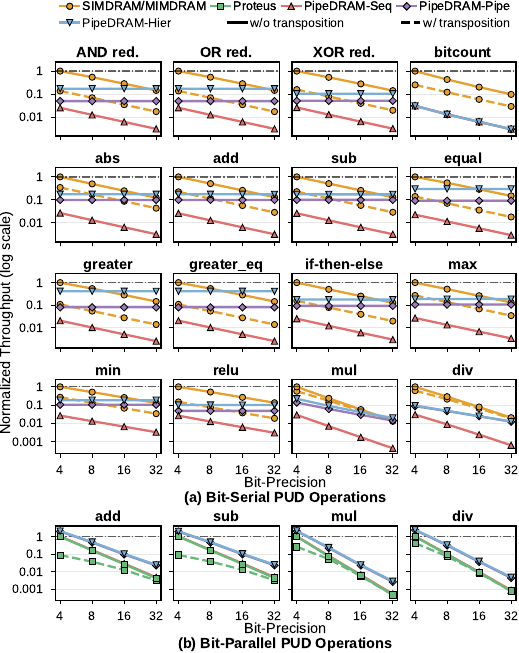}
    \caption{Normalized throughput for 
    (a)~\omcri{sixteen} bit-serial \gls{PuD} operations (norm.\ to SIMDRAM/MIMDRAM at 4-bit precision) and 
    (b)~\omcri{four} bit-parallel operations (norm.\ to \emph{Proteus} at 4-bit precision). Solid lines:
ideal throughput (no transposition); dashed: with transposition overhead.}
    \label{fig:combined_scalability}
\end{figure}

\paratitle{Bit-Serial \gls{PuD} Operations}
Fig.~\ref{fig:combined_scalability}(a) shows the normalized throughput of all 16 bit-serial \gls{PuD} operations for element sizes of 4, 8, 16,
and 32~bits.
We simulate 64M~elements \omcri{in} a single DRAM bank with 32~subarrays.
We compare SIMDRAM/MIMDRAM (with and without data transposition) against the three \prop configurations \omcri{(i.e., \prop-Seq, \prop-Pipe, and \prop-Hier, as described in \cref{sec:methodology})}.
All values are normalized to the throughput of SIMDRAM/MIMDRAM at 4-bit precision.
We make four observations. First, \prop's pipelined execution makes steady-state throughput \emph{independent of bit-precision}: each mat acts as a pipeline stage, and once full, a new batch enters at every initiation interval regardless of \omcri{the bit-precision} $N$. 
In contrast, SIMDRAM/MIMDRAM throughput scales logarithmically \omcri{(i.e., for AND/OR/XOR reduction)}, linearly \omcri{(i.e., for bitcount, abs, addition, subtraction, equal, greater, greater equal, if-then-else, max, min, ReLU)}, or quadratically \omcri{(i.e., for multiplication and division)} with \omcri{bit-}precision depending on the operation class.
Second, with data transposition, SIMDRAM/MIMDRAM's throughput drops 5$\times$ at every \omcri{bit-}precision; \prop-Hier exceeds SIMDRAM/MIMDRAM \omcri{for} throughput above 8 bits ($1.4\times$) and reaches $5.1\times$ \omcri{higher throughput} at 32 bits when data transposition overhead is \emph{realistically} accounted for. 
Third, despite operating on $32\times$ fewer SIMD lanes, \prop-Hier achieves $0.17\times$, $0.30\times$, $0.56\times$, and $1.06\times$ SIMDRAM/MIMDRAM throughput
(without transposition) at 4, 8, 16, and 32 bits, progressively closing the gap as precision increases. Fourth, \prop-Seq \omcri{provides} only
$0.025\times$ \omcri{of} SIMDRAM/MIMDRAM throughput due to the narrower mat-level SIMD width; pipelining recovers this from $40\times$ slower to competitive, confirming that both MABM~\omcri{(\cref{sec:mabm})} and pipelined execution~\omcri{(\cref{sec:pipeline})} are necessary for competitive \gls{PuD} throughput.
\omcri{We conclude that \prop-Hier provides 2.0$\times$ the throughput of SIMDRAM/MIMDRAM with data transposition, on average across all 16 bit-serial \gls{PuD} operations, and matches their transposition-free throughput at 32 bits (1.06$\times$) despite operating on $32\times$ fewer \gls{SIMD} lanes per \gls{PuD} primitive.}

\revdelm{First, \prop's pipelined execution \emph{fundamentally} changes how bit-serial \gls{PuD} throughput scales with bit-precision.
In traditional bit-serial systems (e.g., SIMDRAM/MIMDRAM), all $N$ bit-positions are processed sequentially within a single subarray; thus, the latency of each operation (and consequently its throughput) scales logarithmically (OR-/XOR-/AND-reduction), linearly (abs, addition, bitcount, max, min, ReLU, subtraction, if\_else, equal, greater, greater\_equal), or quadratically (multiplication, division) with the bit-precision.
In contrast, \prop makes steady-state throughput
\emph{independent of bit-precision}.
This is because each mat acts as a pipeline stage for one bit position, and once the pipeline is full, a new batch of elements enters the pipeline at every initiation interval~(II), regardless of how many bit-positions the operation spans. 
The steady-state throughput is determined solely by the per-bit-position latency (i.e., the time to process one pipeline stage's dependent \gls{PuD} primitives plus carry propagation), which does \emph{not} depend on~$N$. 
Second, when worst-case data transposition overhead is
accounted for, SIMDRAM/MIMDRAM's effective throughput drops by ${\sim}$5$\times$ at every precision (e.g., from 1.0$\times$ to 0.21$\times$ at 4~bits, from 0.10$\times$ to 0.02$\times$ at 32~bits).
\prop-Hier, provides higher throughput than SIMDRAM/MIMDRAM when accounting for data transposition overheads for bit-precisions larger than 8 (1.4$\times$) and reaches 5.1$\times$ SIMDRAM/MIMDRAM throughput at 32~bits. 
Third, despite operating on 32$\times$ fewer elements per batch (2,048 vs.\ 65,536 SIMD lanes\omcri{, since each \gls{PuD} primitive in \prop operates on 2,048 data elements, i.e., one 512-column mat in each of the four DRAM chips of a rank, whereas each \gls{PuD} primitive in SIMDRAM/MIMDRAM operates on the 65,536 data elements of a full DRAM row}), \prop-Hier progressively closes the SIMD-width gap with SIMDRAM/MIMDRAM as bit-precision increases.
Averaged across all 16 \gls{PuD} operations, \prop-Hier achieves 0.17$\times$, 0.30$\times$, 0.56$\times$, and 1.06$\times$ the throughput of SIMDRAM/MIMDRAM (without data transposition overhead) at 4, 8, 16, and 32~bits, respectively.
\revdelm{It is important to notice that, at bit-precisions below 32~bits, \prop underutilizes its compute resources, since MABM maps bit~$j$ of each element to a fixed mat, forcing an $N$-bit operation to use only $N$ out of 32~mats per chip; the remaining $32 - N$~mats are idle.
If all 32~mats could be utilized (e.g., by packing multiple independent lower-precision operations into the same pipeline), \prop-Hier would achieve ${\sim}$1.6$\times$ the throughput of
SIMDRAM/MIMDRAM at \emph{every} bit-precision.}
Fourth, \prop-Seq consistently achieves only ${\sim}$0.025$\times$ the throughput of SIMDRAM/MIMDRAM (without data transposition), averaged across all 16 operations and bit-precisions.
This is because \prop-Seq processes bits sequentially across mats, but operates on 32$\times$ fewer elements per batch due to the narrower mat-level SIMD width.
Pipelining is essential, since it increases \prop throughput from 40$\times$ slower than SIMDRAM/MIMDRAM to competitive or faster, depending on the bit-precision.
We conclude that both MABM (eliminating transposition) and pipelined execution (recovering the SIMD-width gap) are necessary to achieve competitive \gls{PuD} throughput at mat-level granularity.}

\paratitle{Bit-Parallel \gls{PuD} Operations}
Fig.~\ref{fig:combined_scalability}(b)
shows normalized
throughput for addition, subtraction, multiplication, and division,
normalized to \emph{Proteus} at 4 bits. We make three observations.
First, \prop-Hier outperforms \emph{Proteus} by $2.2\times$, $3.2\times$,
$4.0\times$, and $5.8\times$ (without transposition) at 4, 8, 16, and 32
bits, because \prop's modulo scheduling overlaps bit-independent
primitives across pipeline stages whereas \emph{Proteus} serializes all
primitives within each bit-position. \prop-Seq performs comparably to
\emph{Proteus} ($1.1\times$), confirming that the substrate change
alone provides only a marginal benefit. 
Second, with 
data transposition \omcri{taken into account}, \emph{Proteus}' throughput drops by 91\% for addition and subtraction at 4-bit precision, where the data transposition cost
dominates; \prop-Hier outperforms \emph{Proteus} with transposition by
$12.4\times$, $7.6\times$, $6.0\times$, and $6.8\times$ at 4, 8, 16,
and 32 bits (geometric mean across all four operations). Third, all
configurations degrade with bit-precision for bit-parallel operations due
to the $O(n \log n)$ parallel-prefix complexity~\omcri{\cite{oliveira2025proteus}}, but the pipelining
benefit grows with precision as more bit-positions create opportunities
for overlapping independent \gls{PuD} primitives.
\revdelmrev{We conclude that \prop's combination of MABM and modulo scheduling outperforms \emph{Proteus} by $2$--$6\times$ without transposition and $6$--$12\times$ with
transposition.}

\revdelm{shows the normalized throughput of the four bit-parallel \gls{PuD} operations (addition, subtraction, multiplication, and division) for element sizes of 4, 8, 16, and 32~bits.
We compare \emph{Proteus} (with and without worst-case data transposition) against the three \prop configurations, using a single DRAM bank with 32~subarrays.
All values are normalized to the throughput of \emph{Proteus} at 4-bit precision. We make three observations.

First, \prop-Hier consistently outperforms \emph{Proteus} across all bit-precisions.
Averaged across all four bit-serial operations, \prop-Hier achieves 2.2$\times$, 3.2$\times$, 4.0$\times$, and 5.8$\times$ the throughput of \emph{Proteus} (without data transposition) at 4, 8, 16, and 32~bits, respectively.
This is because both \emph{Proteus} and \prop execute the same parallel-prefix algorithm, but \prop's modulo scheduling overlaps independent \gls{PuD} primitives across pipeline stages, whereas
\emph{Proteus} serializes all operations within each bit-position.
\prop-Seq, which uses serial scheduling on mats,
performs comparably to \emph{Proteus} (${\sim}$1.1$\times$), confirming that the physical substrate change (mats vs.\ subarrays) alone provides only a marginal benefit.
Second, when worst-case data transposition is accounted for, \emph{Proteus}' throughput drops significantly for operations where in-DRAM compute is fast relative to the data transfer cost.
For addition and subtraction at 4-bit precision, transposition reduces \emph{Proteus}' throughput by 91\%, since the fixed transposition cost dominates the short in-DRAM computation.
At 32-bit precision, the transposition overhead is 26\% for addition and subtraction, and less than 2\% for multiplication and division, where the in-DRAM compute time dominates.
\prop-Hier outperforms \emph{Proteus} with transposition by 12.4$\times$, 7.6$\times$, 6.0$\times$, and 6.8$\times$ at 4, 8, 16, and 32~bits, respectively (geometric mean across all
four operations).
Third, unlike in the bit-serial case, all configurations (including \prop) exhibit throughput degradation with increasing bit-precision for bit-parallel operations. 
This is expected, since the Simplified Kogge--Stone parallel-prefix adder algorithm~\cite{efstathiou2025efficient} used for addition and multiplication has $O(n \log n)$ complexity, and the reduction tree for multiplication and the iterative structure of division scale superlinearly with bit-precision.
However, the pipelining benefit \emph{grows} with precision, since more bit positions create more opportunities for overlapping independent operations across pipeline stages.
We conclude that \prop's combination of MABM and modulo scheduling provides significant throughput benefits for bit-parallel \gls{PuD} operations, outperforming \emph{Proteus} by 2--6$\times$ without
transposition and by 6--12$\times$ when data transposition is realistically accounted for.}

\subsection{Real-World \omcri{Applications}}
\label{sec:eval:real}

\begin{figure}[ht]
    \centering
    \includegraphics[width=\linewidth]{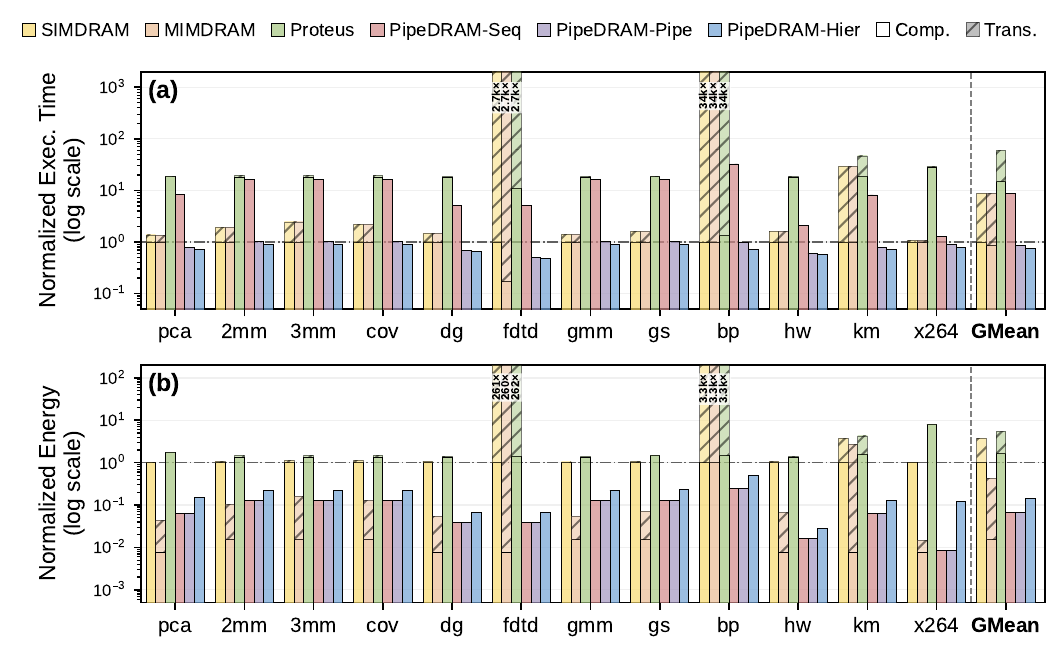}
    \caption{Normalized (a)~execution time and (b)~energy consumption for 12 real-world applications. Values are normalized to SIMDRAM \omcri{\emph{without}} data transposition overhead.}
    \label{fig:workload_normalized}
\end{figure}

\paratitle{Performance Analysis}
Fig.~\ref{fig:workload_normalized}(a) shows the normalized execution time of 12 real-world \omcri{applications}.
All values are normalized to SIMDRAM-ideal (compute-only, \emph{no} transposition overhead) using 32~subarrays per bank and 16~banks. 
We make three observations.
First, data transposition dominates execution time in SIMDRAM and MIMDRAM.
Across all 12 workloads, data transposition accounts for 8--100\% of total execution time (geometric mean \omcri{42.8\%}; 8.76$\times$ \omcri{of} SIMDRAM-ideal).
Two workloads (\texttt{fdtd} and \texttt{bp}) are extreme cases where transposition \omcri{overheads dominate computation time}, because both have very low compute intensity relative to the volume of data that must be transposed (\SI{38.6}{\giga\byte} and \SI{37.1}{\giga\byte}).
\prop completely eliminates this overhead through MABM.
Second, \prop-Hier \omcri{provides} 11.8$\times$ geometric mean speedup over SIMDRAM and MIMDRAM (with transposition) and 80.4$\times$
over \emph{Proteus} across all 12 workloads.
Even excluding the two transposition-dominated outliers, \prop-Hier \omcri{provides} 2.8$\times$ \omcri{higher throughput} over SIMDRAM/MIMDRAM and
27.8$\times$ \omcri{higher throughput} over \emph{Proteus}, demonstrating that the benefit extends beyond transposition elimination.
The speedup over \emph{Proteus} is substantially larger because \omcri{besides suffering from data transposition overheads (as SIMDRAM and MIMDRAM),} \emph{Proteus} \omcri{also} suffers from the absence of \omcri{any scheduling mechanism that can \emph{effectively} pipeline the execution of bit-independent \gls{PuD} primitives across DRAM subarrays during its bit-parallel execution}.
Third, pipelined scheduling is critical.
\prop-Seq (MABM only, no scheduling) \omcri{provides} a geometric mean normalized execution time of 8.62$\times$ \omcri{that of SIMDRAM-ideal}, comparable to SIMDRAM with transposition (8.76$\times$), because the 32$\times$ SIMD-width reduction offsets the transposition savings.
\prop-Pipe \omcri{and \prop-Hier} reduce \omcri{the execution time} to 0.85$\times$ and 0.74$\times$ \omcri{of} SIMDRAM-ideal, \omcri{respectively,} confirming that modulo scheduling and hierarchical data copy are essential to surpass SIMDRAM's compute-only performance despite the narrower SIMD width.
Overall, \prop-Hier is 11.6$\times$ faster than \prop-Seq and 1.1$\times$ faster than \prop-Pipe.
We conclude that \prop-Hier consistently outperforms all baselines in execution time, \omcri{providing} \omcri{11.77$\times$, 11.75$\times$, and 80.4$\times$ lower execution time over SIMDRAM, MIMDRAM, and \emph{Proteus}, respectively,} by combining transposition elimination with pipelined scheduling.

\paratitle{Energy Analysis}
Fig.~\ref{fig:workload_normalized}(b)~shows the normalized energy of the 12 real-world applications for the \omcri{three baselines} \gls{PuD} systems \omcri{and three \prop configurations}.
We make three observations.
First, \prop is the most energy-efficient \gls{PuD} substrate.
\prop-Pipe consumes only \omcri{0.066}$\times$ \omcri{of} the energy of SIMDRAM-ideal (geometric mean), reducing energy by 55.3$\times$ over SIMDRAM, 6.6$\times$ over MIMDRAM, and 82.8$\times$ over \emph{Proteus} \omcri{(with transposition in all three cases)}.
\omcri{Compared to SIMDRAM and \emph{Proteus}, besides eliminating energy consumed during data transposition, \prop energy savings stem} from mat-granularity activation\omcri{:} each \gls{PuD} primitive activates a single 512-bit mat rather than a full 65,536-bit subarray row, reducing activation energy by 128$\times$.
Since pipelining only overlaps existing operations without adding new activations, \prop-Seq and \prop-Pipe have identical energy.
Second, \prop maintains MIMDRAM's fine-grained activation benefits while eliminating transposition energy.
MIMDRAM with transposition consumes 0.43$\times$ SIMDRAM-ideal energy, already significantly lower than SIMDRAM (3.65$\times$) due to \omcri{MIMDRAM's} fine-grained subarray activation.
\prop-Pipe reduces energy further to 0.066$\times$ SIMDRAM-ideal, achieving 6.6$\times$ lower energy than MIMDRAM
(with transposition), because \prop inherits the same fine-grained activation at mat granularity while completely eliminating the transposition energy overhead.
Third, \prop-Hier increases energy by 2.2$\times$ compared to \prop-Pipe (geometric mean), consuming 0.14$\times$ \omcri{of the} SIMDRAM-ideal energy. 
This overhead arises because hierarchical data copy uses a dedicated communication subarray for each carry propagation, the
LISA inter-subarray transfers and relay operations require additional row activations that are absent in \prop-Pipe's direct inter-mat carry path.
Despite this overhead, \prop-Hier still reduces energy by 25.4$\times$ over SIMDRAM, 3.0$\times$ over MIMDRAM (with transposition), and 38.0$\times$ over \emph{Proteus}.
We conclude that \prop is the most energy-efficient \gls{PuD} substrate.

\begin{figure}[ht]
    \centering
    \includegraphics[width=\linewidth]{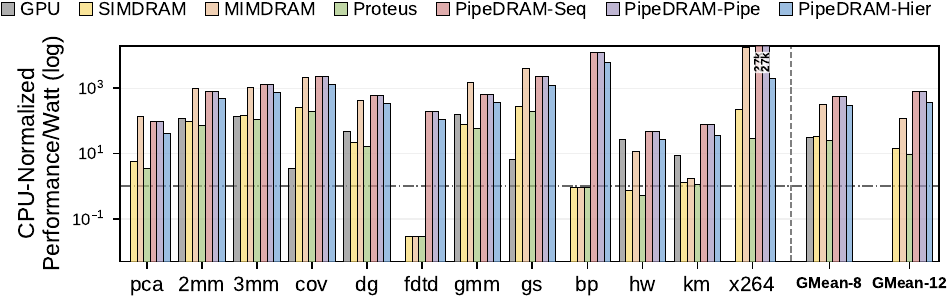}
    \caption{CPU-normalized \omcri{energy efficiency} (performance per Watt) for twelve real-world applications. Phoenix~\cite{yoo_iiswc2009} and SPEC2017~\cite{spec2017} do \emph{not} provide GPU implementations of \texttt{pca} and \texttt{x264}.}
    \label{fig:energy_efficiency}
\end{figure}

\paratitle{Energy Efficiency}
Fig.~\ref{fig:energy_efficiency} shows the CPU-normalized \omcri{\emph{performance per Watt}} across all 12 workloads. We include CPU and GPU results (obtained from~\cite{proteusgit}) for context.\gfcut{\footnote{\omcri{Our analysis of Fig.~\ref{fig:workload_normalized} uses SIMDRAM-ideal as a normalization point and raw execution time and energy as key metrics in order to isolate the impact of data transposition on \gls{PuD} execution. Our analysis of Fig.~\ref{fig:energy_efficiency}, in contrast, aims to compare \prop against both processor-centric and memory-centric systems. Hence, we employ \emph{performance per \omcri{Watt}} as our key metric \omcrii{for Fig.~\ref{fig:energy_efficiency} analysis}, since it simultaneously captures the impact of data movement on an application's performance and energy~\cite{mimdramextended}}.}}
We include GPU results for the \omcri{eight} workloads with available GPU measurements \omcri{(i.e., GMean-8 in the figure)}. We make three observations.
First, \prop-Hier achieves 356$\times$/11.7$\times$/25.4$\times$/3.0$\times$/37.9$\times$ the energy efficiency of the CPU, GPU, SIMDRAM, MIMDRAM, and \emph{Proteus}, respectively, on average across the 12 real-world applications.
\prop-Hier outperforms the GPU in energy efficiency in every
workload with available GPU measurements, with gains ranging from 1.0$\times$ (\texttt{hw}) to 363$\times$
(\texttt{cov}).
Second, data transposition overhead makes SIMDRAM and \emph{Proteus} \emph{less} energy-efficient than the CPU on three workloads (\texttt{fdtd}, \texttt{bp}, and
\texttt{hw}), and MIMDRAM on two (\texttt{fdtd} and \texttt{bp}), where the energy cost of data layout conversion exceeds the benefits from in-DRAM
computation.
\prop avoids this entirely through MABM, maintaining energy
efficiency above 27$\times$ that of the CPU across all 12 workloads.
Third, \prop-Seq and \prop-Pipe achieve identical energy efficiency (776$\times$), since pipelining overlaps existing
operations without introducing new DRAM activations.
\prop-Hier trades a 2.2$\times$ energy increase for its latency benefits due to the additional DRAM row activations required by the hierarchical inter-subarray carry propagation, but still achieves 356$\times$ higher performance per watt than the CPU.
We conclude that \prop is the most energy-efficient substrate when data transposition overheads \omcri{are considered}.

\subsection{\revCommon{Load Balance \& Pipeline Utilization}}
\label{sec:eval:utilization}

\begin{figure}[ht]
    \centering
    \includegraphics[width=0.91\linewidth]{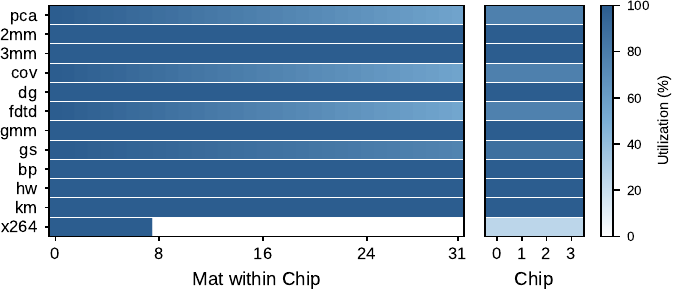}
    \caption{\revCommon{DRAM-mat (left) and DRAM chip (right) utilization of twelve applications during \prop's execution.}}
    \label{fig:mat_utilization}
\end{figure}

\revCommon{\changeCM{\#CQ3}Fig.~\ref{fig:mat_utilization} quantifies the \prop's utilization of each of the 32 mats within a DRAM chip (left) and of each of the four DRAM chips (right) for our twelve real-world applications. A mat's utilization is the fraction of time it is busy; a chip's utilization is the average \omcri{mat utilization across} 32 mats. We make three observations. 
First, \prop achieves high utilization at both mat and chip granularities. 
Averaged across our twelve workloads, intra-chip (mat) utilization is 87\% (93\% across the eleven 32-bit workloads), and the four chips are utilized identically, with inter-chip imbalance below 0.04\%. This happens as a consequence of our two mechanisms acting together: MABM distributes each operation's bit-positions across the 32 mats, \omcri{and \prop's modulo scheduler executes the bit-independent \gls{PuD} primitives of one bit-position in its mat while the carry-dependent primitives of the neighboring bit-position execute in the adjacent mat, so
that every mat stays busy.} 
Chip balance is structural, \omcri{since MABM places data element $i$ on DRAM chip $i \bmod 4$, each group of 2,048 consecutive data elements that one \gls{PuD} primitive operates on has exactly 512 data elements on each of the
four chips. 
Only the last such group of an operation whose element count is \emph{not} a multiple of 2,048 can have more data elements on one chip than on another, by \emph{at most} one data element per chip.} 
Second, the residual intra-mat load imbalance is caused by the deployed bit-serial algorithm, \emph{not} by \prop's pipeline structures. 
Each bit-serial operation has a characteristic per-bit-position work distribution:
\li~multiplication is the most uneven \gls{PuD} operation, since \omcri{a} result \omcri{at bit-position} $m$ accumulates every partial product $A_j\cdot B_k$ with $j{+}k{=}m$, so its per-bit work is maximal at the middle result bits and minimal at the least- and most-significant bits;
\lii~division is monotonic, since its low-order quotient bits operate on a wider intermediate remainder, so per-bit work decreases from the low-order to the high-order quotient bits;
\liii~addition/subtraction and the logic bitwise operations are nearly uniform. 
MABM reduces the multiplication imbalance at \emph{no} cost by assigning bit-positions~$j$
and~$j{+}32$ to the same mat, and because multiplication's work is symmetric \omcri{across} its
midpoint, these paired positions sum to nearly equal per-mat work across all 32 mats. We observe that the seven multiplication-dominated applications (\texttt{2mm},
\texttt{3mm}, \texttt{dg}, \texttt{gmm}, \texttt{bp}, \texttt{hw}, \texttt{km}) reach
100\% \omcri{mat and chip} utilization. 
In contrast, division has exactly 32 bit-positions, so no two bit-positions share a mat; the four division-heavy applications (\texttt{pca}, \texttt{cov}, \texttt{fdtd}, \texttt{gs}) consequently settle at
78--88\% mat utilization. Third, \texttt{x264}'s lower utilization is a property of its kernel, \emph{not} of \prop. 
Its operands are 8-bit, so the algorithm spans only 8 bit-positions and exercises 8 of the 32 mats, each at 100\% utilization. 
We conclude that MABM introduces \emph{no} structural imbalance across chips or mats; the only non-uniformity is the deployed bit-serial algorithm's intrinsic per-bit work, which \prop equalizes across mats.}

\subsection{\revD{\gls{SIMD}-Width Trade-Off Analysis}}
\label{sec:eval:simdwidth}

\begin{figure}[ht]
    \centering
    \includegraphics[width=0.92\linewidth]{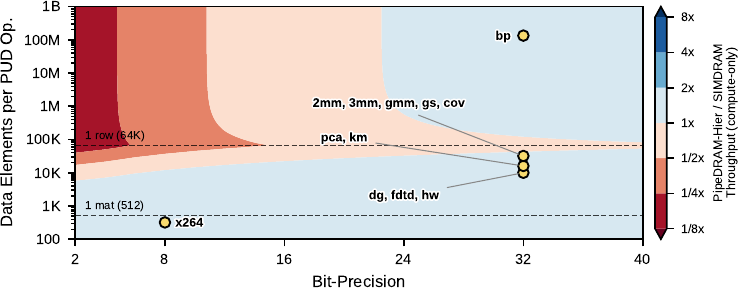}
    \caption{\revD{\prop vs.\ SIMDRAM (no transposition) throughput for varying bit-precision and \omcri{number of} data elements.}}
    \label{fig:precision_tradeoff}
\end{figure}

\revD{\changeD{\#D1}\prop reorganizes \gls{PuD} parallelism rather than reducing it. 
As Table~\ref{table:pudsummary} shows, \prop's peak \gls{SIMD} width is $2{,}048\times\#\text{bit-precision}$, since each bit-position of a \gls{PuD} operation occupies one mat (512 columns across four chips, i.e., 2,048 lanes). 
Therefore, two parameters set \prop's throughput: \li~\emph{bit-precision} fixes how many mats a single \gls{PuD} operation occupies, and thus its width; 
\lii~\emph{number of data elements} fixes how many 2,048-element batches iterate through those mats. 
A single batch occupies one mat at a time, and once enough batches are in flight to fill all \omcri{$\#\text{bit-precision}$} mats, \prop reaches its steady-state width of $2{,}048\times\#\text{bit-precision}$ (65,536 at 32~bits, matching SIMDRAM's row \omcri{granularity}). 
Which organization is faster, \prop's pipeline or SIMDRAM's full row, depends on \emph{both} the bit-precision and the number of data elements per \gls{PuD} operation. Fig.~\ref{fig:precision_tradeoff} quantifies this trade-off by showing \prop-Hier's compute-only throughput relative to SIMDRAM, with \emph{no} data transposition for either system, across both parameters, for multiplication (the dominant operation in seven of our twelve applications); markers show where the twelve applications fall. 
We make three observations. 
First, at large \omcri{operand} sizes, the bit-precision decides the trade-off. 
SIMDRAM's bit-serial latency grows quadratically with precision for multiplication, whereas \prop's steady-state throughput is precision-independent. 
Therefore, \prop-Hier \omcri{has the same performance as SIMDRAM} at 23~bits and is $1.4\times$ faster \omcri{than SIMDRAM} at 32~bits. 
Second, at small \omcri{operand} sizes, both systems are bound by operation latency and SIMDRAM's wide row goes underused, so \prop-Hier is faster at \emph{every} bit-precision (e.g., $1.5\times$ at 2,048 elements and 2~bits). 
SIMDRAM's coarser-granularity benefits appear only when bit-precision is \omcri{below 23~bits \emph{and} per-operation data sizes is above 6K--30K data elements} (the red region). 
\omcri{For a \emph{single} \gls{PuD} operation over $M < 65{,}536$ data elements, both SIMDRAM and \prop leave $1 - M/65{,}536$ of their available \gls{SIMD} lanes unused
(at 32~bits). 
However, differently than SIMDRAM that wastes $65{,}536 - M$ columns that hold no data element, in \prop, a DRAM mat is unused \emph{only} during pipeline fill and drain, and \prop's fine-grained mat activation lets the control unit issue an \emph{independent} \gls{PuD} operation to that mat in those cycles.}
Third, all twelve applications fall in the \prop-favored region \omcri{(even though we do \emph{not} include any data transposition overhead for SIMDRAM)}. 
The matrix-multiplication kernels (\texttt{2mm}, \texttt{3mm}, \texttt{gmm}; 32~bits, 32K~elements per operation) run $1.12\times$ faster than SIMDRAM even with its transposition overhead ignored, and \prop-Hier is $1.35\times$ faster, on average across all twelve applications (Fig.~\ref{fig:workload_normalized}). 
We conclude that \prop's pipelined organization of \gls{PuD} parallelism matches or exceeds full-row \gls{SIMD} at the precision and data sizes of real workloads, even when SIMDRAM's transposition overhead is ignored.} 

\subsection{\revE{Carry-Propagation Overhead Analysis}}
\label{sec:eval:carry}

\begin{figure}[ht]
    \centering
    \includegraphics[width=0.91\linewidth]{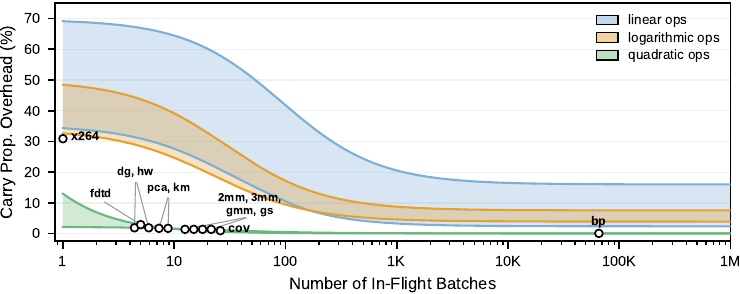}
    \caption{\revE{Exposed carry-propagation overhead of \prop-Hier \omcri{as a fraction of its execution time} vs.\ the number of in-flight batches~$B$.}}
    \label{fig:exposed_carry}
\end{figure}

\revE{\changeE{\#E1}\prop hides carry-propagation latency by overlapping it with 
\li~bit-independent \gls{PuD} primitives within \omcri{a given \gls{PuD}} operation and 
\lii~subsequent batches of elements entering the pipeline. 
Fig.~\ref{fig:exposed_carry} shows the carry-propagation overhead that remains exposed on the critical path as the number of in-flight batches~$B$ (i.e., the elements per \gls{PuD} operation divided by the
2,048-lane \gls{SIMD} width) varies, for the three \gls{PuD} operation scaling classes \omcrii{(i.e., linear, logarithmic, and quadratic)}; markers show our twelve applications at their measured~$B$ and dominant operation. We make three observations. First, the quadratic operations, which dominate the \gls{PuD} execution time of eleven of our twelve applications, are largely insensitive to carry propagation \omcri{latency} even in the $B{=}1$ worst case (2.2--13\% exposed), since each carry round-trip overlaps with the many bit-independent \gls{PuD} primitives that each mat executes per bit-position.
Second, carry-propagation overhead decays rapidly with $B$: once the pipeline is full, the only component that is \emph{never} hidden is the short per-batch \omcri{inter-subarray data movement operation that transfers the carry from the computing to the copying subarray} (\cref{sec:indramcopy}), which \omcri{our} hierarchical \omcri{in-DRAM data copy} scheme makes 15$\times$ cheaper than a direct inter-mat transfer.
\omcri{In steady state, the exposed carry-propagation time accounts for 2.8--16\% of \prop-Hier's execution time for linear operations, 7.5\% for logarithmic operations, and below 0.1\% for quadratic operations.}
Third, eleven applications operate at $B \geq 5$, where exposed carry \omcri{propagation overhead} is at most 2.9\% (\texttt{fdtd}); \texttt{x264} sits at the $B{=}1$ worst case (31\% \omcri{overhead}) yet still executes in 0.78$\times$ the time of SIMDRAM-ideal
(\cref{sec:eval:real}), since SIMDRAM utilizes fewer than 1\% of its 65,536 SIMD lanes at 320~elements per \gls{PuD} operation. 
We conclude that carry-propagation \omcri{latency} is effectively hidden for data-parallel workloads and that, even in the single-batch worst case, exposed carry \omcri{propagation latency} does \emph{not} negate \prop's performance benefits.}

\subsection{Area Analysis}
\label{sec:eval:area}
\paratitle{Area Analysis}
We use CACTI~7.0~\cite{cacti} to evaluate the area overhead of
\prop using a \SI{22}{\nano\meter} technology node, following the
methodology of prior \gls{PuD} works~\cite{mimdramextended, oliveira2025proteus, hajinazarsimdram}.

\textbf{DRAM Chip Area Overhead.}
\prop does \emph{not} introduce any \emph{new} modifications to the DRAM array circuitry.
It builds on three existing DRAM enhancements:
\li~the fine-grained mat access substrate of MIMDRAM~\cite{mimdramextended}, including mat isolation transistors, row decoder latches, mat selectors, and the inter-mat interconnect, which incurs \SI{1.11}{\percent} area overhead on a DRAM chip~\cite{mimdramextended};
\lii~the inter-subarray isolation transistors of LISA~\cite{chang2016low}, which incur \SI{0.6}{\percent} area overhead~\cite{chang2016low, oliveira2025proteus}; and
\liii~the SALP \omcri{circuitry}~\cite{kim2012case}, which \omcri{incurs} 0.15\% area overhead~\cite{kim2012case, oliveira2025proteus}.
In total, \prop incurs 1.86\% area overhead \omcri{in} a DRAM chip.

\textbf{MABM Area Overhead.}
The MABM mechanism (\cref{sec:mabm}) implements the deterministic bit permutation~$\Phi$ and its inverse~$\Phi^{-1}$ on the memory controller's cache line data path.
Because the permutation is a static rewiring of bit positions within each cache line, it is a \emph{fixed bit permutation} that requires \emph{no} logic gates: only wire routing.
To confirm this, we implemented both $\Phi$ and $\Phi^{-1}$ in Verilog~HDL and synthesized them using Yosys~\cite{Yosys}, an open-source synthesis framework, targeting the open-source Nangate \SI{45}{\nano\meter} standard cell library~\cite{nangate45}.
Both permutations synthesize to \emph{zero logic cells}; the mapping is implemented entirely through wire routing, incurring no \omcri{logic gate} area, latency, or power overhead.

\textbf{Memory Controller Area Overhead.}
\prop leverages MIMDRAM's existing control unit~\cite{mimdramextended} to
fetch, decode, and dispatch \bbop instructions.
\prop extends this control unit with a \emph{pipeline schedule table} that stores the precomputed static schedule (Algorithm~\ref{alg:pipedram_scheduler}) for each \uprog.
Each schedule entry maps a \gls{PuD} primitive to a (mat, cycle, stage) tuple, requiring \SI{4}{\byte} per entry.
For our 16 bit-serial operations at 32-bit precision, the largest schedule (division) contains 1,350 entries, requiring \SI{5.3}{\kilo\byte}.
We provision a \SI{16}{\kilo\byte} schedule scratchpad, sufficient
to hold the schedules for all evaluated operations simultaneously.
Using CACTI, we estimate the extended control unit area to be \SI{0.27}{\milli\meter\squared}, comparable to MIMDRAM's control unit (\SI{0.253}{\milli\meter\squared}~\cite{mimdramextended}).

\revdelm{\textbf{Data Transposition Unit Elimination.}
Unlike SIMDRAM~\cite{hajinazarsimdram}, MIMDRAM~\cite{mimdramextended}, and
\emph{Proteus}~\cite{oliveira2025proteus}, \prop does \emph{not} require a
data transposition unit.
Prior \gls{PuD} architectures require a data transposition unit (\SI{0.06}{\milli\meter\squared}~\cite{hajinazarsimdram, hajinazarsimdram, oliveira2025proteus}) in the
memory controller to convert between horizontal and vertical data
layouts at runtime.
\prop's MABM mechanism eliminates the need for runtime data layout
transformation entirely, removing this component from the system.}

\textbf{Total CPU Die Area Overhead.}
Considering the MABM logic (\omcri{negligible} area) and the extended control
unit (\SI{0.27}{\milli\meter\squared}), \prop's total memory controller-side area is \SI{0.27}{\milli\meter\squared}.
This represents 0.05\% of the die area of a state-of-the-art Intel Xeon E5-2697 v3 CPU~\cite{dualitycache}.
\revdelmrev{\prop's CPU-side area is \emph{smaller} than that of \omcri{SIMDRAM,} MIMDRAM\omcri{,} and \emph{Proteus} because it eliminates the \SI{0.06}{\milli\meter\squared} transposition unit that both prior works require.}

\revdelmrev{We conclude that \prop incurs low area cost on top of a DRAM chip (1.86\%) and CPU die (0.05\%), while eliminating the \omcri{data} transposition unit required by all prior \gls{PuD} architectures.}

\paratitle{\revCommon{Implementation Complexity \& Practical Deployability}} \changeCM{\#CQ1} \revCommon{\prop adds \emph{no} arithmetic logic to the DRAM array\omcri{. The cost
and complexity of such logic arise from building it} in density-optimized DRAM~\cite{devaux2019true,kim1996assessing}.
\omcri{\prop's} DRAM-side \omcri{modifications are limited to structural circuitry (i.e., isolation transistors, latches, and narrow interconnects) that reuses the existing sense amplifiers as the compute substrate}. These changes are modular, touching \emph{disjoint} subarray structures, so their area costs simply add (1.11\% + 0.6\% + 0.15\% = 1.86\%), with \emph{no} extra cost from combining them.
We make two key observations regarding \prop's practical deployability. 
First, \prop's area cost and complexity are well within previous area budgets reported by DRAM vendors for commodity DRAM chips~\cite{kwon202125,lee2021hardware,skhynixpim,he2020newton,kim20231} that include non-conventional on-die logic.
For example, 
\omcri{\li~JEDEC standardized per row activation counting (PRAC) in DDR5~\cite{jedec2024ddr5c}, which requires DRAM chips to keep an activation counter for every row and to increment and compare it against a
threshold with in-die logic; MINT~\cite{qureshi2024mint} reports the area
overhead of such per-row counters in an SK~hynix design at approximately 9\% of a DRAM chip; and 
\lii~SK~hynix fabricated} multiply-accumulate units directly in a DRAM chip in Newton~\cite{he2020newton} (productized as GDDR6-AiM~\cite{skhynixpim}), imposing  ${\sim}20\%$ area cost to DRAM (within the imposed 25\% area cost that SK~hynix defined as budget for in-DRAM compute~\cite{he2020newton}). 
Second, \prop's DRAM-side modifications also fit within the three-to-four metal layers in commodity DRAM~\cite{o2021energy, oconnor2017fine}), since the isolation transistors required for fine-grained activation can be rerouted within the already-present metal layers, as Half-DRAM~\cite{zhang2014half} demonstrates using only a wordline shift and a few metal vias~\cite{zhang2014half}. 
The metal-track budget is strained only when a mat's datapath is \emph{widened}~\cite{oconnor2017fine}; \prop instead computes at narrow mat granularity, with an HFF-width inter-mat carry path that adds \emph{no} coarse-pitch wiring tracks and inter-subarray transfers that reuse the existing master-data-line routing~\cite{o2021energy}. 
Therefore, \prop requires \emph{no} circuit of unprecedented complexity and is deployable within commodity DRAM's existing structure and metal budget. 
It is important to note that, if enabled, fine-grained DRAM~\omcri{\cite{cooper2010fine,udipi2010rethinking,zhang2014half,ha2016improving,lee2017partial,olgun2022sectored,o2021energy,oconnor2017fine, olgun2024sectored}}, SALP~\omcri{\cite{kim2012case}}, and LISA~\omcri{\cite{chang2016low}} would independently provide performance and energy benefits beyond assisting \prop, as they were originally proposed to improve the efficiency of commodity DRAM chips broadly. 
Our evaluation does \emph{not} include these independent benefits of the three schemes; so the cost of these structures that support \prop would actually be amortized further if they are enabled for general deployment.
}
\section{Related Work}


\gfcut{\omcri{To our knowledge, \prop is the first \omcrii{processing-using-DRAM (PUD)} architecture that executes bit-serial/bit-parallel \gls{PuD} operations directly over horizontally laid-out data \omcrii{without requiring \emph{any} data transposition overheads. 
\prop architecture has three major novel aspects:}
\li~the bit-positions of each data element within a cache line are interleaved across the DRAM mats of a single DRAM chip, one bit-position per mat, \emph{without} incurring runtime data transposition overheads; 
\lii~the mats of a DRAM chip operate as pipeline stages that process successive bit-positions concurrently under a statically computed \omcrii{iterative} modulo schedule; and
\liii~carry propagation between pipeline stages is decoupled from computation through an inter-subarray data copy path. 
We highlight \prop's key contributions by contrasting them with state-of-the-art \gls{PIM}
designs.}}

\paratitle{Processing-Using-DRAM} 
Prior works propose different ways of implementing bit-serial/-parallel \gls{PuD} operations (e.g.,~\cite{seshadri2017ambit,xin2020elp2im,deng2018dracc,gao2019computedram,angizi2019graphide,hajinazarsimdram,li2018scope,kim2019d,li2017drisa,zhou2022flexidram, mimdramextended, oliveira2025proteus}). 
Such works could benefit from \prop's key mechanisms to enable \gls{PuD} operations over horizontally laid-out data using their specific \gls{PuD} primitive.
\revdelmrev{Among bit-serial approaches, MIMDRAM~\cite{mimdramextended} introduces fine-grained mat-level access and inter-/intra-mat data movement to enable MIMD execution of data-independent \gls{PuD} operations across DRAM mats. 
\prop builds on MIMDRAM's fine-grained \gls{PuD} substrate but repurposes it for a \emph{fundamentally} different goal, i.e., rather than exploiting \emph{data-level parallelism} across \emph{multiple} independent \gls{PuD} operations, \prop exploits \emph{bit-level parallelism} within a \emph{single} \gls{PuD} operation by treating each mat as a pipeline stage for one bit-position of a bit-serial/bit-parallel computation. 
\emph{Proteus}~\cite{oliveira2025proteus} distributes the  bit-positions of a data word across multiple DRAM subarrays within a bank, leveraging LISA~\cite{chang2016low} to propagate intermediate (e.g., carry) values between subarrays and enabling \omcri{\emph{bit-parallel}} \gls{PuD} arithmetic. 
\prop exploits a similar form of bit-level parallelism but at a finer granularity (mats instead of subarrays) and deploys a modulo scheduling-based pipeline execution model that overlaps bit-independent \gls{PuD} primitives across pipeline stages, sustaining high throughput with simple statically-defined control logic. 
In contrast, \emph{Proteus} serializes all primitives within each bit-position and requires simultaneously operating multiple subarrays in a bank (constrained by the DRAM power-delivery limits, e.g., \texttt{tFAW}), without providing a scheduler that can keep high-throughput \gls{PuD} execution across bit-positions. 
Our evaluation shows that \prop outperforms both MIMDRAM (by 11.8$\times$) and \emph{Proteus} (by 80.4$\times$) in execution time, on average across twelve real-world applications, while consuming 25.4$\times$ and 38.0$\times$ less energy, respectively.}

\revdelmrev{\paratitle{NVM-Based \gls{PuM} Architectures} 
Several processing-using-NVM works~\omcri{\cite{truong2022adapting, truong2021racer,Gupta2018FELIXFA,leitersdorf2022partitionpim,leitersdorf2023aritpim,leitersdorf2024pypim,joshi2025lut,wong2026darth,seiler2026cross,truong2026memory}} exploit the concept of \emph{partitions} (i.e., dividing a memory array into smaller, independently operable sections) to accelerate in-memory computation.
FELIX~\cite{Gupta2018FELIXFA} segments ReRAM crossbar bitlines with transistor switches to enable multiple stateful logic gates to execute concurrently within the same row. 
PartitionPIM~\cite{leitersdorf2022partitionpim} formalizes this mechanism into serial, parallel, and semi-parallel execution modes, where semi-parallel mode enables inter-partition communication for carry propagation across bit-positions. 
AritPIM~\cite{leitersdorf2023aritpim} generalizes the partition concept into a technology-agnostic framework, showing that partitions can reduce bit-parallel addition execution time by assigning each partition to a different bit-position of the same operand. 
These approaches exploit partitions for spatial bit-level parallelism, but do \emph{not} temporally overlap the execution of successive operations across partitions: once an operation completes, the next begins from scratch with \emph{no} pipelined steady state.   
RACER~\cite{truong2022adapting, truong2021racer} goes further by organizing ReRAM tiles into a pipeline with bit-striped data placement and inter-tile ReRAM buffers that serve as pipeline registers, propagating micro-ops tile-to-tile via hardware queues. 
This is conceptually closest to \prop's pipelined execution across DRAM mats. However, RACER relies on fixed hardware micro-op queues rather than a software modulo scheduler, limiting its ability to exploit operation-specific parallelism (e.g., overlapping bit-independent primitives that do \emph{not} lie on the carry chain). 
A key distinction that separates \prop from the NVM-based \gls{PuM} architectures is that \prop targets commodity DRAM, which serves as the system's main memory. 
Because DRAM is shared between the host processor and \gls{PuD} execution, it \emph{must} preserve the horizontal data layout that modern processors rely on for cache line abstraction, memory interleaving, and high memory throughput. 
NVM-based \gls{PuM} systems \omcri{often} operate as dedicated accelerators with custom data layouts and do \emph{not} face this constraint. 
\revdelm{In contrast, \prop must reconcile efficient in-DRAM computation with the existing memory system interface, which is precisely the challenge that its MABM
mechanism and pipelined execution model address. 
By deterministically reorganizing bits within each cache line to distribute them across DRAM mats, \prop enables pipelined bit-serial/bit-parallel \gls{PuD} execution directly over horizontally laid-out data, without any runtime data transposition or format conversion, bridging the gap between the processor's data layout expectations and \gls{PuD}'s computational requirements at the main memory level. }}

\paratitle{\revA{Avoiding Dual-Data Layout in \gls{PnM} Systems}} \revA{\changeA{\#A1} DRAM-based \gls{PnM} systems \omcri{(e.g.,~\cite{upmem,gomez2022benchmarking,devaux2019true,zhao2024pim,seo2025facil})} often suffer from similar dual-data layout issues we identify for \gls{PuD} systems (\cref{sec_motivation}), where data words need to be localized within a DRAM chip (or DRAM bank) prior to computation using data transformation routines, leading to added performance overheads, programming, and system complexities~\omcri{\cite{upmem,gomez2022benchmarking,devaux2019true,zhao2024pim,seo2025facil,zhao2026gumpim}}. 
Prior works (e.g.,~\cite{zhao2024pim,seo2025facil}) aim to mitigate such a dual-data layout issue in \gls{PnM} systems either via 
\li~DIMM-based data relayout hardware units that filter out continuous cache line blocks of data before DRAM--CPU data transfer~\omcri{\cite{zhao2024pim,zhao2026gumpim}}, or 
\lii~specialized huge-page based data allocation routines that align \gls{PnM} data within a DRAM bank of a DRAM chip~\cite{seo2025facil}. 
Even though such solutions are effective \omcri{at} alleviating the dual-data layout \omcri{issue} in \gls{PnM} systems, they fall short when applied to \gls{PuD} systems for two main reasons. 
First, \gls{PuD} systems impose a more restrictive environment where data must be properly aligned across all levels of the DRAM hierarchy since computation happens \emph{in-situ}. 
This requirement means that \omcri{employing} flexible DRAM interleaving and data allocation routines (as in~\omcri{\cite{seo2025facil,zhao2024pim,zhao2026gumpim}}) \emph{cannot} fully mitigate the dual-data layout \omcri{issue} in \gls{PuD} systems, since neither the memory controller nor the \gls{OS} has enough information regarding internal DRAM organization (i.e., how the address maps to a DRAM subarray and DRAM mat~\omcri{\cite{kim2020revisiting, orosa2021deeper,yauglikcci2022understanding}}). 
\prop addresses this issue by leveraging MIMDRAM's data allocation routine, which combines huge pages and reverse-engineered DRAM address mapping to align pages further into the DRAM hierarchy (across subarrays and mats)~\cite{mimdramextended,oliveira2024puma}. Second, while \gls{PnM} systems operate under a conventional horizontal data layout (spanning a single DRAM row), \gls{PuD} systems operate over a vertical data layout (spanning multiple DRAM rows). This means that moving between \omcri{a} \gls{PuD}-friendly data \omcri{layout} and the CPU-friendly data \omcri{layout} \emph{fundamentally} requires reading \omcri{\emph{multiple DRAM rows}} to compose a single cache line. 
A DRAM-side data re-layout accelerator~\omcrii{(e.g.,~\cite{akin2015data,akin2015hamlet})} could potentially improve such a process, but it \emph{cannot} eliminate the data transposition overhead caused by multiple DRAM row activations. 
\prop \emph{fundamentally} eliminates the need for data transposition by employing a \gls{PuD}-friendly horizontal data layout using its \gls{MABM} interleaving mechanism \omcri{(\cref{sec:mabm})}.} 

\glsresetall

\section{Conclusion}
\label{sec:conclusion}

We introduce \prop, a \gls{PuD} architecture that eliminates the need for runtime data layout transformation, enabling bit-serial/bit-parallel \gls{PuD} execution \emph{directly} over horizontally laid-out data\omcri{\omcrii{, which} is friendly to and \omcrii{conventional} in compute-centric (CPU, GPU, TPU) systems}. 
\revdelmrev{\prop combines three  \omcri{new} key mechanisms: 
\li~a \emph{mat-aware bit mapping (MABM)} mechanism that \emph{deterministically} reorganizes bits within each cache line to enable \gls{PuD}-friendly data placement in a horizontal layout\gfcut{, without incurring any additional data movement, latency, or area \omcri{overheads}}; 
\lii~a \emph{software-assisted \omcri{pipelined} execution model} that statically schedules \gls{PuD} primitives across DRAM mats to overlap bit-dependent and bit-independent computation; and 
\liii~a \emph{hierarchical in-memory data copy scheme} that separates latency-critical carry propagation from non-critical intermediate data transfers, ensuring that inter-mat communication does \emph{not} dominate \gls{PuD} execution time.} 
Our \omcri{detailed} evaluation demonstrates that \prop significantly outperforms three state-of-the-art PUD systems in both performance and energy efficiency, while incurring low area cost on top of a DRAM chip and CPU die.
\revdelmrev{We conclude that \prop effectively bridges the gap between the horizontal data layout of modern memory systems and the vertical layout required by \gls{PuD} architectures, unlocking the full performance and energy efficiency potential of in-DRAM computation.
\omcri{To enable further research\omcrii{,} design\omcrii{, and adoption} of \gls{PuD} systems, we freely open-source the \prop infrastructure at \url{https://github.com/CMU-SAFARI/PipeDRAM}.}}

\section*{\gfcr{Acknowledgments}}

\gfcr{We thank the anonymous reviewers of MICRO 2026 for their encouraging feedback. 
We thank the SAFARI Research Group members for providing a stimulating intellectual environment. 
We acknowledge the generous gifts from our industrial partners, including Google, Huawei, Intel, and Microsoft. This work is supported in part by the ETH Future Computing Laboratory (EFCL), Semiconductor Research Corporation, AI Chip Center for Emerging Smart Systems (ACCESS), sponsored by InnoHK funding, Hong Kong SAR, and European Union’s Horizon programme for research and innovation [101047160 - BioPIM].}

\balance
{
  \bstctlcite{IEEEexample:BSTcontrol}
  \let\OLDthebibliography\thebibliography
  \renewcommand\thebibliography[1]{
    \OLDthebibliography{#1}
    \setlength{\parskip}{0pt}
    \setlength{\itemsep}{0pt}
  }
  \bibliographystyle{IEEEtran}
  \bibliography{refs}
}

\newpage
\nobalance
\appendix
\section{Artifact Appendix}

\subsection{Abstract}

This artifact contains the analytical \gls{PuD}, instrumented application workloads, immutable paper inputs, plotting programs, MABM RTL, and validation tools needed to reproduce \prop's evaluation results. 
It compares SIMDRAM, MIMDRAM, \emph{Proteus}, and three \prop configurations while preserving the submitted timing, energy, batching, and transposition methodology. We provide two workflows. 
\li~The \emph{quick} workflow regenerates and
validates all figures from frozen paper CSVs. \lii~The \emph{full} workflow rebuilds and executes every redistributable paper-scale workload before regenerating the figures. 
The artifact also supports one-command reproduction of an individual figure and a guided notebook with separate cells for every experiment. 

\subsection{Artifact Checklist (Meta-Information)}

{\footnotesize
\begin{itemize}[leftmargin=*,itemsep=1pt,topsep=2pt]
  \item \textbf{Algorithm:} Per-primitive PUD simulation, dependency-graph construction, modulo scheduling, \prop hierarchical carry transfer, MABM bit permutation.
  \item \textbf{Program:} Python, C/C++, Verilog, and shell scripts.
  \item \textbf{Compilation:} GCC/G++, GNU Make, Yosys, and Icarus Verilog.
  \item \textbf{Model:} Analytical DRAM latency/energy model with fixed
    submitted parameters; instrumented application-level PUD cost model.
  \item \textbf{Data sets:} Included frozen CSVs; deterministically generated
    kmeans input; optional checksum-verified heartwall video; frozen x264
    results without restricted SPEC source or media.
  \item \textbf{Run-time environment:} Ubuntu 24.04 x86-64, Python~3.12.
  \item \textbf{Hardware:} Quick mode requires a conventional x86-64 system.
    Full mode is recommended on a 32-thread host with 64\,GB RAM. No GPU is
    required.
  \item \textbf{Execution:} Automated quick, full, area-only, and
    per-figure workflows.
  \item \textbf{Metrics:} Latency, throughput, energy, normalized
    performance/W, transposition overhead, mat utilization, carry overhead,
    crossover point, and area.
  \item \textbf{Output:} One-page PDF figures, auxiliary PNGs, regenerated
    CSVs, JSON metrics, and textual validation results.
  \item \textbf{Experiments:} Figures~4 and~9--14, application workload
    evaluation, dataset verification, and area/RTL evaluation.
  \item \textbf{Disk space:} Approximately 1\,GB for setup and quick mode;
    reserve 10\,GB for full mode and generated inputs.
  \item \textbf{Preparation time:} Approximately 5--15 minutes after system
    packages are available.
  \item \textbf{Experiment time:} Quick mode normally completes in under two
    minutes. Full mode is hardware-dependent; allow up to 24 hours on an
    evaluator machine.
  \item \textbf{Available to evaluators:} Tokenized Zenodo preview; permanent
    record DOI: \url{https://doi.org/10.5281/zenodo.21530459}.
  \item \textbf{Code licenses:} MIT for project-authored code; retained
    third-party benchmark notices and licenses.
  \item \textbf{Data licenses:} Frozen numerical results are included.
    Restricted SPEC source/media and the heartwall video are excluded.
  \item \textbf{Workflow automation:} Shell scripts and a Jupyter notebook.
  \item \textbf{Archived:} DOI:
    \href{https://doi.org/10.5281/zenodo.21530459}
    {10.5281/zenodo.21530459}.
\end{itemize}
}

\subsection{Description}

\subsubsection{How to Access}

The artifact is archived under
\url{https://doi.org/10.5281/zenodo.21530459}.  Extract the ZIP and enter the \artifact{} directory. The directory is
independently distributable and does not read files from its parent
directory. The root \texttt{README.md} is the primary entry point.
\texttt{PipeDRAM\_AE.ipynb} provides a guided, figure-by-figure interface.

\subsubsection{Hardware Dependencies}

Quick reproduction and area verification require no specialized hardware.
A few CPU cores, 2\,GB RAM, and roughly 1\,GB free disk space are sufficient.
The full workflow uses paper-scale problem dimensions. We recommend 32
hardware threads, 64\,GB RAM, swap space, and 10\,GB free disk. Large
PolyBench allocations rely on the operating system's virtual-memory support. CPU RAPL and GPU measurements are immutable paper inputs and are not
remeasured because they depend on the original evaluation platforms.

\subsubsection{Software Dependencies}

The supported environment is Ubuntu~24.04 x86-64. Install:

\begin{lstlisting}
sudo apt update
sudo apt install -y build-essential python3-venv \
  texlive-extra-utils poppler-utils \
  fonts-liberation yosys iverilog
\end{lstlisting}

The artifact pins NumPy, Matplotlib, adjustText, SciPy, Pillow, and their
Python dependencies in \texttt{requirements.txt}. The setup script creates
an in-artifact virtual environment and checks GCC, G++, Make, PDF tools,
Yosys, Icarus Verilog, and \texttt{vvp}.









\subsection{Installation}

From \artifact{}:

\begin{lstlisting}
./scripts/setup.sh
\end{lstlisting}

Successful setup ends with \texttt{Artifact environment is ready}. No
administrator privileges are needed after the prerequisite packages have
been installed.

\subsection{Experiment Workflow}

\subsubsection{Quick Reproduction (Recommended First)}

\begin{lstlisting}
./scripts/reproduce.sh quick
./scripts/validate.sh quick
\end{lstlisting}

This regenerates Figures~4 and~9--14 plus the area comparison from immutable
paper inputs. It then checks numerical claims, PDFs, labels, rasterized
renderings, reference checksums, MABM RTL, and path isolation.

\subsubsection{Full Paper-scale Reproduction}

\begin{lstlisting}
./scripts/reproduce.sh full
./scripts/validate.sh full
\end{lstlisting}

When the verified heartwall video is available:

\begin{lstlisting}
./scripts/reproduce.sh full \
  --heartwall-input /path/to/test_4k.avi
./scripts/validate.sh full
\end{lstlisting}

Full mode rebuilds and runs 2mm, 3mm, covariance, doitgen, fdtd-apml,
gemm, gramschmidt, backprop, PCA, and kmeans. It also builds heartwall and
runs it when the verified video is supplied. x264 always uses its frozen
result. Generated files are placed under \texttt{generated/}; immutable
references are never overwritten.

\subsubsection{Individual Figures}

Each figure supports a lightweight replot and a full experiment path:

\begin{lstlisting}
./scripts/reproduce_figure.sh 10 replot
./scripts/reproduce_figure.sh 10 full
./scripts/reproduce_figure.sh area full
\end{lstlisting}

Figures~4, 10, and~11 consume application CSVs; their full mode regenerates
those results. Figures~9 and~12--14 execute their analytical simulator
experiments directly. The notebook caches full workload preparation so
multiple data-dependent figure cells do not repeat it.

\subsubsection{Area and RTL}

\begin{lstlisting}
./scripts/reproduce_area.sh
\end{lstlisting}

This verifies that $1.11+0.60+0.15=1.86\%$, confirms that the 1,350-entry
schedule requires 5,400 bytes and fits in 16\,KiB, synthesizes MABM to zero
logic cells using Yosys, and checks the permutation/inverse round trip using
Icarus Verilog.

\subsection{Evaluation and Expected Results}

A successful quick or full validation prints:

\begin{lstlisting}
Numerical, PDF, label, and rendering validation passed
Source and path isolation scan passed
quick validation passed
\end{lstlisting}

The final line reports \texttt{full validation passed} in full mode.
Expected outputs are summarized below. \texttt{expected\_metrics.json} contains plotted workload values,
geometric means, crossover points, mat utilization, and exposed-carry
fractions. 

\begin{table}[h]
\centering
\caption{Artifact outputs.}
\begin{tabular}{@{}cl@{}}
\toprule
Paper item & Output PDF \\
\midrule
Fig.~4 & \path{transposition_reuse_combined.pdf} \\
Fig.~9 & \path{combined_scalability.pdf} \\
Fig.~10 & \path{workload_normalized_combined.pdf} \\
Fig.~11 & \path{energy_efficiency_cpu_normalized.pdf} \\
Fig.~12 & \path{precision_datasize_tradeoff.pdf} \\
Fig.~13 & \path{mat_utilization.pdf} \\
Fig.~14 & \path{exposed_carry.pdf} \\
Area & \path{area_comparison.pdf} \\
\bottomrule
\end{tabular}
\end{table}

\subsection{Experiment Customization}

The notebook exposes:

\begin{lstlisting}
FIGURE_MODE = "replot"  # or "full"
HEARTWALL_INPUT = None
\end{lstlisting}

DRAM organization, timing, energy, precision, scheduling, transposition,
subarray, and bank parameters can be explored using the simulator CLI.
Customization is encouraged for sensitivity analysis, but changed parameters
are not expected to pass the paper-reference validator.

\subsection{Troubleshooting and Notes}

\begin{itemize}[leftmargin=*,itemsep=2pt]
  \item If setup reports a missing command, install the Ubuntu packages
    listed under software dependencies.
  \item A heartwall checksum failure means the video differs from the exact
    paper-scale input. Omit the argument to use the frozen result.
  \item A kmeans checksum failure indicates a missing, corrupt, or
    non-reference generated input. Remove it and rerun the generation script.
  \item Rendering mismatches usually indicate an unpinned Python environment
    or missing Liberation Sans fonts. Remove \texttt{.venv}, rerun setup, and
    reproduce again.
  \item Full workloads intentionally use paper-scale dimensions and may
    allocate several gigabytes. Quick mode is the recommended path on a
    constrained system.
  \item The package excludes development logs, caches, build products, and
    large generated inputs. Workflow scripts recreate them as needed.
\end{itemize}

\subsection{Methodology}

Artifact-review and badging methodology:
\begin{itemize}[leftmargin=*]
  \item \url{https://www.acm.org/publications/policies/artifact-review-and-badging-current}
  \item \url{https://ctuning.org/ae}
\end{itemize}

\end{document}